\documentclass[%
 reprint,
 amsmath,
 amssymb,
 aps,
 superscriptaddress,
]{revtex4-2}
\usepackage{amsmath}
\usepackage{amssymb}
\usepackage{amsfonts}
\usepackage{graphicx}
\usepackage{dcolumn}
\usepackage{mathtools,abraces}
\usepackage{bm}
\usepackage{hyphenat}
\usepackage[hidelinks]{hyperref}
\usepackage{physics}
\usepackage[capitalise]{cleveref}
\usepackage{nicefrac}
\usepackage[acronym]{glossaries}
\usepackage[binary-units]{siunitx}
\usepackage{color}
\usepackage{tikz}
\usepackage[caption=false]{subfig}
\usepackage{multirow}

\newcommand{\nodagger}{{\vphantom{\dagger}}}
\newcommand{\noprime}{{\vphantom{\prime}}}
\newcommand{\mathcomma}{\;\text{,}}
\newcommand{\mathperiod}{\;\text{.}}
\newcommand{\LandauO}[1]{\mathrm{O}\left(#1\right)}
\newcommand{\imagunit}{\mathsf i}
\newcommand{\enumber}{\mathsf e}
\DeclareMathOperator{\hc}{h.c.}
\DeclareMathOperator{\const}{const.}
\newcommand{\blackline}{\raisebox{2pt}{\tikz{\draw[-,black,solid](0,0) -- (5mm,0);}}}
\newcommand{\dashedline}{\raisebox{2pt}{\tikz{\draw[dashed,black](0,0) -- (5mm,0);}}}
\newcommand{\dottedline}{\raisebox{2pt}{\tikz{\draw[dotted,black](0,0) -- (5mm,0);}}}
\newcommand{\blueline}{\raisebox{2pt}{\tikz{\draw[blue,thick](0,0) -- (5mm,0);}}}
\DeclareSIUnit\permille{\text{\textperthousand}}
\newcommand{\qop}[1]{\hat{#1}}
\newcommand{\matrixid}{\qop 1}
\newcommand{\lldots}{\mathinner{{\ldotp}{\ldotp}}}

\newacronym{1d}{1D}{one\hyp dimensional}
\newacronym{2d}{2D}{two\hyp dimensional}
\newacronym{3d}{3D}{three\hyp dimensional}
\newacronym{tll}{TLL}{Tomonaga\hyp Luttinger liquid}
\newacronym{mf}{MF}{mean\hyp field}
\newacronym{ee}{EE}{energy expansion}
\newacronym{qmc}{QMC}{Quantum Monte Carlo}
\newacronym{sse}{SSE}{stochastic series expansion}
\newacronym{vqmc}{vQMC}{variational Quantum Monte Carlo}
\newacronym{pbc}{PBC}{periodic boundary condition}
\newacronym{obc}{OBC}{open boundary condition}
\newacronym[plural = MPSes, firstplural = matrix product states (MPS)]{mps}{MPS}{matrix product state}
\newacronym{mpo}{MPO}{matrix product operator}
\newacronym[longplural = {reduced density matrices}]{rdm}{RDM}{reduced density matrix}
\newacronym{dmrg}{DMRG}{density matrix renormalisation group}
\newacronym{tdvp}{TDVP}{time dependent variational principle}
\newacronym{tebd}{TEBD}{time evolving block decimation}
\newacronym{metts}{METTS}{minimally entangled typical thermal state}
\newacronym{ram}{RAM}{random access memory}
\newacronym{afm}{AFM}{antiferromagnet}
\newacronym{tn}{TN}{tensor network}
\newacronym{irrep}{irrep}{irreducible representation}
\newacronym{sso}{SSO}{single site operator}
\newacronym{hcb}{HCB}{hardcore boson}
\newacronym{svd}{SVD}{singular value decomposition}
\newacronym{qsl}{QSL}{quantum spin liquid}
\newacronym{dmft}{DMFT}{dynamical mean field theory}
\newacronym{chdmft}{chDMFT}{chain dynamical mean field theory}
\newacronym{peps}{PEPS}{projected entangled\hyp pair states}
\newacronym{afqmc}{AFQMC}{free auxiliary\hyp field Quantum Monte Carlo}
\newacronym{qr}{QR}{QR decomposition}
\newacronym{dimpy}{DIMPY}{(C$_7$H$_{10}$N)$_2$CuBr$_4$}
\newacronym{dtn}{DTN}{NiCl$_2$-$4$SC(NH$_2$)$_2$}

\begin{document}
\preprint{APS/123-QED}
\title{Ground states of quasi\hyp\acrlong{2d} correlated systems via \acrlong{ee}}
\newcommand{\hwaffiliation}{Institute of Photonics and Quantum Sciences, Heriot\hyp Watt University, Edinburgh EH14 4AS, United Kingdom}
\newcommand{\cnsraffiliation}{Laboratoire de Physique Th\'{e}orique, CNRS and Universit\'{e} de Toulouse, 31062 Toulouse, France}
\newcommand{\genevaaffiliation}{Department of Theoretical Physics, University of Geneva, Quai Ernest-Ansermet 24, 1211 Geneva, Switzerland}
\author{Sam Mardazad}
\email{s.mardazad@hw.ac.uk}
\affiliation{\hwaffiliation}
\author{Nicolas Laflorencie}
\email{nicolas.laflorencie@cnrs.fr}
\affiliation{\cnsraffiliation}
\author{Johannes Motruk}
\email{johannes.motruk@unige.ch}
\affiliation{\genevaaffiliation}
\author{Adrian Kantian}
\email{a.kantian@hw.ac.uk}
\affiliation{\hwaffiliation}
\date{\today}
\begin{abstract}
	We introduce a generic method for computing groundstates that is applicable to a wide range of spatially anisotropic \acrshort{2d} many-body quantum systems.
	By representing the \acrshort{2d} system using a low-energy \acrshort{1d} basis set, we obtain an effective \acrshort{1d} Hamiltonian that only has quasi-local interactions, at the price of a large local Hilbert space.
	%
	We apply our new method to three specific \acrshort{2d} systems of weakly coupled chains: \acrlongpl{hcb}, a spin-$1/2$ Heisenberg Hamiltonian, and spinful fermions with repulsive interactions.
	In particular, we showcase a non\hyp trivial application of the \acrlong{ee} framework, to the anisotropic triangular Heisenberg lattice, a highly challenging model related to \acrshort{2d} spin liquids.
	Treating lattices of unprecedented size, we provide evidence for the existence of a quasi\hyp\acrshort{1d} gapless spin liquid state in this system.
	We also demonstrate the \acrlong{ee}-framework to perform well where external validation is possible.
	For the fermionic benchmark in particular, we showcase the \acrlong{ee}-framework's ability to provide results of comparable quality at a small fraction of the resources required for previous computational efforts.
\end{abstract}
\maketitle
\section{\label{sec:intro}Introduction}
Quantum systems comprised of weakly coupled \gls{1d} subsystems (c.f.~\cref{fig:2d-chain}) represent an important area in the study of correlated many-body physics.
Not only are three different groups of repulsively\hyp mediated superconductors in this class (the Bechgaard and Fabré salts~\cite{Bourbonnais2007,Jerome2024}, the ``telephone\hyp number'' compounds~\cite{Nagata1998,Dagotto1999} and chromium pnictide~\cite{Bao2015,Watson2017}), but so are Cs$_2$CuCl$_4$~\cite{Coldea1996,Coldea2001,Coldea2002} and Cs$_2$CuBr$_4$~\cite{Tanaka2002,Ono2003}, paradigmatic materials in the ongoing search for \gls{2d} spin liquids~\cite{Alicea2006,Weng2006,Yunoki2006,Fjaerestad2007,Starykh2007,Hayashi2007,Pardini2008,Heidarian2009,Starykh2010,Cong2011,Weichselbaum2011,Reuther2011,Thesberg2014,PhysRevB.93.085111,Tutsch2019,Gonzalez2020,Szasz2021,Gonzalez2022,Yu2023}.
As for their more isotropic siblings (e.g. cuprates for repulsively mediated superconductivity~\cite{Anderson2002,Scalapino2012a,Stewart2017}, {$\kappa$-(ET)$_2$Cu$_2$(CN)$_3$} and EtMe$_3$Sb[Pd(dmit)$_2$]$_2$ for spin liquids~\cite{Shimizu2003,Itou2008,Itou2009}), progress on understanding the correlated phases of these materials at low temperatures has been famously slow, due to the outstanding difficulty of obtaining reliable theory for a great number of \gls{2d} correlated quantum systems.
\par
At the same time, the spatial anisotropy of such materials has long been recognized as a significant potential advantage over their isotropic cousins, namely due to the uniquely powerful theoretical methods for the \gls{1d} sub\hyp units that make up these materials.
The use of analytical \gls{tll} theory~\cite{BookGiamarchi2003} and \gls{mps} numerics~\cite{SCHOLLWOCK201196,Paeckel2019a} has yielded insights for \gls{1d} correlated systems that are unmatched for higher\hyp dimensional ones.
Thus, for anisotropic \gls{2d} and \gls{3d} models as well as for materials made up from weakly coupled \gls{1d} sub\hyp units, one strategy has been to leverage this unparalleled understanding of the \gls{1d} physics to the full, higher\hyp dimensional system~\cite{Schulz1996,Giamarchi1999,Sandvik1999,Dupont2018}.
A striking early example of this are the \gls{3d} model materials for repulsively\hyp mediated high\hyp $T_c$ superconductivity formed out of many doped two\hyp leg Hubbard ladders~\cite{BookGiamarchi2003,Karakonstantakis2011}.
Combining a \gls{tll}\hyp description of the isolated ladders with a static \gls{mf} treatment of the weak inter\hyp ladder tunneling yielded a conclusive description for the possibility of high\hyp $T_c$ superconductivity down to the level of repulsively mediated microscopic pairing, which remains an unrealized goal for isotropic \gls{2d} and \gls{3d} models to this day.
More recently, this line of work has been further expanded on with the \acrshort{mps}+\acrshort{mf} treatment of fermions, which allowed the first quantification of superconducting properties in these model materials, as well as resolving the competition with alternative insulating phases~\cite{Bollmark2023,Bollmark2025}.
This \acrshort{mps}+\acrshort{mf} framework for fermions had developed out of earlier \acrshort{mps}+\acrshort{mf} approaches for anisotropic systems of either spins or bosons, for which the low\hyp cost theoretical modeling of the magnetic BPCB compound had been a particularly noteworthy achievement~\cite{ruegg2008,Klanjsek2008,thielemann2009,Bouillot2011,Bollmark2020a}. 
This compound is one of numerous remarkable quasi-1D spin materials that exhibit non-trivial dimensional crossover effects, such as the $S=1/2$ ladder \gls{dimpy}~\cite{jeong2017,schmidiger2012,jeong2013,Furuya2016}, or the $S=1$ chain \gls{dtn} compound~\cite{mukhopadhyay2012,blinder2017}.
\par
Yet the treatment of anisotropic correlated \gls{2d} systems by \gls{mf}\hyp decoupling the interactions between the \gls{1d} sub\hyp units has significant limitations.
For all three existing classes of anisotropic repulsion\hyp mediated superconducting materials, as well as for the spin liquid candidate materials Cs$_2$CuCl$_4$ and Cs$_2$CuBr$_4$, such decoupling is either inadmissable or ill\hyp defined.
One approach to exploit spatial anisotropy for these systems beyond static \acrlong{mf} was \gls{chdmft}~\cite{Biermann2001,Berthod2006}.
It would treat a single \gls{1d} sub\hyp unit -- Hubbard chains with various repulsive short\hyp range interactions -- explicitly and would represent the rest of the system via a self\hyp consistent  hybridization function.
But due to the underlying \gls{qmc} solver, this approach battles the exponential scaling to solution with decreasing temperature known as the sign problem, and would not be useful for any of the near\hyp\gls{1d} anisotropic superconducting materials.
For spin systems such as those realized in Cs$_2$CuCl$_4$ and Cs$_2$CuBr$_4$, even setting up a \gls{chdmft} algorithm would be difficult due to the inherent geometric frustration.
\par
Brute\hyp force applications of \gls{mps}\hyp based numerics such as the \gls{dmrg} to compute the ground states of \gls{2d} systems are generally also severely limited, especially when spatial anisotropy is large~\cite{Weichselbaum2011}.
When these algorithms are confined to shared\hyp memory parallelism, typically only long, narrow strips are feasible, as entanglement entropy grows rapidly with the width of the strip~\cite{Eisert2010,Stoudenmire2012}.
Distributed\hyp memory parallelism as implemented in p\gls{dmrg} can treat significantly wider systems, especially when combined with exploitation of the spatial anisotropy~\cite{PhysRevB.100.075138}.
Nevertheless, even these high\hyp performance implementations will rapidly struggle once the width of the strip approaches a significant fraction of its length.
They also presuppose copious computational time and simultaneous access to many nodes of a high\hyp performance supercomputer with fast inter\hyp node connections.
\par
In addition to multiple recent and prolific numerical developments~\cite{Sandvik2010,lauchli2010numerical,Bauer_2011,GAENKO2017235,becca2017quantum}, there are two other classes of algorithms geared at \gls{2d} correlated systems at low or zero temperature that are more generic, in that they do not presuppose a spatial anisotropy, and that aim to reduce resource requirements at the price of introducing approximations.
These are \gls{peps}, which represents a class of variational ansatz states build on higher\hyp order tensor networks~\cite{Cirac2021}, and constrained\hyp path or phase\hyp \gls{afqmc}, which builds on the standard \gls{afqmc} for lattice fermions and purposefully removes the fermionic sign problem in different ways~\cite{He2019}.
While these algorithms can be powerful, the level of their approximations can individually depend on the specific physical system.
An approach such as \gls{peps}, which is generally limited to low bond dimensions will also generally struggle to retain enough entanglement to describe an anisotropic system accurately in the strong\hyp coupling direction.

What would be called for is a numerical technique that leverages the particular advantage inherent to spatially anisotropic problems, while also exhibiting only controlled approximation errors and being equally applicable to bosonic and fermionic systems.
\par
In the present work, we introduce a generic numerical technique to calculate the many\hyp body ground states of a variety of spatially anisotropic \gls{2d} quantum systems that are comprised of weakly coupled \gls{1d} sub\hyp units.
The essence of this approach is to represent the system in the low\hyp energy basis of the individual \gls{1d} sub\hyp units.
This representation yields a new effective \gls{1d} Hamiltonian with a large local Hilbert space, spanned by those low\hyp energy many\hyp body states of the \gls{1d} sub\hyp units that are being retained, as sketched in~\cref{fig:2d-chain}.
The error introduced by limiting the energy\hyp width of the \gls{1d} basis set is a controlled one, in that the approach starts from the lowest lying excitations and systematically incorporates more and more higher\hyp energy states;
if all states could be retained, the full Hilbert\hyp space of the \gls{2d} system would be recovered.
As long as the energetic widths of the retained \gls{1d}\hyp spectra exceed the  coupling range of the interactions between the \gls{1d} sub\hyp units, this approach is able to capture the physics of the 2d ground state well and with surprisingly low computational effort.
We demonstrate the algorithm's resulting ability to reach lattice sizes that would be infeasible with other controlled\hyp approximation many\hyp body numerics for both repulsively interacting fermions as well as frustrated spin systems.
We point out that as the present work was concluded, we have become aware of a similar approach for arrays of coupled~\gls{1d} continuum systems, but constrained to integrable models for which the~\gls{1d} basis is known analytically~\cite{James2013,James2018}.
\par
The paper is organised as follows.
In~\cref{sec:method} we outline the main technical components of this approach, and discuss the use of symmetries to speed up calculations, how observables are computed, and the main sources of error. 
In~\cref{sec:model-results}, we then show the concrete application of the \gls{ee}\hyp approach to obtain ground state properties for \gls{2d} Hamiltonians of spinless bosons, unfrustrated and frustrated spins, as well as repulsively interacting spinful fermions in a doped regime.
Where possible, here we also benchmark against alternative approaches in order to validate the \gls{ee}\hyp technique. 
Finally, in~\cref{sec:conclusions}, we summarize the main strengths and drawbacks of the \gls{ee}\hyp technique, and discuss the outlook for its application to other concrete systems, as well as for further improving its performance and widening its scope.
\section{\label{sec:method}Method}
\begin{figure}
	\centering
	\includegraphics[width=0.45\textwidth]{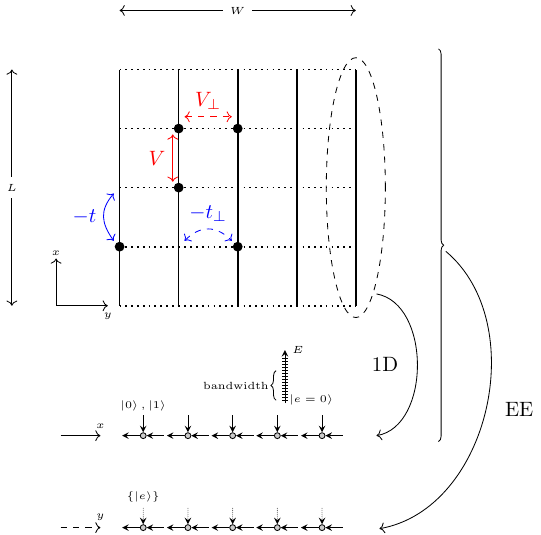}
	\caption{%
		\label{fig:2d-chain}%
		Weakly interacting chains where particles can hop with transition probability $t =1$ (\protect\blackline) or $t_\perp < t$ (\protect\dottedline).
		All models we investigate also exhibit nearest neighbour repulsion with strength $V$ along the chains, and some also contain such repulsion in the the perpendicular direction, of strength $V_\perp$.%
	}
\end{figure}

Without loss of generality, we can divide a \gls{2d} many\hyp body system into chains running along the $x$\hyp axis, as shown e.g. in~\cref{fig:2d-chain}.
The Hamiltonian is split into terms which act only within the chains, labled by the $y$\hyp coordinate, which are subsumed into $\qop H_1$.
The residual terms connecting different chains are collected in $\qop H_\perp$ respectively as also visualised in \cref{fig:2d-chain}, i.e.
\begin{align}
	\label{eq:hamiltonian-splitting}
	\qop H &= t \qop H_1 + t_\perp \qop H_\perp \mathperiod
\end{align}
We assume that the magnitude of the matrix elements of $\qop H_1$ and $\qop H_\perp$ are of order 1.
\Gls{ee} is applicable when the interchain coupling in \cref{eq:hamiltonian-splitting} is small $t_\perp/t \ll 1$.
Since $\qop H_1$ is purely \gls{1d} we can employ \gls{dmrg} to efficiently compute its low\hyp energy subspace~\cite{SCHOLLWOCK201196}.
We split it into the terms for the individual chains:
\begin{align}
	\label{eq:H1-decomp}
	\qop H_1
	=
	\sum_{y = 1}^W
	\left(
		\matrixid_1
		\otimes
		\matrixid_2
		\otimes
		\ldots
		\otimes
		\qop h_y
		\otimes
		\ldots
		\otimes
		\matrixid_W
	\right)
	\mathperiod
\end{align}
We generally suppress index $y$ when focusing on a single chain.
Our overall approach is to re\hyp express the total Hamiltonian~\cref{eq:hamiltonian-splitting}, using the low\hyp lying spectrum of the chains:
\begin{align}
	\label{eq:1d-eigenvalue}
	\qop h \ket{e} &= E_{e} \ket{e} \mathperiod
\end{align}
Here, the index $e$ is counting the excitations and $e = 0$ denotes the ground state.
We obtain this spectrum  sequentially, running a series of \gls{dmrg}\hyp based ground state searches, that incorporate orthogonality constraints for all previously found eigenstates (c.f.~\cref{appendix-subsec:1d-eigensystem}).
In general, the state $\ket{e=0}$ has entanglement entropy scaling with surface area~\cite{Eisert2010}, however,
most of the subunits we look at are gapless and thus have a log correction to the area law~\cite{Vidal2003,Calabrese2004}.
This makes the \gls{mps} ansatz efficient, but we also find the same to hold true for the low\hyp lying eigenstates.
In practice, we can go up to $\numrange[range-phrase=\text{ to }]{60}{100}$ energy eigenstates in the most relevant quantum number sectors, keeping less than $\num{20}$ states in other sectors.
On a further practical note, we remark that for the relatively short chains with \glspl{pbc} for which all \gls{1d} basis sets in the work have been obtained, a particularly efficient way to perform the calculations, is to implement the boundary condition with a single long-range coupling term from the first to the last site of the chain.
Though this ``naive'' implementation of \glspl{pbc} substantially boosts entanglement entropy and thus computational cost to obtain eigenstates, the increase of these costs for chains of modest length can be handled by any efficient, modern implementation of \gls{mps}\hyp based eigensolvers.
Crucially, the increase is more than offset by the very significant boost in stability and well\hyp behaved convergence as one works ones way up the ladder of excited states, as the greatly increased entanglement entropy of the states turns out to be a key accelerant in avoiding false minima during the constrained eigenstate search.

Once the \gls{1d} subspace is obtained, we span a new \gls{2d} many\hyp body basis using the tensor product
\begin{align}
	\ket{\bm e} = \ket{e_1,\ldots,e_W} \mathcomma
\end{align}
where $W$ denotes the number of chains, arrayed in $y$\hyp direction, as shown in~\cref{fig:2d-chain}.
We now express~\cref{eq:hamiltonian-splitting} in terms of the \gls{1d} eigenstates by computing the matrix elements
\begin{align}
	\mel
	{
		\bm e^\noprime
	}
	{
		\qop H
	}
	{
		\bm e^\prime
	}
	&= t
	\mel
	{
		e^\noprime_W\ldots e^\noprime_1
	}
	{
		\qop H_1
	}
	{
		e^\prime_1\ldots e^\prime_W
	}
	+ t_\perp
	\mel
	{
		\bm e^\noprime
	}
	{
		\qop H_\perp
	}
	{
		\bm e^\prime
	}
	\\
	\label{eq:2d-ee-generic}
	&= t
	\sum_{y^\prime}
	\left(
		\prod_
		{
			y^\noprime = 1
		}
		^W
		\delta_
		{
			e^\noprime_{y^\noprime}
			,
			e^\prime_{y^\noprime}
		}
		E_{e^\noprime_{y^\prime}}
	\right)
	+ t_\perp
	\mel
	{
		\bm e^\noprime
	}
	{
		\qop H_\perp
	}
	{
		\bm e^\prime
	}
	\mathcomma
\end{align}
using the eigenvector properties of the single\hyp chain basis $\ket{e}$.
The first term in~\cref{eq:2d-ee-generic} is a sum of identities on the subspace of the chosen energy eigenbasis of the chains. 
For each chain, which we will interchangeably denote a ``supersite'' in the following whenever represented by its low-energy basis, these identities are weighted with the energies of each basis state
\begin{align}
	\label{eq:ee-diag-part}
	\qop E_y
	&=
	\matrixid_1
	\otimes
	\ldots
	\otimes
	\sum_{e_y}
	E_{e_y}
	\outerproduct
	{
		e_y
	}
	{
		e_y
	}
	\otimes
	\ldots
	\otimes
	\matrixid_W
	\mathcomma
\end{align}
thereby effectively diagonalizing $\hat{H}_1$ in the \gls{1d} low-energy subspace.
The nature of the second term in~\cref{eq:2d-ee-generic} depends on the specifics of the perpendicular part of the model and will thus be discussed in~\cref{sec:model-results} for each of the models individually.
Note that if all eigenstates of the chains are retained, the basis transformation would be exact.
\par
\begin{figure}
	\centering
	\includegraphics[width=0.5\textwidth]{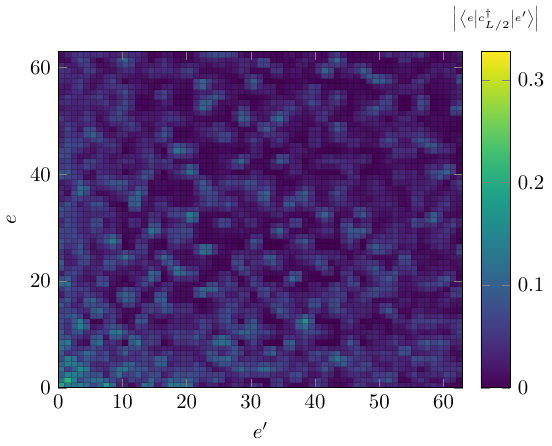}
	\caption{%
		\label{fig:creator-matrixelem}%
		Matrix elements of the creation operator in a \gls{hcb} system of size $L = 16$ and nearest neighbour repulsion $V = 2$ going from $N = 8$ to $N = 9$ particles in the center of the chain.%
	}
\end{figure}
Generally, this work focuses on interactions between chains that can be written as
\begin{align}
	&\mel
	{
		\bm e^\noprime
	}
	{
		\qop H_\perp
	}
	{
		\bm e^\prime
	}
	=
	\sum_{xy}
	\mel
	{
		e^\noprime_W\ldots e^\noprime_1
	}
	{
		\qop A_{x^\noprime y}
		\
		\qop B_{x^\prime y+1}
	}
	{
		e^\prime_1\ldots e^\prime_W
	}
	\label{eq:matrix_el}
	\\
	&=
	\sum_{xy}
	\left(
		\delta_
		{
			e^\noprime_1,e^\prime_1
		}
		\ldots
		\mel
		{
			e^\noprime
		}
		{
			\qop A_{x^\noprime}
		}
		{
			e^\prime
		}
		_y
		\mel
		{
			e^\noprime
		}
		{
			\qop B_{x^\prime}
		}
		{
			e^\prime
		}
		_{y+1}
		\ldots
		\delta_
		{
			e^\noprime_W,e^\prime_W
		}
		\label{eq:matrix_el_supsi}
	\right)
	\mathcomma
\end{align}
where $\qop A_{xy}$ and $\qop B_{xy}$ denote operators acting locally around site $x$ of chain $y$.
For the purposes of the present work, we assume every chain to have the same Hamiltonian and all couplings between neighbouring chains to be the same.
We thus shift the $y$\hyp index from the operators and basis states onto the overlaps, as already done when going from~\cref{eq:matrix_el} to~\cref{eq:matrix_el_supsi}.
However, the \gls{ee}\hyp framework does not depend on this assumption and it would be straightforwardly generalised to heterogeneous systems.
\par
Heuristically, we find amplitudes $\mel{e^\noprime}{\qop A_x}{e^\prime}_y$ to be maximally of order $\LandauO{\num{e-1}}$ for all studied systems.
The actual values and magnitudes depend strongly on $|{e^\noprime}\rangle$ and $|{e^\prime}\rangle$, as exemplified in~\cref{fig:creator-matrixelem} for the creation operator in a \gls{hcb} system.
The main weight is accumulated in the bottom left corner, with clusters of transitions betweeen the energetically higher excited states appearing as well.
\par
Now that the full Hamiltonian~\cref{eq:2d-ee-generic} is expressed in terms of single\hyp chain quantities that we can access, we can pick a method of choice for calculating the ground state of the resulting \gls{2d} Hamiltonian.
We chose to employ \gls{dmrg} again, which yields a Hamiltonian with nearest\hyp neighbour coupling at most.
Different to many other applications of the \gls{dmrg}, the \gls{ee}\hyp framework has to wrestle with the large local Hilbert space $|{e^\noprime}\rangle$ that we retain for each chain.
We have found a sensible cut off to be $d = 264$ for the bosonic or spin systems and $d = 384$ for the fermionic systems, which will be discussed further in~\cref{sec:model-results,subsec:error-sources}.
Thus, for the ability to perform actual calculations, the exploitation of conserved quantum numbers becomes even more essential than in most other applications of the \gls{dmrg}.
\subsection{Symmetries}
\label{subsec:syms}
All studied models exhibit $U(1)$ symmetries.
The bosonic and spin models preserve particle number or $z$\hyp spin, respectively, while the fermionc model conserves both these quantum numbers simultaneously.
This applies to Hamiltonian~\cref{eq:hamiltonian-splitting} as well as the individual chain Hamiltonians $\hat{h}_y$.
\par
Therefore, we can make use of the usual decomposition of the \acrlong{tn} entries into symmetry protected blocks when obtaining the \gls{1d} basis set~\cite{SCHOLLWOCK201196,ediss21348,PhysRevA.82.050301,PhysRevB.83.115125}.
This allows us to perform the orthogonalisation procedure described in~\cref{eq:orthogonality-condition} for states of the same sector only.
It also eliminates many redundant entries in the \gls{ee} Hamiltonian by restricting the number of elements to compute, as
\begin{align}
	\mel
	{
		e
	}
	{
		\qop A
	}
	{
		e^\prime
	}
	&=
	\mel
	{
		e
	}
	{
		\qop A
	}
	{
		e^\prime
	}
	\delta_{\hat \sigma(e^\prime),\hat \sigma(e^\noprime)+T}
	\mathcomma
\end{align}
with $\qop A$ an arbitrary single\hyp chain operator and $T$ its transformation behaviour (i.e. the change in quantum number(s) effected by $\qop A$).
Furthermore, $\hat\sigma$ denotes an operator mapping all quantum numbers of the respective energy eigenstate to their scalar eigenvalues $\sigma$, e.g.
\begin{align}
	\hat\sigma(e)
	=
	(\{\sigma\})
	=
	(n_e,z_e)
	\mathperiod
\end{align}
In this example, $\left\{\sigma\right\}$ denotes particle number $n_e$ and the $z$\hyp spin component $z_e$ of the energy eigenstate $e$.
For the overlaps of creation and annihilation operators it is $T = +1$ and $T = -1$, respectively, in the particle number sector, while for particle number operators $T = 0$.
In practice, this means that any symmetry protected \gls{dmrg} code can be repurposed for the \gls{ee} framework after adjusting the \acrlongpl{irrep} for the local basis and reading in the new operator weights generated from the \gls{1d} basis set.
\subsection{Measurements}
\label{subsec:measurements}
In order to extract quantities beyond the energy and entanglement entropies from the \gls{ee}\hyp\gls{dmrg} wavefunction, we need to establish a method to compute observables.
However, working with the supersite tensor network, we have to be deliberate in defining our local operators to perform correct measurements.
Matrix elements which act on one chain only, i.e. which have the same $y$\hyp component for all involved operators can be written as
\begin{align}
	\label{eq:single-supersite-obs}
	\mel{\psi}{\qop O_{\bm x,y}}{\psi} &=
	\sum_{\sigma_y e^\noprime_y e^\prime_y} M^{\sigma_y e^\noprime_y\dagger} \cdot M^{\sigma_y e^\prime_y} \mel{\sigma e^\noprime}{\qop O_{\bm x}}{\sigma e^\prime}_y
	\mathperiod
\end{align}
\par
Note that the notation $\bm x$ implies the operator acting on any number of chain sites, i.e. $\qop O_{\bm x,y}$ might be a particle number operator $\qop n_{x,y}$ or the entry of a correlator $c^\dagger_{x^\noprime,y} c^\nodagger_{x^\prime,y}$ on any chain sites $x^\noprime$ and $x^\prime$.
However, we have to measure the full operator on the \gls{1d} system and cannot create it from more elementary operators, e.g.
\begin{align}
	\notag
	\mel{\sigma e^\noprime}{\qop n_x}{\sigma e^\prime}
	&=
	\mel{\sigma e^\noprime}{\qop c^\dagger_x \qop c^\nodagger_x}{\sigma e^\prime}
	\\
	&\neq
	\sum_{e^{\prime\prime} = ^\noprime 0}^{\tilde{d}-1}
	\mel{\sigma e^\noprime}{\qop c^\dagger_{x^\noprime}}{\sigma^{\prime\prime} e^{\prime\prime}}
	\mel{\sigma^{\prime\prime} e^{\prime\prime}}{\qop c^\nodagger_x}{\sigma e^\prime}
	\mathcomma
\end{align}
since we usually truncate our system in \gls{1d} excitations $\tilde d < d$, as is described in~\cref{subsec:error-sources}.
Furthermore, $M^{n_ye^\noprime_y}$ is the tensor on supersite $y$ for the $e_y$\hyp th \gls{1d} state in sector(s) $\sigma_y$.
\par
We made use of the fact that the \gls{mps} can be canonically normalised and therefore overlaps of local \gls{1d} supersites are identities for all but supersite $y$.
In evaluating~\cref{eq:single-supersite-obs}, we also exploit that our symmetry protected framework does not allow for changes in the global quantum numbers when computing observables.
The quantity $M^{\sigma_y e^\noprime_y \dagger} M^{\sigma_y e^\prime_y}$ that appears in~\cref{eq:single-supersite-obs} is the \gls{rdm} of chain $y$.
It contains all the information/probabilities for this chain extractable from the 2D wavefunction in the \gls{ee} basis. 
The matrix elements are the individual weights of the \gls{1d} modes.
Therefore, we only need to measure the transition elements $\mel{\sigma e^\noprime}{\qop O_{x}}{\sigma e^\prime}$ permitted by symmetry in \gls{1d} and can proceed with them as ``local operators''.
Then measurements for supersite operators acting on single\hyp chain Hilbert spaces can be executed in the ordinary way.
\par
We also study observables acting on two chains, but the procedure can be generalised to more sites with some additional bookkeeping.
Assuming an observable $\qop O_{\bm x,y^\noprime y^\prime}$ acting on two chains $y^\noprime$ and $y^\prime$ and an arbitrary number of sites within the chains $\bm x$ we can write
\begin{align}
	\label{eq:two-supersite-obs}
	&\mel
	{
		\psi
	}
	{
		\qop O_{\bm x,y^\noprime y^\prime}
	}
	{
		\psi
	}
	=
	\sum_
	{
		\substack
		{
			i_{y^\noprime} i_{y^\prime}
			\\
			j_{y^\noprime} j_{y^\prime}
		}
	}
	\rho^
	{
		j_{y^\noprime} j_{y^\prime}
	}
	_
	{
		i_{y^\noprime} i_{y^\prime}
	}
	\mel
	{
		i_{y^\noprime} i_{y^\prime}
	}
	{
		\qop O_{\bm x, y^\noprime y^\prime}}
	{
		j_{y^\noprime} j_{y^\prime}
	}
	\\
	&
	\rho^
	{
		j_{y^\noprime} j_{y^\prime}
	}
	_
	{
		i_{y^\noprime} i_{y^\prime}
	}
	=
	\notag
	\sum_
	{
		i_{y^\noprime+1}\ldots i_{y^\prime-1}
	}
	\left(
		M^
		{
			i_{y^\prime} \dagger
		}
		M^
		{
			i_{y^\prime-1} \dagger
		}
		\ldots
		M^
		{
			i_{y^\noprime+1} \dagger
		}
		M^
		{
			i_{y^\noprime} \dagger
		}
		\right.
	\\
	&
	\qquad\qquad\qquad\qquad\qquad
	\cdot
	\left.
		M^
		{
			j_{y^\noprime}
		}
		M^
		{
			i_{y^\noprime+1}
		}
		\ldots
		M^
		{
			i_{y^\prime-1}
		}
		M^
		{
			j_{y^\prime}
		}
	\right)
	\mathcomma
\end{align}
where have defined the two chain density matrix $\rho$ and assumed $y^\noprime < y^\prime$ without losing generality.
Note that we introduced the abbreviation $i_{y^\noprime} = (\sigma^\noprime e^\noprime)_{y^\noprime}$, as the quantum numbers of one of the two chains might now change along with an adjoint change $j_{y^\prime} = (\sigma^\prime e^\prime)_{y^\prime}$ on the other chain.
At the same time, the two\hyp chain density\hyp matrix $\rho$ is never explicitly computed, as its size will scale as $\tilde d^4$.
Generally, $\rho$ it is a non\hyp local quantity as one needs to contract all tensors in\hyp between the chains $y^\noprime$, $y^\prime$.
However, many observables of interest factorise in the different chain\hyp Hilbert spaces and therefore we can write
\begin{align}
	\mel
	{
		i_{y^\noprime} i_{y^\prime}
	}
	{
		\qop O_{\bm x, y^\noprime y^\prime}
	}
	{
		j_{y^\noprime} j_{y^\prime}
	}
	&=
	\mel
	{
		i_{y^\noprime} i_{y^\prime}
	}
	{
		\qop O_{\bm x, y^\noprime}
		\qop O_{\bm x, y^\prime}
	}
	{
		j_{y^\noprime} j_{y^\prime}
	}
	\\
	\label{eq:two-chain-ee-matrixelem}
	&=
	\mel
	{
		\sigma^\noprime e^\noprime
	}
	{
		\qop O_{\bm x}}{\sigma^\prime e^\prime
	}_y
	\mel
	{
		\sigma^\noprime e^\noprime
	}
	{
		\qop O_{\bm x}
	}
	{
		\sigma^\prime e^\prime
	}_{y^\prime}
	\mathperiod
\end{align}
This simplifies calculation tremendously since it means we can measure the operators for all basis states of the \gls{1d} set and proceed with them as local operators again.
Two\hyp chain measurements are thus reduced to regular expectation values, using the \gls{ee} wavefunction and these supersite operators in the conventional manner: apply the local gate on site $y^\prime$ and contract the tensor train up to site $y^\noprime$ where another local gate is then applied.
\subsection{Error sources}
\label{subsec:error-sources}
Practical application of the \gls{ee} framework beyond very small systems entails some unavoidable yet controlled approximations. 
The derivation of the first term of~\cref{eq:2d-ee-generic} only requires the application of $\qop H_1$ to the basis state being proportional to the state itself and the \gls{1d} states to be orthonormal.
The former can be controlled by computing the basis state's variance and the later by checking the overlaps with other states and imposing a threshold which we chose to be $\delta = \num{e-12}$
\begin{align}
	\mel
	{
		e^\noprime
	}
	{
		\qop H^2_1
	}
	{
		e^\noprime
	}
	&-
	\mel
	{
		e^\noprime
	}
	{
		\qop H_1
	}
	{
		e^\noprime
	}^2
	\leq \delta
	\qquad
	\forall e^\noprime = 0,\ldots,d-1
	\mathcomma
	\\
	\braket{
		e^\noprime
	}
	{
		e^\prime
	}
	\leq \delta
	&
	\qquad
	\forall e^\noprime , e^\prime \text{ such that } \hat \sigma(e^\noprime) = \hat \sigma(e^\prime)
	\mathperiod
\end{align}
The second term in~\cref{eq:2d-ee-generic} can be computed exactly for many Hamiltonians by evaluating matrix elements of the \gls{1d} system, as desribed in~\cref{subsec:measurements}.
Yet, the number of single\hyp chain basis states $\ket{e}$ that are retained will always have to be a small slice of the exponentially growing Hilbert space of the single-chain eigenbasis.
\par
Crucially, only a limited number of this large set is necessary to describe the system accurately when inter\hyp chain coupling $t_\perp$ is small.
In the limit ${t_\perp \rightarrow 0}$ the system will minimise its energy by occupying the lowest energy mode on all chains, i.e. forming a product state.
As interchain coupling grows from zero, the system can lower its energy by coupling \gls{1d} eigenstates different from $|0\rangle$ to each other, but the spread of energies across which this coupling can take place will still be restricted by $t_\perp$.
We thus do not expect to observe arbitrarily high \gls{1d} excitations, as these are especially costly in energy when $t_\perp$ is small.
\par
We find that analysing the energy spread of \gls{1d} states that appear with sufficient weight on the diagonal of the single\hyp chain density matrix (c.f. discussion after~\cref{eq:single-supersite-obs}) is a suitable method for estimating the range of single\hyp chain eigenstates that need to be retained, as demonstrated in~\cref{fig:spin-hdb-rdm-probabilities}.
Assuming continuity with increasing excitation number, we suppress a mode if its excitation probability drops below a certain value.
Sectors in which all excitation probabilities are below the truncated weight of the \gls{ee}\hyp\gls{mps} are dropped by the \gls{svd} automatically.
Of course, the above is no exact criterion because a mode $i_{y^\noprime}$ may only be suppressed on all chains when the excitation probability $P(i_1\ldots i_y\ldots i_W)=\abs{\braket{i_1\ldots i_y\ldots i_W}{\psi}}^2$ of this mode is negligible for all other $i_{y^\prime}$.
For all studied systems we find the diagonal of the single\hyp chain density matrix to have a strongly peaked distribution around the central sector, as can be seen in~\cref{fig:spin-hdb-rdm-probabilities}.
Usually, the ground state of $\qop H_1$ takes up more than $\SI{95}{\percent}$ of the weight for weak coupling and some of the adjacent sectors lowest\hyp energy states make up most of the remaining weight.
This is followed by a long tail of states which are marginally important but which are still retained to obtain the highest precision possible for observables.
\par
Another way of probing whether sufficiently many single\hyp chain eigenstates have been included is whether a given set has the capacity to describe the expected physics of the \gls{2d} system correctly.
For example, depending on lattice structure and the type and range of interactions, a \gls{2d} tight\hyp binding bosonic model at zero temperature can exhibit Bose\hyp Einstein condensation.
One signature behaviour of this are single particle correlators decaying to a constant at long distances.
In contrast to that, the ground states of the individual $\qop H_1$\hyp Hamiltonians can at best exhibit algebraically decaying correlations.
Therefore, we investigate the matrix elements of different excited \gls{1d} states, ${\langle e^\prime |\hat{b}^\dagger_x \hat{b}_y| e^\noprime \rangle}$ (c.f.~\cref{appendix:fig:1d-overlaps}) and observe when and how strong they decay with distance.
We found that once we start to retain $\numrange[range-phrase=\text{ to }]{10}{20}$ eigenstates in the most relevant quantum number sectors we pick up many cross\hyp state correlators which no longer show decay with distance ${|x-y|}$.
\par
Finally, one practical criterion for deciding whether the local Hilbert space of the individual \gls{1d} subunits has been chosen judiciously consists of checking the stability of observables as the size of this space is varied.
We discuss a concrete example of this towards the end of~\cref{subsubsec:square-spins}.
\par
The second source of error stems from the truncated weight of the \gls{2d} \gls{dmrg} wavefunction obtained for the Hilbert space spanned by the $\ket{\bf e}$\hyp vectors.
This error is not only well known and perfectly controllable via the bond dimension, but its impact on chain\hyp local observables is known~\cite{RevModPhys.77.259}.
By virtue of the entanglement ``hidden'' in the supersite basis, the bond dimension required to achieve low truncated weights in the \gls{ee}\hyp framework remains much below those values that would be necessary to reach comparable discarded weights with conventional brute\hyp force \gls{dmrg} calculations for \gls{2d} models~\cite{PhysRevB.100.075138}.
Furthermore, whenever the truncated weight of the \gls{ee}\hyp\gls{dmrg} was above that of the required accuracy, we used the standard extrapolation in bond dimension for chain\hyp local observables in order to improve precision for those quantities.
\par
In the following section, we demonstrate the application of the \gls{ee} framework to models of hardcore bosons, unfrustrated and frustrated spins, as well as to doped, repulsively interacting fermions.
This includes comparisons to other controlled\hyp approximation techniques where possible, including \gls{qmc} and massively parallelized brute\hyp force \gls{dmrg}.
\section{Models \& Results}
\label{sec:model-results}
\subsection{\Acrlong{hcb} on square lattice}
\label{subsec:hcb-rect}
The first model for which we benchmark the \gls{ee} framework are \acrlongpl{hcb} on a square lattice with rectangular geometry, $L \neq W$.
These systems can be treated with great accuracy using \gls{qmc} techniques, which makes them perfect for a first round of benchmarking.
We study lattices of size $L\times W$ with nearest\hyp neighbour repulsion $V$ on the same chain but not between chains, which corresponds to $V_\perp = 0$ in~\cref{fig:2d-chain}.
The Hamiltonian elements of~\cref{eq:hamiltonian-splitting} read
\begin{align}
	\label{eq:hcb-rect-h1-hamiltonian}
	\qop H_1
	&=
	-
	\sum_{xy}
	\left(
		\qop b^\dagger_{x,y}
		\right.
		\qop b^\nodagger_{x+1,y}
		\left.
		+
		\qop b^\dagger_{x+1,y} \qop b^\nodagger_{x,y}
	\right)
	+
	V / t
	\sum_{xy}
	\qop n_{x,y} \qop n_{x+1,y}
	\\
	\label{eq:hcb-rect-hperp-hamiltonian}
	\qop H_\perp &=
	- \sum_{xy}
	\left(
		\qop b^\dagger_{x,y} \qop b^\nodagger_{x,y+1}
		+
		\qop b^\dagger_{x,y+1} \qop b^\nodagger_{x,y}
	\right)
	\mathperiod
\end{align}
Here we apply \glspl{pbc} in the $x$\hyp direction of strong coupling, and \glspl{obc} in the $y$\hyp direction of weak coupling $t_\perp$. 
We note that it would be possible to choose \glspl{pbc} in both directions, at the cost of additional computational complexity.
We highlight that the coupling constants have been factored in these Hamiltonians such that they are consistent with~\cref{eq:hamiltonian-splitting}.
Applying the \gls{ee} procedure according to~\cref{eq:2d-ee-generic} and with the definition from~\cref{eq:ee-diag-part} implies
\begin{align}
	\label{eq:hcb-rect-ee-hamiltonian}
	\qop H &= 
	t \sum_y \qop E_y
	-
	t_\perp \sum_{xy}
	\left(
		\qop \beta_y(x)^\dagger \qop \beta_{y+1}(x)^\nodagger
		+
		\qop \beta_{y+1}(x)^\dagger \qop \beta_{y}(x)^\nodagger
	\right)
	\mathcomma
\end{align}
where non\hyp diagonal operators have been defined according to~\cref{eq:two-chain-ee-matrixelem}
\begin{align}
	\label{eq:ee-creator}
	\notag
	\qop \beta_y(x)^\dagger
	=
	\qop 1_1
	\otimes
	&\ldots
	\qop 1_{y-1}
	\\
	\notag
	&\otimes
	\sum_
	{
		ne^\noprime e^\prime
	}
	\mel
	{
		n+1, e^\noprime
	}
	{
		\qop b^\dagger_x
	}
	{
		n, e^\prime
	}_y
	\outerproduct
	{
		n+1, e^\noprime
	}
	{
		n, e^\prime
	}_y
	\\
	&
	\otimes
	\qop 1_{y+1}
	\otimes
	\ldots
	\otimes
	\qop 1_W
	\\
	\label{eq:ee-annihilator}
	\notag
	\qop \beta_y(x)^\nodagger
	=
	\qop 1_1
	\otimes
	&\ldots
	\qop 1_{y-1}
	\\
	\notag
	&\otimes
	\sum_
	{
		ne^\noprime e^\prime
	}
	\mel
	{
		n, e^\noprime
	}
	{
		\qop b^\nodagger_x
	}
	{
		n+1, e^\prime
	}_y
	\outerproduct
	{
		n, e^\noprime
	}
	{
		n+1, e^\prime
	}_y
	\\
	&\otimes
	\qop 1_{y+1}
	\otimes
	\ldots
	\otimes
	\qop 1_W
	\mathperiod
\end{align}
Here, $n$ is the particle number supplied by the application of the symmetry operator $\hat\sigma(e) = n$.
The first term in~\cref{eq:hcb-rect-ee-hamiltonian} acts analogous to a chemical potential for the occupation of the \gls{1d} modes, while the second term re\hyp expresses interchain hopping using the \gls{1d} basis states.
We point out that the flow of quantum numbers due to application of a local operator is independent of the position within the chain $x$, which only affects the numerical values of the matrix elements.
This helps to simplify the implementation of the framework.
\par
The number of particles is chosen to be half the number of total orbitals, i.e. $N = LW/2$.
This yields our choice for the \gls{1d} modes included in the local basis
\begin{align}
	\notag \underline{\abs{n_y-L/2} = 4} &: & 0 \leq e \leq 5 \mathcomma
	\\
	\notag \underline{\abs{n_y-L/2} = 3} &: & 0 \leq e \leq 9 \mathcomma
	\\
	\notag \underline{\abs{n_y-L/2} = 2} &: & 0 \leq e \leq 19 \mathcomma
	\\
	\label{eq:boson-local-Hilbert-space}
	\underline{\abs{n_y-L/2} = 1 \text{ or } 0} &: & 0 \leq e \leq 63 \phantom{\mathcomma}
\end{align}
by means of the heuristics outlined in~\cref{subsec:error-sources}, as well as checking for the convergence of observables such as the ground state energy and correlation functions with the size of the basis $\ket{e}$.
For the system sizes studied here we thus settle on total local dimension of $d = 264$.
This results in significant memory requirements, reaching up to $\SI{2}{\tera\byte}$ of \acrshort{ram} for a two\hyp site \gls{dmrg} update at a bond dimension of $m = 256$ and a real\hyp valued wavefunction.
The technical details to this are discussed in~\cref{appendix-section:bottleneck-analysis}.
\par
\begin{figure}
	\centering
	\subfloat
	[%
		\raggedright
		\label{fig:rect-energies-over-V0}%
	]
	{\includegraphics[width = 0.45\textwidth]{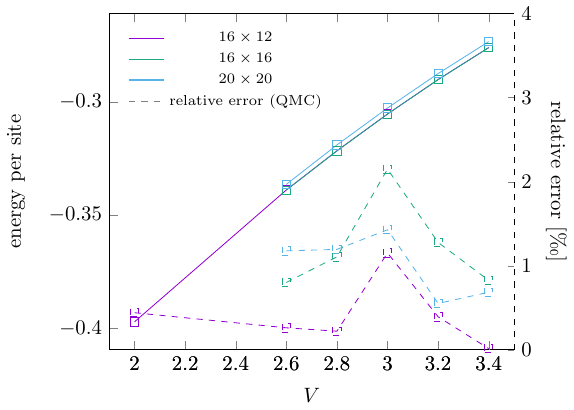}}%
	\\
	\subfloat
	[%
		\raggedright
		\label{fig:rect-energies-over-tp}%
	]
	{\includegraphics[width = 0.45\textwidth]{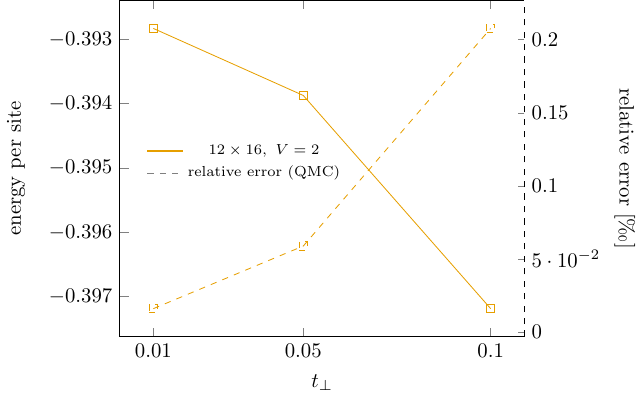}}%
	\\
	\subfloat[%
		\raggedright
		\label{fig:rect-bos-corr}%
	]
	{\includegraphics[width = 0.45\textwidth]{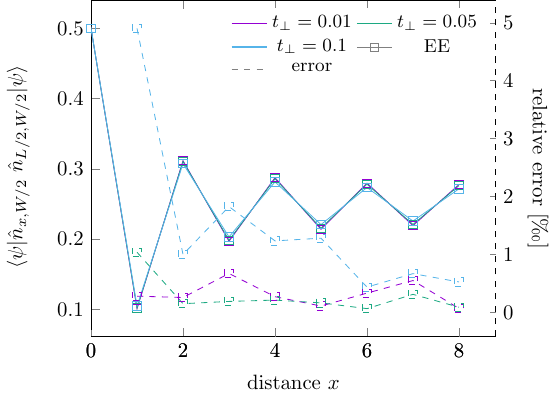}}%
	\caption{%
		\label{fig:rect-energies-comparison}%
		\Cref{fig:rect-energies-over-V0,fig:rect-energies-over-tp} show the energies of the system scaling with intrachain density repulsion and interchain coupling, respectively.
		In the former case, $t_\perp$ was fixed to be $0.1$ while in the latter $V = 2$.
		\Cref{fig:rect-bos-corr} shows the correlator~\cref{eq:rect-hcb-density-density-corr} along the strong coupling direction for different intra\hyp chain couplings $t_\perp$ of a $16\times 12$ strip with repulsion $V = 2$.
		The relative error compared to exact \gls{qmc} data is shown on the right\hyp hand $y$\hyp axis (\protect\dashedline).%
	}
\end{figure}

The first validation of the \gls{ee}\hyp framework against zero\hyp temperature \gls{qmc} calculations (which provide data that are numerically exact within statistical
errors fully controlled by the number of Monte Carlo steps~\cite{Sandvik2010}), is done for the ground state energy.
The results are shown in~\cref{fig:rect-energies-comparison}.
We observe correct physical behaviour as $V$ and $t_\perp$ are changed, the former increasing the energy per site as it grows, while the latter lowers it.
Even more importantly, we obtain good agreement with the \gls{qmc} gold standard for this type of model as shown in~\cref{fig:rect-energies-comparison}.
With the truncated weight of the final wave function being $\LandauO{\num{e-12}\lldots\num{e-8}}$, we could dispense with the extrapolation to infinite bond dimension in this case.
The same holds true for the measurement of a simple density\hyp density correlation
\begin{align}
	\label{eq:rect-hcb-density-density-corr}
	\mel
	{
		\psi
	}
	{
		\qop n_{x, W/2}
		\
		\qop n_{L/2,W/2}
	}
	{
		\psi
	}
\end{align}
which again yields relative errors of the order of permills compared to exact \gls{qmc}, as is displayed in~\cref{fig:rect-bos-corr}.
\subsection{Spins on square \& triangular lattice}
\label{subsec:spins}
\begin{figure}
	\centering
	\includegraphics[width = 0.45\textwidth]{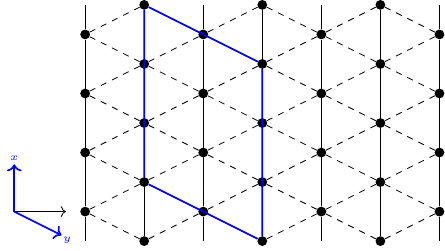}
	\caption
	{
		\label{fig:2d-triangular}%
		Triangular lattice made up of weakly coupled chains where spins interact with coupling strength $J = 1$ (\protect\blackline) or $J_\perp$ (\protect\dashedline).
		The blue lines (\protect\blueline) show the choice of our elementary vectors.
		The unfinished lines in $x$\hyp direction indicate \glspl{pbc} while the horizontal direction has \glspl{obc}.%
	}
\end{figure}

Here, we study the performance of the \gls{ee} framework for \gls{2d} spin\hyp $1/2$ systems.
First, we test the ability of this new approach to capture ordered many\hyp body quantum states for a case that is still amenable to exact validation via \gls{qmc}, the anisotropic Heisenberg quantum \gls{afm} in the absence of frustration.
Building on that, we then showcase the power of the \gls{ee} framework for a notoriously difficult problem in the domain of strongly correlated \gls{2d} systems, the conjectured spin liquid states on the anisotropic triangular lattice.
With the ability to address significantly larger \gls{2d} lattices than other many\hyp body numerics, we demonstrate the ability to sort through the predictions of several competing theories for the conjectured ground states of this model.
In both cases, we study a spin\hyp $1/2$ Heisenberg Hamiltonian
\begin{align}
	\label{eq:spin-hamiltonian}
	\qop H &= J \sum_{\langle i,j \rangle_x} \qop{\bm S}_i \qop{\bm S}_j + J_\perp \sum_{\langle i,j \rangle_\perp} \qop{\bm S}_i \qop{\bm S}_j
	\mathcomma
\end{align}
with antiferromagnetic coupling ${J,J_\perp > 0}$.
The notation ${\langle i,j \rangle_x}$ implies nearest neighbours in $x$\hyp direction only, while ${\langle i,j \rangle_\perp}$ denotes nearest neighbours in the perpendicular direction, a set of bonds that will be different
between the square and the triangular lattice (c.f.~\cref{fig:2d-chain} and~\cref{fig:2d-triangular}).
The operators are defined in the standard way, ${\qop{\bm S}_i = \left(\qop S^x_i,\qop S^y_i,\qop S^z_i\right)^\intercal}$ being a vector of spin\hyp $1/2$\hyp operators on site ${i=(x,y)}$.
In this subsection, we also always consider a square-shaped geometry, i.e. $L = W$.
\subsubsection{Square lattice: anisotropic Heisenberg \acrlong{afm}}
\label{subsubsec:square-spins}
We exploit that the Heisenberg Hamiltonian in the $S_z = 0$ subspace can be mapped to hardcore bosons at half\hyp filling, as recapped in~\cref{appendix-subsec:spin-to-hcb}.
While the spin\hyp spin\hyp coupling in the $z$\hyp direction leads to an additional density\hyp density interaction term in the direction of transverse coupling compared to~\cref{subsec:hcb-rect}, we can keep using the local basis~\cref{eq:boson-local-Hilbert-space}.
We therefore require comparable resources, except for the growth of the \gls{ee}\hyp \gls{mpo} bond dimension.
\par
Analogous to the mapping that yields~\cref{eq:hcb-rect-ee-hamiltonian} from~\cref{eq:hcb-rect-h1-hamiltonian} and~\cref{eq:hcb-rect-hperp-hamiltonian}, for the square lattice we thus obtain 
\begin{align}
	\label{eq:spin-hcb-sql-ee-hamiltonian}
	\notag
	\qop H_\square =
	J \sum_y \qop E_y
	&+
	J_\perp
	\sum_{xy}
	\left(
		\qop \beta_y(x)^\dagger \qop \beta_{y+1}(x)^\nodagger
		+
		\hc
	\right)
	\\
	&+
	2J_\perp
	\sum_{xy}
	\qop \eta_y(x) \qop \eta_{y+1}(x)
\end{align}
from the original Hamiltonian, \cref{eq:spin-hamiltonian}.
The definition of the supersite operators $\hat{\beta}$ and $\hat{\beta}^\dagger$ is the same as in~\cref{eq:ee-creator,eq:ee-annihilator}, exploiting the standard mapping of spin\hyp $1/2$ operators to \gls{hcb}, i.e. ${\hat{b}^\dagger:=\hat{S}^x+i\hat{S}^y}$ (c.f.~\cref{appendix-subsec:spin-to-hcb}).
Complementing these operators is the local density operator expressed in the \gls{1d} basis of the chains:
\begin{align}
	\qop \eta_y(x) &=
	\qop 1_1 \otimes \ldots
	\sum_{nee^\prime}
	\mel{n,e^\noprime}{\qop n_x}{n,e^\prime}_y
	\outerproduct{n,e^\noprime}{n,e^\prime}_y
	\ldots\otimes\qop 1_W
	\mathperiod
\end{align}
This operator appears again during the mapping of spins to \gls{hcb}, ${\hat{n}:=\hat{S}^z+1/2}$.
\par
Our primary aim in studying the ground state of $\qop H_\square$
is to quantify the capacity of the \gls{ee} framework to describe ordered states of \gls{2d} many\hyp body systems.
To achieve this goal, we exploit that the underlying Hamiltonian~\cref{eq:spin-hamiltonian} on a square lattice can be treated with high efficiency using \gls{qmc} techniques, such as the \gls{sse}~\cite{Syljuasen2002}.
This technique scales linearly in the number of lattice sites and inverse temperature, with the statistical error decreasing with the square root of the independent Monte Carlo samples.
For the range of parameters that we target here, \gls{sse}\hyp \gls{qmc} renders a benchmark that is exact for all practical purposes.
\par
\begin{figure*}
	\centering
	\subfloat
	[%
		\label{fig:spin-hcb-ee-correlator-J-sql}%
	]
	{\includegraphics[width = 0.5\textwidth]{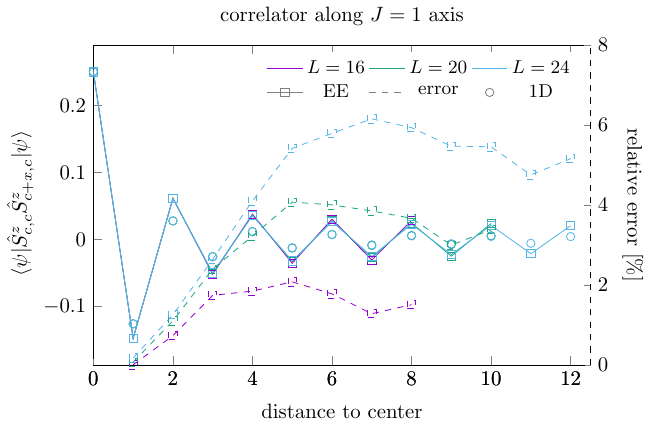}}%
	\hspace{0.4cm}
	\subfloat
	[%
		\label{fig:spin-hcb-ee-correlator-J-triangular}%
	]
	{\includegraphics[width = 0.45\textwidth]{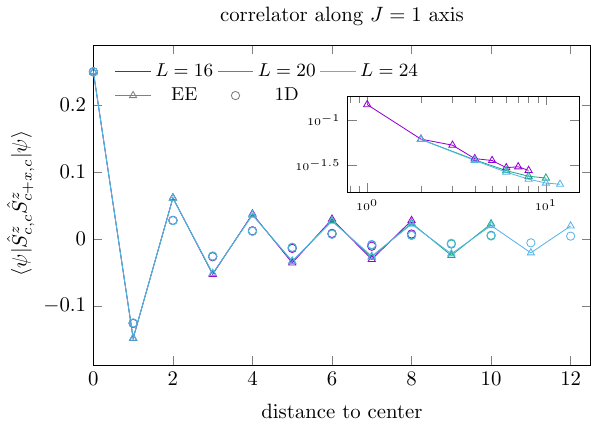}}%
	\\
	\subfloat
	[%
		\label{fig:spin-hcb-ee-correlator-Jperp}%
	]
	{\includegraphics[width = 0.52\textwidth]{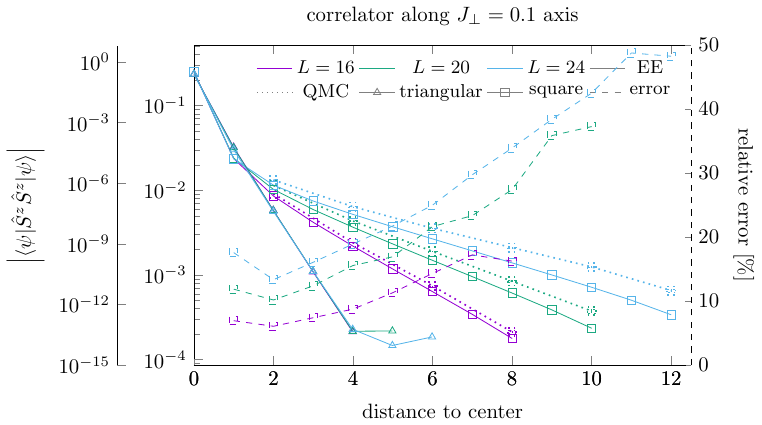}}%
	\hspace{1.2cm}
	\subfloat
	[%
		\label{fig:spin-hcb-ee-correlation-length}%
	]
	{\includegraphics[width = 0.4\textwidth]{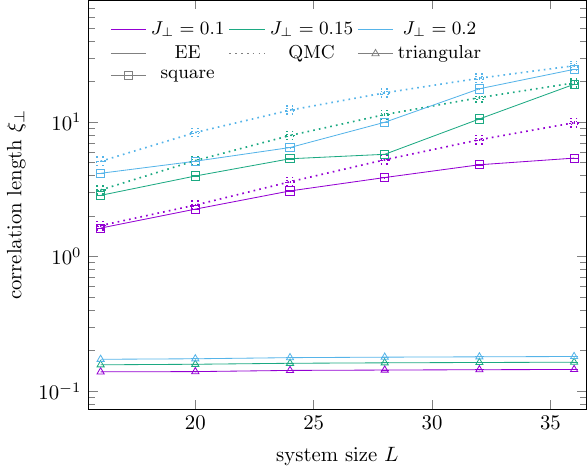}}%
	\caption{%
		\label{fig:spin-hcb-ee-correlator}%
		\Cref{fig:spin-hcb-ee-correlator-J-sql,fig:spin-hcb-ee-correlator-J-triangular} show the algebraic decay of the $S^zS^z$\hyp correlator for $L = 16,20,24$ along the strong coupling direction and with an interchain coupling of $J_\perp / J = 0.1$ for the square ($\square$) and the triangular ($\triangle$) system, respectively (c.f. discussion below~\cref{eq:spin-hcb-perp} and~\cref{eq:spin-hcb-perp_triangular}).
		The inset of~\cref{fig:spin-hcb-ee-correlator-J-triangular} shows $|C(x)|$ on a double logarithmic scale.
		\Cref{fig:spin-hcb-ee-correlator-Jperp} shows $|C_\perp(y)|$ at $J_\perp = 0.1$ (c.f. discussion in main text).
		\Cref{fig:spin-hcb-ee-correlation-length} shows the perpendicular correlation length $\xi_\perp(L)$ for multiple values of $J_\perp = 0.1, 0.15, 0.2$.%
	}
\end{figure*}

To gauge the performance of the \gls{ee}\hyp technique against the \gls{qmc} gold standard, we focus on the spin\hyp spin correlation functions both along one of the two central chains as well as along the perpendicular coupling direction:
\begin{align}
	\label{eq:spin-hcb-alinged}
	C(x) &=
	\mel{
		\psi
	}
	{
		\qop S^z_{c,c}
		\qop S^z_{c+x,c}
	}
	{
		\psi
	}
	\mathcomma
	\\
	\label{eq:spin-hcb-perp}
    C_\perp(y) &=
	\mel{
		\psi
	}
	{
		\qop S^z_{c,c}
		\qop S^z_{c,c+y}
	}
	{
		\psi
	}
	\mathcomma
\end{align}
where $c = L/2 = W/2$ picks out one of the four central sites in our chosen lattices, for which both $L$ and $W$ are even.
As shown in~\cref{fig:spin-hcb-ee-correlator-J-sql}, the \gls{ee} framework matches the \gls{qmc} benchmark very well in the strong\hyp coupling $x$\hyp direction, the error in the correlator even at maximal distance reaching at most $\SI{6}{\percent}$ for the largest lattice (${L = W = 24}$).
In the transverse direction, where correlations are naturally significantly weaker, the \gls{ee}\hyp based calculations see a stronger increase in error as the system size grows, even at the lowest $J_\perp$\hyp value, as illustrated in~\cref{fig:spin-hcb-ee-correlator-Jperp}.

Nevertheless, the \gls{ee} approach still captures the existence of an ordered anisotropic \gls{afm} state~\cite{Schulz1996,Sandvik1999,Furuya2016}.
Indeed, in the regime ${J_\perp\ll J}$ for finite lattices, the transverse correlations are found to decay with a $W$-dependent exponentially decaying envelope,
\begin{align}
	\label{fig:spin-hcb-perp-corrlength}
	|C_\perp (y)|
	& \propto
	\enumber^
	{
		-y/\xi_\perp(W)
	}
	\mathperiod
\end{align}
 but the length\hyp scale $\xi_\perp(W)$ associated with this decay itself grows at least linearly with the value of $W$, recovering the ordered \gls{afm} in the thermodynamic regime.
As summarized in~\cref{fig:spin-hcb-ee-correlation-length}, both our \gls{ee} framework and the \gls{qmc} benchmarks correctly capture this physical behaviour for a range of small $J_\perp/J$\hyp values.
\par
While the \gls{ee} approach is capable of reproducing the correct physics, it does underestimate the magnitude of $\xi_\perp(W)$ systematically compared to the \gls{qmc} benchmark.
This systematic error grows with lattice size and the value of $J_\perp/J$. 
This trend is to be expected, for two reasons:
(1) As $L$ grows, the size of the retained single\hyp chain basis $\ket{e}$ would need to increase exponentially if every state within the retained bandwidth of energy contributed equally.
In practice we observe that at the level of the weights on the diagonal of the single\hyp chain \gls{rdm}, more than $\SI{90}{\percent}$ of the weight is in the ground state of $\hat{H}_1$, with some additional percentages for the first excited state of the quantum number sector of that ground state, as well as to the ground states of the adjacent sectors.
Beyond that, in every relevant quantum number sector, the higher in energy a state $\ket{e}$ is relative to that sector's ground state $\ket 0$, the less it contributes to the \gls{2d} wave function in general.
At the same time, there is also a significant variation, with a long tail of states with spiking contributions that will be increasingly difficult to incorporate as $L$ is increased, as illustrated in~\cref{fig:spin-hdb-rdm-probabilities}.
Still, we find that observables appear to converge in the size of the retained basis\hyp set $\ket{e}$ of the chains, for which examples are shown in~\cref{appendix:fig:d-scaling}.
(2) Even though the targeted state is ordered, the entanglement entropy of the \gls{2d} wavefunction scales with subsystem boundary.
Consequently, the truncated weight increases strongly when the maximal bond dimension is fixed to the \acrshort{ram}\hyp limited value of $256$, as shown in~\cref{fig:spin-hcb-trunc-weight}.
Toward the upper range of $L$ and $J_\perp$, we observe that the truncated weight begins to saturate, clearly due to an insufficient bond dimension.
\par
There are ways to mitigate the \acrshort{ram} bottleneck and increase the bond dimension, as discussed in~\cref{appendix:subsec:eedmrg}.
%
%
As the present study of the \gls{afm} ordered state on the anisotropic Heisenberg model on the square lattice is a proof of principle, we do not pursue these alternatives further here.
\subsubsection{Triangular lattice: quasi\hyp\gls{1d} spin liquid}
\label{subsubsec:triangular-spins}
The Heisenberg model, \cref{eq:spin-hamiltonian}, placed on the anisotropic triangular lattice (c.f.~\cref{fig:2d-triangular}) has been subject to a number of previous studies.
These are motivated by the realization of this model - or a close variant of it - in Cs$_2$CuCl$_4$ and Cs$_2$CuBr$_4$, which are believed to host quantum spin liquid states~\cite{Coldea2002}.
Previous theoretical work yielded conflicting predictions for the ground or low\hyp temperature state of this model in the physically relevant regime of $J_\perp\ll J$~\cite{Alicea2006,Weng2006,Yunoki2006,Fjaerestad2007,Starykh2007,Hayashi2007,Pardini2008,Heidarian2009,Starykh2010,Cong2011,Weichselbaum2011,Reuther2011,Thesberg2014,PhysRevB.93.085111,Gonzalez2020,Gonzalez2022,Yu2023}.
This is due to the substantial technical difficulty in obtaining solutions for larger lattices.
Not only are any \gls{qmc}\hyp based techniques incapable of addressing even moderately sized lattices at low or zero temperature due to the sign\hyp problem, which is caused by the frustrated nature of nearest-neighbour coupling in this type of lattice.
Also, attempts at brute\hyp force application of the \gls{dmrg}\hyp algorithm to these \gls{2d} systems are made difficult exactly by the anisotropic nature of the lattice: a low ratio of $J_\perp/J$ causes a failure to converge in any standard zig\hyp zag mapping in the $y$\hyp direction~\cite{Weichselbaum2011}.
The alternative, which would be to combine a mapping in the $x$\hyp direction with p\gls{dmrg}, would require simultanous access to many compute notes over a sustained period, and even then it is possible that this approach would not be able to treat many chains~\cite{PhysRevB.100.075138}.
\par
The \gls{ee}\hyp approach however is tailor\hyp made for this regime, as the frustrated nearest\hyp neighbour coupling can be expected to reduce entanglement between chains, and thus the bond dimension required to efficiently approximate the true \gls{2d} ground state with the underlying choice of \gls{mps}\hyp representation (c.f.~\cref{fig:2d-chain}).
Using the lattice and coordinate definitions shown in~\cref{fig:2d-triangular}, we obtain the \gls{ee}\hyp Hamiltonian $\qop H_\triangle$ for this lattice by adding to~\cref{eq:spin-hcb-sql-ee-hamiltonian}:
\begin{align}
	\qop H_\triangle = \qop H_\square +
	J_\perp&
	\sum_{xy}
	\left(
	\qop \beta_y(x)^\dagger \qop \beta_{y+1}(x+1)^\nodagger
	+
	\hc
	\right)
	\nonumber\\
	&+
	2J_\perp
	\sum_{xy}
	\qop \eta_y(x) \qop \eta_{y+1}(x+1)
	\label{eq:spin-hcb-triangular-ee-hamiltonian}
\end{align}
Along with this, we also use a modified definition of the transverse correlator,
\begin{align}
	\label{eq:spin-hcb-perp_triangular}
    C_\perp(y) &=
    \mel{
		\psi
	}
	{
		\qop S^z_{c,c}
		\qop S^z_{c+y,c+2y}
	}
	{
		\psi
	}
    \mathcomma
\end{align}
which takes the triangular structure of the lattice into account, where equivalent sets of sites in the direction perpendicular to the $x$\hyp axis will be found only on every second chain.

\begin{figure}
	\centering
	\includegraphics
	[
		width = 0.45 \textwidth
	]
	{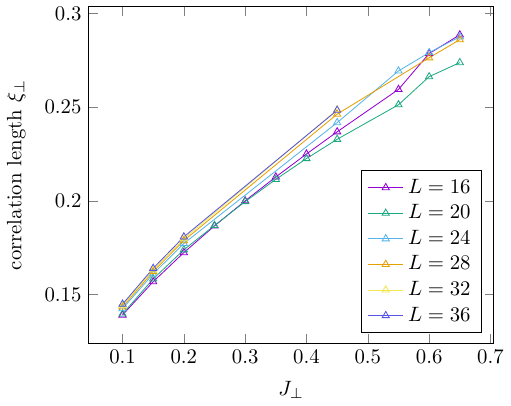}
	\caption
	{%
		\label{fig:spin-hcb-z-corr-length-over-jp}%
		Length-scale $\xi_\perp$ for transverse correlations as a function of $J_\perp$ for triangular lattices of different sizes.%
	}
\end{figure}

Our results conclusively show the suitability of the \gls{ee} framework for these anisotropic frustrated systems.
While correlations $C(x)$ in the strong\hyp coupling direction are actually slightly enhanced over the isolated case as shown in~\cref{fig:spin-hcb-ee-correlator-J-triangular}, the correlator $C_\perp(y)$ is massively depressed when compared to the anisotropic square lattice, as shown in~\cref{fig:spin-hcb-ee-correlator-Jperp}.
More importantly, the exponential decay of these correlations with distance appears to exhibit almost no dependence on the size of the system.
Therefore, the correlation length $\xi_\perp$, that can be extracted from fitting, almost does not change with $L$ and $W$ as summarized in~\cref{fig:spin-hcb-ee-correlation-length}, and increases only sub\hyp linearly with $J_\perp/J$, c.f.~\cref{fig:spin-hcb-z-corr-length-over-jp}.
This is accompanied by a drastic decrease in the truncated weight as shown in~\cref{fig:spin-hcb-trunc-weight}, while maintaining the same value of the bond dimension, $m=256$, as for the square lattice.
Complementing this is the observation that even more of the weight of the single\hyp chain \gls{rdm} is shifted to the ground state of the isolated chain than in the case of the square lattice, as shown in~\cref{fig:spin-hdb-rdm-probabilities}.
\par
\begin{figure}
	\centering
	\includegraphics[width = 0.42\textwidth]{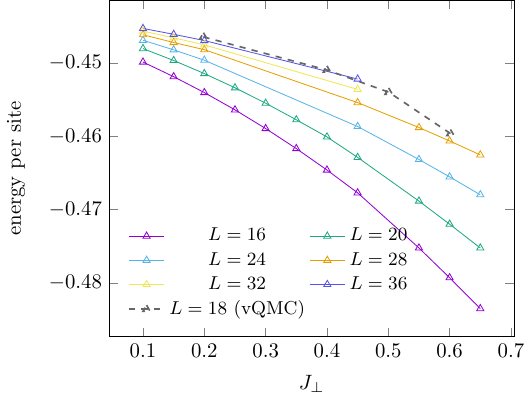}
	\caption{%
		\label{fig:triangular-energies}%
		Energies for different system sizes and intrachain couplings $J_\perp$ of the triangular lattice compared to \acrshort{vqmc} data extracted from~\cite{PhysRevB.93.085111} for a $18\times 18$ sites system.%
	}
\end{figure}

The ability of the \gls{ee}\hyp technique to treat this lattice at unprecedented sizes with controlled\hyp approximation many\hyp body numerics allows to draw inferences about the possible nature of the ground state of this model, which is the subject of intense and conflicting research.
The strong suppression and size\hyp independence of $C_\perp(y)$, coupled with the apparently algebraic behaviour of $C(x)$ appears to support those works that predict this model to realize a gapless quasi\hyp\gls{1d} spin liquid at low $J_\perp/J$~\cite{Weng2006,Yu2023}.
Thus, our results seem incompatible with work that indicated an \gls{afm} at low $J_\perp/J$ due to higher\hyp order processes~\cite{Starykh2007}.
At the same time we cannot rule out with absolute certainty that the \gls{afm} order predicted in~\cite{Starykh2007} isn't so weak that it cannot be resolved even by the unprecedented lattice sizes accessible to our framework.
Likewise, we find no sign that this model exhibits a phase transition around $J_\perp/J\approx 0.6$.
Such predictions had been based on ground state energies obtained from \gls{vqmc} calculations~\cite{PhysRevB.93.085111}, as well as series expansion~\cite{Gonzalez2022}.
However, our \gls{ee}\hyp calculated energies substantially improve upon those obtained from \gls{vqmc}, as shown in~\cref{fig:triangular-energies}.
Not only do we observe no change in the slope of the energy per site with changing $J_\perp/J$ around $0.6$ (c.f.~\cref{fig:triangular-energies}), but the entanglement entropy, another possible indicator of a quantum phase transition, also does not exhibit distinguishing features for any $J_\perp/J\leq 0.65$.
\par
The low truncated weights that we find at small $J_\perp/J$ for the triangular lattice and the large bandwidth of energies spanned by the single\hyp chain basis states that we incorporate (c.f.~\cref{fig:spin-hcb-trunc-weight} and~\cref{fig:spin-hdb-rdm-probabilities}, respectively) makes it likely that the \gls{ee}\hyp framework approximates the ground state wavefunction well in these cases.
But despite the large lattices amenable to the~\gls{ee}\hyp framework, we are prevented from drawing definite conclusions about the actual ground state properties of the model at this point, and will return to this question in future work.
That is because we have encountered persistent problems in converging the wavefunction once a finite flux through the periodic direction is inserted into the Hamiltonian in order to locate states with spiral ordering that could compete with the quasi\hyp\gls{1d} spin liquid~\cite{Weichselbaum2011}.
%
%
This flux\hyp insertion needs to be performed in two places, namely both in the single\hyp chain Hamiltonian, and in the effective \gls{ee}\hyp Hamiltonian.
The former is done by taking the first term of~\cref{eq:spin-hamiltonian}, where the nearest\hyp neighbour spin\hyp flipping terms are modified, i.e.
\begin{align}
	\notag
	\qop{S}^+_{x,y}
	\qop{S}^-_{x+1,y}
	+
	&
	\qop{S}^+_{x+1,y}
	\qop{S}^-_{x,y}
	\\
	&\rightarrow
	\enumber^{\imagunit\phi}
	\qop{S}^+_{x,y}
	\qop{S}^-_{x+1,y}
	+
	\enumber^{-\imagunit\phi}
	\qop{S}^+_{x+1,y}
	\qop{S}^-_{x,y}
	\mathperiod
\end{align}
From the resulting isolated\hyp chain Hamiltonian $\hat{H}_1(\phi)$ a new basis\hyp set $\ket{e^\noprime(\phi)}$ needs to be constructed for every $\phi$\hyp value of interest.
A matching flux also needs to be inserted into $\qop{H}_\perp$ in the \gls{ee}\hyp representation, which we do by modifying the diagonal tunneling terms in~\cref{eq:spin-hcb-triangular-ee-hamiltonian}:
\begin{align}
	\notag
	\hat{\beta}_{y}(x)^\dagger
	&
	\hat{\beta}_{y+1}(x+1)
	+
	\hat{\beta}_{y+1}(x+1)^\dagger
	\hat{\beta}_{y}(x)
	\\
	&
	\rightarrow
	\enumber^{\imagunit\phi}
	\hat{\beta}_{y}(x)^\dagger
	\hat{\beta}_{y+1}(x+1)
	+
	\enumber^{-\imagunit\phi}
	\hat{\beta}_{y+1}(x+1)^\dagger
	\hat{\beta}_{y}(x)
	\mathperiod
\end{align}
In principle, searching for the flux\hyp value that minimizes the energy of the \gls{2d} ground state would allow to test how the model transitions to a state with generically incommensurate spiral order as $J_\perp/J$ grows.
Such spiral order has to appear at some point, either continuously or due to a quantum phase transition~\cite{Weihong1999,Weng2006,Fjaerestad2007,Pardini2008,Weichselbaum2011,Gonzalez2022,Yu2023}.
However, in practice we have found it difficult to obtain converged wavefunctions at $\phi\neq0$.
We have not yet identified the cause of this challenge, which grows pronouncedly with increasing $J_\perp/J$.
Mitigating action, such as switching to single\hyp site \gls{dmrg}\hyp updates is discussed in more detail in~\cref{appendix:subsec:eedmrg}.
This would be done with the aim of being able to increase bond dimension significantly within the constraints of \acrshort{ram}, in order to pull free of the manifold of metastable states that appears at finite flux.
However, in this instance we have found no benefit from this approach so far, with energies only decreasing marginally on what in the overwhelming number of cases are still clearly metastable excited states of the effective \gls{2d} Hamiltonian.
%
%
\subsection{Fermions on square lattice}
Our final system are spin\hyp $1/2$ fermions on a square lattice with rectangular geometry, $L \neq W$.
In general, stiff systems of weakly coupled \gls{2d} Fermi\hyp Hubbard chains with repulsive interactions are minimal models for superconductivity as well as competing insulating phases in the organic Bechgaard and Fabré salts, and have been studied as such~\cite{PhysRevB.100.075138,Bourbonnais2011}.
Generally, this is outstandingly difficult for controlled\hyp approximation many\hyp body calculations, for the same reasons as for the isotropic \gls{2d}\hyp Hubbard model of the high\hyp $T_c$ superconductors:
the sign\hyp problem for \gls{qmc} techniques, superlinear growth of entanglement\hyp entropy with subsystem surface area for \gls{mps}\hyp based algorithms~\cite{PhysRevX.10.031016}.
Here, we want to provide a proof\hyp of\hyp concept study for the capacity of the \gls{ee}\hyp framework to obtain results of comparable quality to those obtained from massively parallelized p\gls{dmrg} at a small fraction of the computational effort, which moreover can treat far wider lattices, and thereby actually address the \gls{2d} limit.
The $U-V$-model of the Bechgaard and Fabré salts has the following form:
\begin{align}
	\notag
	\label{eq:ferms-rect-h1-hamiltonian}
	\qop H_1
	=
	-\sum_{xy,\tau=\uparrow\downarrow}
	\left(
	\hspace{-0.15cm}
	\vphantom{
		\qop c^\dagger_{\tau,xy}
		\qop c^\nodagger_{\tau,x+1y}
		+
		\qop c^\dagger_{\tau,x+1y} \qop c^\nodagger_{\tau,xy}
	}
	\right.
	&
	\left.
		\qop c^\dagger_{\tau,xy}
		\qop c^\nodagger_{\tau,x+1y}
		+
		\qop c^\dagger_{\tau,x+1y} \qop c^\nodagger_{\tau,xy}
	\right)
	\\
	&
	+ U/t
	\sum_{xy} \qop n_{\uparrow\downarrow,xy}
	+
	V/t
	\sum_{xy} \qop n_{xy} \qop n_{x+1y}
	\mathcomma
	\\
	\qop H_\perp
	=
	-\sum_{xy}
	\left(
	\hspace{-0.15cm}
	\vphantom{
		\qop c^\dagger_{\tau,xy}
		\qop c^\nodagger_{\tau,xy+1}
		+
		\qop c^\dagger_{\tau,xy+1} \qop c^\nodagger_{\tau,xy}
	}
	\right.
	&
	\left.
		\qop c^\dagger_{\tau,xy}
		\qop c^\nodagger_{\tau,xy+1}
		+
		\qop c^\dagger_{\tau,xy+1} \qop c^\nodagger_{\tau,xy}
	\right)
	\mathcomma
	\label{eq:ferms-rect-hperp-hamiltonian}
\end{align}
where $c^{(\dagger)}_{xy}$ are now the fermionic creation and annihilation operators on the respective site and chain, $\qop n_{\uparrow\downarrow,x,y}$ counts the number of doublons, and generally ${U\gg V >t\gg t_\perp}$
\par
In principle, we can apply the same \gls{ee} as in~\cref{eq:2d-ee-generic} to~\cref{eq:ferms-rect-h1-hamiltonian,eq:ferms-rect-hperp-hamiltonian} and obtain a new supersite Hamiltonian.
However, additional care must be taken due to the fermionic nature of the basis spanning the Hilbert space, i.e. we transform~\cref{eq:ferms-rect-hperp-hamiltonian} to be
\begin{widetext}
	\begin{align}
		\mel
		{
			\bm i^\noprime
		}
		{
			\qop H_\perp
		}
		{
			\bm i^\prime
		}
		&=-
		\sum_{xy,\tau}
		\left[
			\mel{i^\noprime_W\ldots i^\noprime_1}{\qop c^\dagger_{\tau,xy} \qop c^\nodagger_{\tau,xy+1}}{i^\prime_1\ldots i^\prime_W}
			+
			\hc
		\right]
		\\\label{eq:UVsso}
		&=-
		\sum_{xy,\tau}
		\left[
			\prod_{j = 1}^{y-1}
			\left(
				(-1)^{n^\noprime_j} (-1)^{n^\prime_j}
				\delta_{n^\noprime_j n^\prime_j}
			\right)
			(-1)^{n^\prime_y}
			\mel{i^\noprime}{\qop c^\dagger_{\tau,x}}{i^\prime}_y
			\mel{i^\noprime}{\qop c^\nodagger_{\tau,x}}{i^\prime}_{y+1}
			\prod_{j = y+2}^{W}
			\delta_{n^\noprime_j n^\prime_j}
			+
			\hc
		\right]
		\mathperiod
	\end{align}
\end{widetext}
Here, $i = (n,z,e)$ again counts the \gls{1d} eigenstate basis divided into its subsector through particle number and spin.
Furthermore, we recognize the regular Jordan\hyp Wigner transformation of ``local'' operators where the active site is mediating between the local orbitals and introduces signs.
However, now the mediation happens by means of matrix elements of the \gls{1d} basis.
Apart from that, sites to the right of the active site are identities and sites to the left of it are parity operators counting the total occupation per chain.
Squaring the parity gives the identity which effectively makes all the sites left of $y$ trivial.
Nevertheless, writing it in this way helps when implementing supersite local operators in code.
Thus,  creation and annihilation operators read
\begin{align}
	\notag
	\qop \gamma_y(\tau,x)^{(\dagger)}
	=
	\qop P_1 \otimes&\ldots\otimes \qop P_{y-1}
	\\
	\notag
	&\otimes
	\sum_{i^\noprime i^\prime}
	\mel{
		i^\noprime
	}
	{
		\qop c^{(\dagger)}_{\tau,x}
	}
	{
		i^\prime
	}
	\outerproduct{
		i^\noprime
	}
	{
		i^\prime
	}
	\\
	&\otimes
	\qop 1_{y+1}
	\otimes\ldots\otimes
	\qop 1_W
\end{align}
with the parity operator $\qop P_y = (-1)^{\qop n_y}$, which is generalized to the total number\hyp operator ${{\qop n_y}=\sum_x {\qop n_{x,y} }}$ for chain $y$ instead of the number of particles on a site.
Note that by exploiting symmetries it yields that the only non zero matrix elements are
\begin{align}
	&
	\mel
	{
		n+1,z+\tau,e^\noprime
	}
	{
		\qop c^\dagger_{\tau,x}
	}
	{
		n,z,e^\prime
	}
	\\
	&=
	\mel
	{
		n,z,e^\prime
	}
	{
		\qop c^\nodagger_{\tau,x}
	}
	{
		n+1,z+\tau,e^\noprime
	}
	^\ast
	\mathperiod
\end{align}
With this definition, the perpendicular Hamiltonian reduces drastically to
\begin{align}
	\label{eq:ferm-rect-ee-hamiltonian}
	\qop H_\perp
	&=
	-
	\sum_{xy,\tau}
	\left[
		\qop \gamma_y (\tau,x)^\dagger
		\qop \gamma_{y+1} (\tau,x)^\nodagger
		+
		\qop \gamma_{y+1} (\tau,x)^\dagger
		\qop \gamma_y (\tau,x)^\nodagger
	\right],
\end{align}
which we recognize as the hopping between supersites.
\par
We now want to investigate $L\times W$ strips again, this time with \glspl{obc} in both directions to ease comparison with prior work that was done in the same manner.
Due to the known density of itinerant charge in the quasi\hyp \gls{1d} Bechgaard and Fabré salts we focus on $N = LW/2$~\cite{Jerome2024}, which means we can use the particle number sectors $0 \leq n \leq L$ for the \gls{1d} basis set again.
However, technically sectors $n > L$ are also possible, they are just heavily suppressed due to the strong repulsive interactions.
This also means we can restrict ourselves to supersite occupation numbers $0 \leq n_y \leq L$, even though, we restrict this number in practice even further, as is discussed in~\cref{subsec:ferm-1d-basis}.
Running a quick \gls{ee}\hyp\gls{dmrg} with low bond dimension and evaluating the chain \gls{rdm} in the centre of the system yields the symmetry sectors and respective number of excitations in~\cref{table:ferm-1d-basis}.
The interchain coupling is chosen to be $t_\perp/t = 0.1$, a typical value for the Bechgaard and Fabré salts~\cite{Jerome2024}.
For the electron interactions, we used ${U/t= 4}$, ${V/t=1}$.
While these values are not quite representative for those obtained from actual materials, they allow us to compare to pre\hyp existing data.
\par
\begin{figure}
	\centering
	\includegraphics[width = 0.45\textwidth]{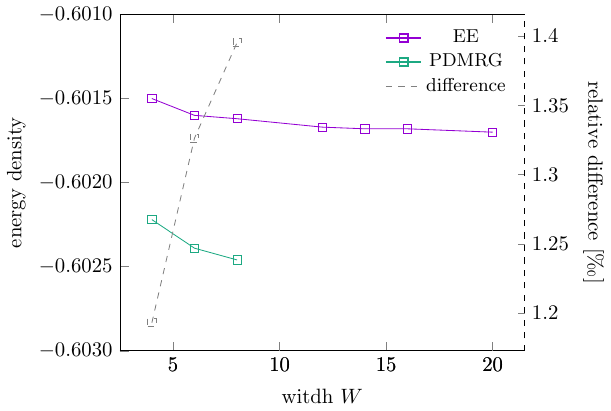}
	\caption
	{%
		\label{fig:rect-ferm-energies-comparison}%
		\gls{ee} groundstate energy of an $20\times W$ strip of fermions with on\hyp site and nearest neighbour repulsion $U = 4$ and $V = 1$, respectively.
		Both directions have \glspl{obc} and the system is at quarter filling $N = LW/2$.
		The chains are coupled with $t_\perp = 0.1$.
		The \gls{ee} results are obtained with bond dimension $m = 512$ while the multi\hyp node \gls{dmrg} from~\cite{PhysRevB.100.075138} used a bond dimension of $m = \SI{18}{\kilo{}}$.
		The relative difference on the right $y$\hyp axis (\protect\dashedline) is of the order of a permill.
	}
\end{figure}

For the benchmark, we again compare both energies per site and correlation functions to previous results obtained from brute force  distributed-memory \gls{dmrg} calculations at high bond dimension.
The comparison of the energy\hyp densities is shown in~\cref{fig:rect-ferm-energies-comparison}.
The chains coupled with a bond dimension of $m = 512$ are around a permill higher in energy than those of the brute\hyp force calculations performed with $m = \SI{18}{\kilo{}}$.
Whether this difference originates from the lower interchain entanglement or the truncated local Hilbert space will be the subject of future work.
In any case, further mitigation would demand more \acrshort{ram} to be accessible.
\par
Since the hopping in perpendicular direction is weak, the mapping of the \gls{2d} system to \gls{mps} in~\cite{PhysRevB.100.075138} had to be done in the strong\hyp coupling direction in order to achieve convergence.
In practice, this leads to a hopping\hyp range of length $L$ for the perpendicular tunneling for each site, necessitating the high bond dimension to converge the system.
This makes it very demanding to scale the system in the perpendicular direction  using the brute\hyp force approach.
In contrast, \gls{ee} can readily scale in this direction even up to square geometry since the supersite Hamiltonian~\cref{eq:ferm-rect-ee-hamiltonian} is local.
As we write those parts of the \acrlong{tn} to disk which are not involved in the local \gls{dmrg} optimisation, the influence of the system's width on memory consumption is constant.
We found this to be around $\SI{1.8}{\tera\byte}$ for our maximum bond dimension of $m = 512$.
Regarding the runtime, the scaling is linear as the \gls{mpo} bond dimension is constant ($\sim \num{80}$ for our benchmark), as we just need to update more supersites.
A detailed discussion of the computational requirements can be found in~\cref{appendix:subsec:eedmrg}.
\par
Energies and other operators with extensively scaling operator norms might be notoriously difficult observables for \gls{ee} to match up to high numerical accuracy.
For the chosen parameters of the \gls{2d} $U-V$ Hubbard Hamiltonian, at first glance the energies obtained seem to suggest the system to behave as effectively uncoupled \gls{1d} chains.
Specifically, the isolated\hyp chain ground state energy from \gls{dmrg} is
\begin{align}
	E_\mathrm{1D} = \num{-12.022 406 40}
    \mathcomma
\end{align}
while the slope of the \gls{ee} energy shown in~\cref{fig:rect-ferm-energies-comparison} is
\begin{align}
	E_\mathrm{EE}/W =
	\num{-12.035 131 86}
	\pm
	\num{7.64e-06}
    \mathperiod 
\end{align}
To illustrate that we actually probe the \gls{2d} regime, we analyse chain\hyp local quantities such as correlators.
Concretely, we follow~\cite{PhysRevB.100.075138} in defining the nearest neighbour spin\hyp singlet operators
\begin{align}
	\qop D_{x,y,a,b}
	&=
	\qop c_{x,y,\uparrow}
	\qop c_{x+a,y+b,\downarrow}
	-
	\qop c_{x,y,\downarrow}
	\qop c_{x+a,y+b,\uparrow}
\end{align}
that connect the two central chains.
From these, we build the $d_{xy}$\hyp correlation function (c.f.~\cref{fig:dxy}), as this order has been identified as the most likely instability for these parameters in reference~\cite{PhysRevB.100.075138}:
\begin{figure}
	\centering
	\includegraphics[width = 0.4\textwidth]{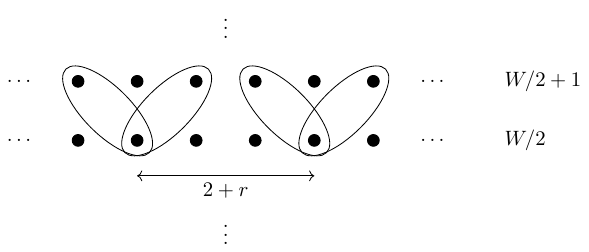}
	\caption
	{
		\label{fig:dxy}%
		Singlets on neighbouring chains are connected by the $d_{xy}$ correlator defined in~\cref{eq:dxy}.%
	}
\end{figure}
\begin{align}
	\label{eq:dxy}
	\notag
	d_{xy}(r)
	=
	\bra
	{
		\psi
	}
	{
		&\left(
			\qop D_
			{
				L/2+1 , W/2 , -1 , 1
			}
			-
			\qop D_
			{
				L/2+1 , W/2 , 1 , 1
			}
		\right)^\dagger
		\\
		&\cdot
		\left(
			\qop D_
			{
				L/2+3+r , W/2 , -1 , 1
			}
			-
			\qop D_
			{
				L/2+3+r , W/2 , 1 , 1
			}
		\right)
	}
	\ket
	{
		\psi
	}
	\mathperiod
\end{align}
In evaluating this correlator in \gls{ee} we need to multiply out all 16 terms in~\cref{eq:dxy} and obtain operator strings which always consists of two creation and two annihilation operators.

\begin{figure}
	\centering
	\includegraphics[width = 0.45 \textwidth]{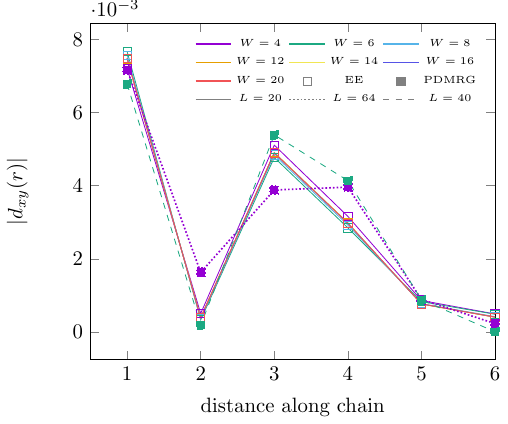}
	\caption{%
		\label{fig:ferm-correlator}%
		$d_{xy}$\hyp correlator for a \gls{2d} $U-V$ Hubbard model with $U/t=4$, $V/t=1$ at quarter filling $N = LW/2$, for different numbers of chains $W$, coupled with $t_\perp = 0.1$. 
        \gls{ee}\hyp results for chains of length $L = 20$ (\protect\blackline) are compared to results computed using distributed\hyp memory p\gls{dmrg} for four or six chains of length $L = 64$ (\protect\dottedline) and $L = 40$ (\protect\dashedline) respectively~\cite{PhysRevB.100.075138}.%
	}
\end{figure}

The comparison between the \gls{ee}\hyp calculated result and the brute\hyp force approach is shown in~\cref{fig:ferm-correlator}, for a system with $L = 20$.
We see excellent agreement of the absolute values of the correlators in comparison to the p\gls{dmrg} data, in particular given their small amplitudes.
The strong enhancement to the short\hyp range correlations of the $d_{xy}$ singlets that is exclusive to $W\geq 4$ and which was described in~\cite{PhysRevB.100.075138} is a mark of the \gls{2d} nature of the model at these parameters.
The \gls{ee}-framework reproduces this effect with a small fraction of that computational effort.

\section{Conclusion}\label{sec:conclusions}
The present work introduces a new many\hyp body numerical algorithm for calculating the ground state of correlated \gls{2d} systems with spatial anisotropy.
It exploits that modern \gls{mps}\hyp based numerics can obtain a substantial number of the low\hyp lying many\hyp body eigenstates of the \gls{1d} sub\hyp units making up the systems, especially if symmetries are exploited.
These states are then used to span the low\hyp energy manifold of the full \gls{2d} system, within which a second \gls{dmrg} algorithm is deployed to obtain the ground state.
The \gls{ee}\hyp approach is thus able to efficiently exploit the spatial anisotropy, provided that the energy\hyp spectrum spanned by the \gls{1d} basis sets is at least as wide as the effective coupling mediated by the interactions between the sub\hyp units.
Where possible, we validate the algorithm either against \gls{qmc} numerics (for hardcore bosons on the square lattice), or brute\hyp force calculations carried out with distributed-memory \gls{dmrg} (\gls{2d} $U$\hyp $V$ Hubbard model at $N/L = 0.5$).
\par
We demonstrate that the \gls{ee}\hyp approach can correctly detect the \gls{afm} emerging for localized spins on the anisotropic square lattice, as validated by \gls{qmc} numerics.
We then deploy the \gls{ee}-framework to spins on the anisotropic triangular lattice, which has been studied intensely for its potential to realize spin liquids.
Treating lattices of unprecedented size, we provide evidence for the existence of a gapless quasi-\gls{1d} spin liquid in this model at the lowest couplings that we study.
At the same time, our data appears to be at variance with earlier predictions that higher-order processes might stabilize an \gls{afm} at low coupling, or of a quantum phase transition occurring at intermediate transverse coupling.
At present, we are precluded from studying the expected transition to spiral ordered phases by persistent convergence problems in the presence of finite fluxes.
We also show the ability of the approach to treat repulsively interacting fermions away from integer filling for large systems that are not confined to narrow strips.
Here, we demonstrate that the \gls{ee}-technique can provide results of comparable quality to those obtained by brute-force approaches, but with far lower resource requirements.
\par
At the technical level, the present work invites future studies into the cause for the persistent gaps -- in those cases where they can be quantified -- between the observables computed from the \gls{ee}-approach and the true value of the observable, as can be seen e.g. in~\cref{appendix:fig:d-scaling}.
As the size of the \gls{1d} basis\hyp set is increased, observables saturate to values that are generally different from their true value.
We hypothesize that this persistent gap, which is small at the lower end of the system sizes and inter\hyp chain couplings covered in this work and continuously grows from there, is caused either individually or collectively by the vast manifold of \gls{1d} basis states that fall below the user-defined cut\hyp off, as discussed in~\cref{subsec:error-sources}.
While each on their own may contribute only near\hyp infinitesimally to an observable, the object of future work would be to quantify whether and how their exponentially growing number may result in an overall contribution that scales with system size and inter-chain coupling, as we find to be the case.
\par
At present, the \gls{ee}\hyp framework as shown here has to deal with large intermediate memory\hyp requirements.
These are caused by the interplay of the contractions when building up the Krylov\hyp subspace vectors in the standard formulation of the \gls{dmrg}\hyp algorithm, with the large local Hilbert\hyp spaces spanned by the basis sets of the \gls{1d} sub\hyp units.
For the larger systems for which we demonstrate the \gls{ee}\hyp approach here, this limits the deployment of the method to large\hyp \acrshort{ram} machines at present.
However, as discussed in~\cref{appendix-section:bottleneck-analysis}, there would be alternative versions of the \gls{dmrg} algorithm that could trade an increase in computational time for a drastic decrease of the intermediate memory requirements.
\par
Another aspect to be explored in future work would be the extension of the \gls{ee}\hyp framework to systems at finite temperatures or when evolving out\hyp of\hyp equilibrium.
The most common way of deploying \gls{mps}\hyp based techniques to finite temperatures, state purification, could be applied by the use of the \acrshort{tdvp}\hyp technique for imaginary time evolution, but \acrshort{metts} might also be an attractive possible alternative, given the potentially large local Hilbert spaces of our new method.
Of course, computing finite\hyp temperature properties within the \gls{ee}\hyp approach presupposes that the temperature would be small compared to the spread of energies covered by the \gls{1d} basis sets.
Analogously, the study of out\hyp of\hyp equilibrium dynamics within the \gls{ee}\hyp framework would need to focus on processes that evolve slowly enough to be captured within the bandwidth of \gls{1d} basis states.
Depending on the range of coupling in\hyp between \gls{1d} subunits, it could be carried out by running either a \acrshort{tebd}\hyp algorithm for real time evolution (if coupling is strictly nearest\hyp neighbour), or \acrshort{tdvp} (if any of the couplings would be longer\hyp ranged).
If spectral functions related to perturbing operators $\hat{O}_{1,2}$ were desired, these could be obtained from the Fourier\hyp transform of the matrix element
$
\ev{
	e^{\imagunit E_\mathrm{GS}t}
	\hat{O}_2
	e^{-\imagunit\hat{H}t}
	\hat{O}_1
}
{
	\psi
}
$%
, where $\hat{H}$ and
$\ket{\psi}$ represent the Hamiltonian and 2D wavefunction, as approximated by the \gls{ee}\hyp framework, respectively.
Spectral functions could also be computed directly in frequency\hyp space using $\hat{H}$ and $\ket{\psi}$,
$
\ev
{
	\hat{O}_2
	(\omega+\imagunit\eta-\hat{H})^{-1}
	\hat{O}_1
}
{
	\psi
}
$
as long as $\omega$ is significantly smaller than the spread of energies covered by the \gls{1d}-basis $\ket{e}$~\cite{Holzner2011}.
\par
Finally, another interesting line of future work would lie in comparing the \gls{ee}\hyp technique against other algorithms that can treat large correlated \gls{2d}\hyp systems, but are not tailor-made for the spatially anisotropic regime,
such as \acrshort{peps} and phase\hyp free or constrained\hyp path \gls{afqmc}.
In this setting, it can be difficult to estimate the degree of approximation induced by these methods and
the \gls{ee}\hyp approach could serve to validate calculations with these alternative algorithms.

\begin{acknowledgments}
S.M. and A.K. would like to thank Thierry Giamarchi for helpful discussions.
This work was supported by an ERC Starting Grant from the European Union’s Horizon 2020 research and innovation programme under grant agreement No. 758935; and the UK’s Engineering and Physical Sciences Research Council [EPSRC; grant number EP/W022982/1]. 
J.M. acknowledges funding through SNSF Swiss Postdoctoral Fellowship, grant no. 210478.
The computations were enabled by resources provided through multiple EPSRC ``Access to HPC'' calls (Spring 2023, Autumn 2023, Spring 2024 and Autumn 2024) on the ARCHER2, Peta4-Skylake and Cirrus compute clusters, as well as by compute time awarded by the National Academic Infrastructure for Supercomputing in Sweden (NAISS).
This work was supported by a grant from the Swiss National Supercomputing Centre (CSCS) under project ID s1307 on Alps.
S.M. gratefully acknowledges the computing time provided on the high\hyp performance computers noctua2 at the NHR Center PC2 in Paderborn. These are funded by the Federal Ministry of Education and Research and the state governments participating on the basis of the resolutions of the GWK for the national highperformance computing at universities.
All calculations were carried out with the publicly available containers of the SyTen toolkit~\cite{hubig:_syten_toolk}.
N.L. acknowledges the
ANR research grant ManyBodyNet No. ANR-24-CE30-
5851, the support of the Fondation Simone et Cino Del
Duca, and the use of HPC
resources from CALMIP (grants 2023-P0677) and GENCI
(project A0150500225).
\end{acknowledgments}
\appendix
\section{\Gls{ee}\hyp\gls{dmrg} finite bond dimension error}
\begin{figure}
	\centering
	\includegraphics[width = 0.45\textwidth]{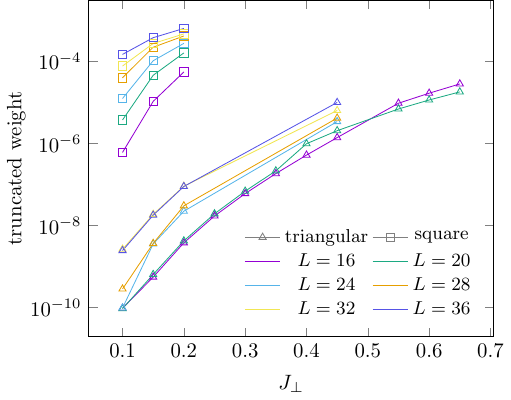}
	\caption{%
		\label{fig:spin-hcb-trunc-weight}%
		Truncated weights for the final sweep of the different system sizes and intrachain couplings $J_\perp$ in~\cref{subsec:spins}.%
	}
\end{figure}

The error arising from finite bond dimension can be estimated used the truncated weight $w$, i.e. the sum of discarded singular values $s_i$
\begin{align}
	w = \sum_
	{
		i = m+1
	}
	^
	{
		M
	}
	s^2_i
	\mathcomma
\end{align}
where $m$ is the number of states kept and $M$ is the bond dimension before truncation.
Often the truncated weight for the investigated model is constant for different parameter settings and therefore mentioned in the main text.
For the systems discussed in~\cref{subsec:spins} it depends on the lattice geometry and coupling and therefore is displayed in~\cref{fig:spin-hcb-trunc-weight}.
\section{Single chain excitation probabilities}
\begin{figure*}
	\centering
	\includegraphics[height = 0.8\textheight]{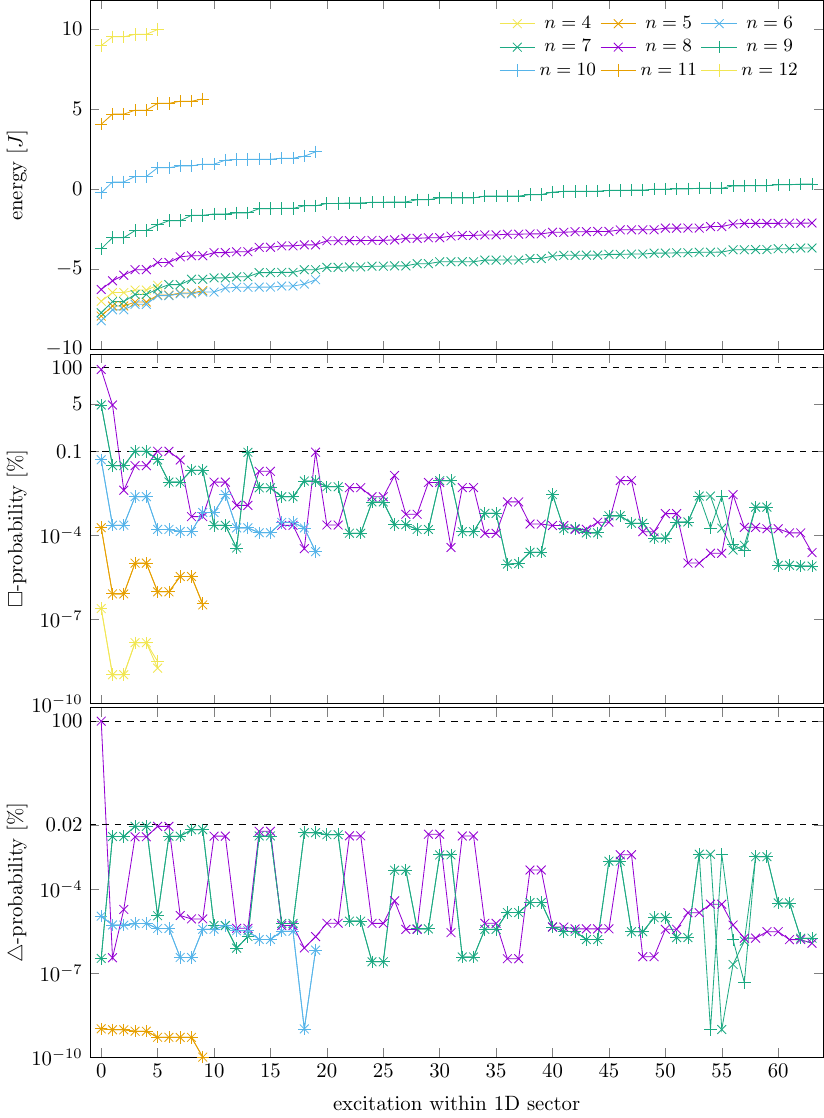}%
	\caption{%
		\label{fig:spin-hdb-rdm-probabilities}%
		Energy bandwidth of the \gls{1d} basis states and excitation probability in the center of the \gls{ee} wavefunction for the system described in~\cref{eq:spin-hcb-sql-ee-hamiltonian,eq:spin-hcb-triangular-ee-hamiltonian}.
		All data shown here is for chains of length $L = 16$ and in the center of a $W = 16$ supersites system with $J_\perp = 0.1$ and the local basis configuration described in~\cref{eq:boson-local-Hilbert-space}.
		Changing the lattice geometry from square to triangular introduces frustration so efficient that the chains transition from a weakly coupled system to one of nearly independent chains, i.e. a product wavefunction of the \gls{1d} ground states.%
	}
\end{figure*}

In order to estimate the importance of a particular chain energy mode within the \gls{2d} system, we plot the single chain excitation probability
\begin{align}
	\label{eq:single-chain-rdm}
	\qop\rho_y
	&=
	\tr_
	{
		y^\prime \neq y^\noprime
	}
	\outerproduct
	{
		\psi
	}
	{
		\psi
	}
	\mathperiod
\end{align}
We trace out everything except for the respective chain's energy eigenmodes.
The diagonal elements of this matrix give us the excitation probability for the respective mode which can be seen in~\cref{fig:spin-hdb-rdm-probabilities} together with the \gls{1d} energy bandwidth.
\section{Fermionic \gls{1d} basis}
\label{subsec:ferm-1d-basis}
Due to the fact that fermions have spin degeneracy for each particle number sector, we need to give a vector of \gls{1d} energy eigenstates kept in \gls{ee} for each of those sectors.
This scheme is depicted in~\cref{table:ferm-1d-basis}.
\begin{table}
	\centering
	\begin{tabular}{l|l|r}
		particle number $n$ & spin $\abs{z}$ & \gls{1d} states
		\\
		\hline
		$L/2-3$ & $1/2$ & $e = 0$
		\\
		\hline
		$L/2-2$ & $\leq 2$ & $0 \leq e \leq 9$
		\\
		\hline
		\multirow{2}{*}{$L/2-1$} & $5/2$  & $e = 0$
		\\
		& $\leq 3/2$  & $0 \leq e \leq 15$
		\\
		\hline
		\multirow{3}{*}{$L/2$} & $3$ & $e = 0$
		\\
		& $\leq 2$ & $0 \leq e \leq 19$
		\\
		& $0$ & $0 \leq e \leq 31$
		\\
		\hline
		\multirow{2}{*}{$L/2+1$} & $7/2,\ 5/2$ & $e = 0$
		\\
		& $\leq 3/2$ & $0 \leq e \leq 15$
		\\
		\hline
		\multirow{2}{*}{$L/2+2$} & $4,\ 3$ & $e = 0$
		\\
		& $\leq 2$ & $0 \leq e \leq 9$
		\\
		\hline
		\multirow{2}{*}{$L/2+3$} & $3/2 \leq \abs{z} \leq 9/2$ & $e = 0$
		\\
		& $1/2$ & $0 \leq e \leq 5$
	\end{tabular}
	\caption{%
		\label{table:ferm-1d-basis}%
		Single chain sectors incorporated in fermionic wavefunctions and their associated number of excitations.
		Note that some of the sectors might not be necessary as judged by the \gls{rdm}, but still need to be included due to the creator (or annihilator) mapping from a state which has high enough weight to this low weight state.
		In this case, we added the ground state of this sector such that the block has size one and the \acrlong{irrep} is present in the wavefunction.%
	}
\end{table}
\section{Key algorithmic steps}
\label{appendix-section:bottleneck-analysis}
\subsection{\acrshort{1d} eigensystem}
\label{appendix-subsec:1d-eigensystem}
We obtain the \gls{1d} eigenbasis of the chain by using conventional \gls{dmrg}~\cite{SCHOLLWOCK201196}.
\gls{dmrg} has proven its utility and accuracy in efficiently computing eigenstates of \gls{1d} systems at high accuracy.
Its performance stems from the small entanglement entropy of low\hyp energy states scaling only very weakly with subsystem size in general.
Once the ground state of the system $\ket{e = 0}$ is found, we can compute low-lying excitations by minimising the energy in \gls{dmrg} with the additional constraint of orthogonality to all previously computed states~\cite{SCHOLLWOCK201196}.
\begin{align}
	\label{eq:orthogonality-condition}
	\min_
	{
		\ket{e}
	}
	\left[
		\ev
		{
			\qop h
		}
		{
			e
		}
		-
		E_e \innerproduct{e}
		-
		\sum_
		{
			e^\prime = 0
		}
		^
		{
			e^\noprime
		}
		\lambda_
		{
			e^\prime
		}
		\innerproduct
		{
			e^\prime
		}
		{
			e^\noprime
		}
	\right]
	\mathperiod
\end{align}
Algorithmically, this consists in keeping the states against which the current excitation should be orthogonalised in the usual canonical form and computing their overlaps with the \gls{mps} site tensor in the local diagonalization step of \gls{dmrg}.
This part of the algorithm is shown in~\cref{appendix:fig:1d-ortho-procedure}.
\begin{figure}
	\centering
	\includegraphics
	[
		width = 0.4\textwidth
	]
	{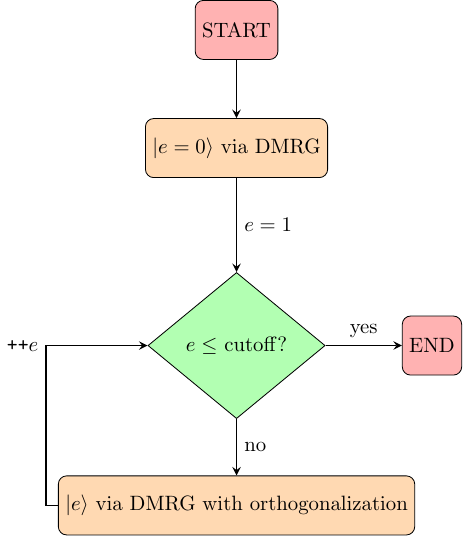}
	\caption
	{%
		\label{appendix:fig:1d-ortho-procedure}%
		Schematic depiction of the orthogonalization procedure to obtain the low lying excitations of the \gls{1d} Hamiltonian $\qop H_1$.
		Note that we can perform this scheme for each symmetry sector separately and thus in parallel.%
	}
\end{figure}
Depending on the boundary conditions, the exact geometry of the \gls{1d} lattice, the Hamiltonian parameters and the exchange statistics of the particles, the typical bond dimension for convergence of such a state can range from $m \sim \mathcal{O}\left(\num{100}\lldots\num{1000}\right)$.
The typical \gls{mps} size on disk then reaches from below $\SI{1}{\mega\byte}$ to several hundreds.
The calculation itself uses $\LandauO{\SI{10}{\giga\byte}}$ of \acrshort{ram} and takes between minutes and hours.
Since all lower-lying excitations are necessary to compute an excited state, there is no room for parallelisation.
However, in addition to the usual parallelisation over tensor blocks we can search in the different quantum number sectors in parallel, as described in~\cref{subsec:syms}.
\subsection{Matrix elements}
\label{appendix:subsec:matrixelems}
Once we have obtained a sufficient amount of \gls{1d} energy eigenstates to describe our system accurately, we need to extract the necessary matrix elements for the transformation of~\cref{eq:hamiltonian-splitting} into the \gls{1d} energy eigenbasis.
For each state $\ket{e,\sigma}$ labelled by its sector $\sigma$ and excitation number within that sector $e$, we need to compute the overlap with all the other states $\{\ket{e^\prime,\sigma+T}\}_{e^\prime}$.
Here, $T$ is the transformation behaviour of the operator - the change in quantum number(s) that applying it induces - that we are measuring and the target sector is $\sigma+T$.
The whole scheme is depicted in~\cref{appendix:fig:matrixelems-procedure}.
\begin{figure}
	\centering
	\includegraphics
	[
		width = 0.4\textwidth
	]
	{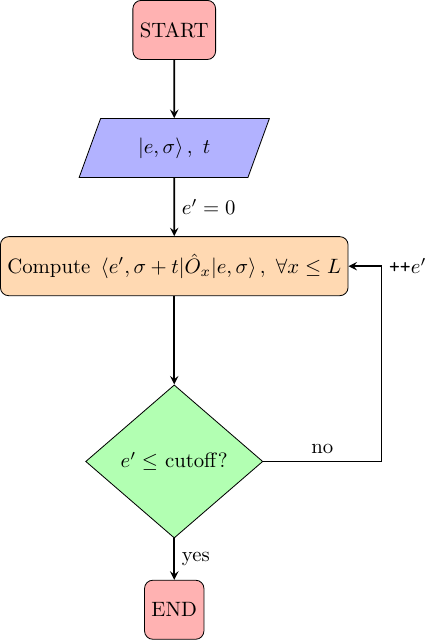}
	\caption
	{%
		\label{appendix:fig:matrixelems-procedure}%
		Computing the matrix elements for the local operators needed in \gls{ee}, typically $c^{(\dagger)}_x$ and $n_x$.
		For each operator and each eigenstate computed in \gls{1d} we need to compute the overlap with all eigenstates in the sector which we obtain by combining the original sector and the transformation behaviour of the operator at hand.
		Note that we can do this in parallel for all states and all operators.%
	}
\end{figure}
The typical size of these matrices is $\LandauO{10}\times\LandauO{10}$, i.e. the computational cost is negligible.
The overlap performance on two \glspl{mps} typically needs less than $\SI{10}{\giga\byte}$ of \acrshort{ram} and under one minute.
Again, we have the advantage that we can parallelize the computation over tensor blocks and, in addition to that, by computing all matrix elements in parallel since we do not have to wait for lower lying elements in the spectrum to finish.
\subsection{\acrshort{ee}\hyp\acrshort{dmrg}}
\label{appendix:subsec:eedmrg}
\begin{figure}
	\centering
	\subfloat
	[
		\raggedright
		\label{fig:ee-algo1}%
		Applying the active two\hyp site tensor to the left environment.
		This step greatly increases memory usage due to two open legs with large local Hilbert space $\tilde{d}$ size and an additional free \gls{mpo} leg.%
	]
	{\includegraphics[width = 0.4\textwidth]{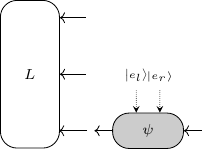}}%
	\\
	\subfloat
	[
		\raggedright
		\label{fig:ee-algo2}%
		Applying the \gls{mpo} site tensors to the composite object.
		This step still involves the high \acrshort{ram} requirement as the number of open legs remains the same and we contract with objects scaling as $\tilde{d}$ and the \gls{mpo} bond dimension.%
	]
	{\includegraphics[width = 0.4\textwidth]{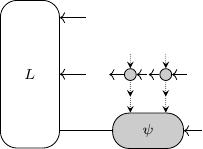}}%
	\\
	\subfloat
	[
		\raggedright
		\label{fig:ee-algo3}%
		Finally, this step gives us the updated \gls{mps} tensor and the memory requirement again drops to the well controllable amounts described in the main text.%
	]
	{\includegraphics[width = 0.4\textwidth]{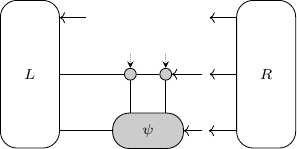}}%
	\caption
	{
		\label{fig:ee-algo}
		Key algorithmic steps of the local two\hyp site update in \gls{ee}\hyp\gls{dmrg}.
	}
\end{figure}

Here, we describe the most costly computational step of the algorithm.
Since our Hamiltonian represents each chain as a large effective supersite, the typical bond dimension of $\qop H_\perp$ is rather modest.
However, the \gls{mpo} bond dimension does scale with the length of the chains $L$, as the summation contains an element for each site $x$.
We typically found a bond dimension $\mathcal{O}(10)$ for the systems investigated in this work ranging from $\sim 30$ for  bosons on the square lattice to $\sim 80$ for spins on triangular lattices and the fermionic systems we looked at.
Together with the bond dimension accessible to \gls{ee} in two\hyp site \gls{dmrg}, this typically resulted in sizes of the left\hyp~and right\hyp environments being $\mathcal{O}(\SI{100}{\mega\byte})$.
Furthermore, the \gls{mps} tensors themselves would have a size of the order $\mathcal{O}(\SI{10}{\mega\byte}\lldots\SI{50}{\mega\byte})$.
We define the filling fraction of a tensor as the ratio between the amount of memory needed when the tensor is symmetry protected~\cite{ediss21348} against the hypothetical total amount of memory necessary would one want to store the object without blocks.
This yields a filling fraction of $\sim\SI{17}{\percent}$ for the \gls{mps} site tensors.
Using the same filling fraction to compute the size of the objects computed in~\cref{fig:ee-algo}, we arrive at around $\SI{500}{\giga\byte}$ just to store the partial contractions in \acrshort{ram}.
We note that the filling fraction is usually not a constant throughout a tensor network and hence the objects might actually be larger.
However, the magnitude matches the memory consumption experienced by us during simulations.
Upon multiplication of the site tensor into the left boundary the memory requirement for performing the further multiplications of a state with bond dimension $m = 256$ jumps to $\SI{2}{\tera\byte}$.
With the highest\hyp\acrshort{ram} machines available to us, we were able to perform this calculation, as we cache the other elements of the tensor network not needed for the local update to disk.
\par
There are two potential remedies for the high\hyp\acrshort{ram} requirements.
We outline these here, but pursuing them in detail would constitute a separate project.
The first is the use of single\hyp site \gls{dmrg} instead of two\hyp site optimisation~\cite{PhysRevB.91.155115}.
The advantage of the single\hyp site approach is that it requires only a single index running over the \gls{1d} basis states, of size $\tilde{d}$, during the optimization approach in~\cref{fig:ee-algo}, instead of the two indices, each of size $\tilde{d}$, for two\hyp site \gls{dmrg}.
For instance, a test calculation performed on the system in~\cref{subsec:spins} yielded a decreased \acrshort{ram} consumption by a factor of ten while keeping the bond dimension constant.
Given that we use $\tilde{d} = 264$ single\hyp chain basis states in production, this is an appropriate improvement.
In order to improve accuracy, we then increase bond dimension within the single\hyp site \gls{dmrg} algorithm up to $m = 1024$, where our original \acrshort{ram} restriction of $\SI{2}{\tera\byte}$ kicks in.
However, systems with weak interchain coupling would often converge around bond dimensions of about $m=500$ and not saturate the available number of states, in particular for $J_\perp < 0.5$.
The improved performance of single\hyp site \gls{dmrg}, however, entails potential problems regarding the convergence of the wavefunction.
Usually, single site \gls{dmrg} can run into convergence problems due to the limited size of the local variational Hilbert space available during the application.
Algorithms exist that attempt to mitigate this by mixing the optimised state tensor with additional data, such that the algorithm  might be able to explore a larger variational manifold~\cite{PhysRevB.91.155115}.
The truncation then automatically takes care of unnecessary degrees of freedom.
However, there is no guarantee for the expansion data to actually contain the necessary states for the \gls{mps} to reach a true minimum in the energy optimisation landscape.
Running these codes thus entails a case by case decision on whether single\hyp site \gls{dmrg} has found a state that converged equally well or even better than the one found by two-site \gls{dmrg} with fewer resources, or whether one might be stuck in a metastable state.
Within the \gls{ee}\hyp framework as deployed in the present work, there is the advantage that all supersite ``interactions'' are either local or at most nearest\hyp neighbour.
Hence, the single\hyp site \gls{dmrg} might not be obstructed by long range interactions, which is a fairly common problem for this algorithm.
For the triangular test systems the difference between single\hyp site and two\hyp site energies would be around $\SI{e-2}{\permille}$, with the single site \gls{dmrg} being lower in energy.
Key observables, e.g. the correlators in~\cref{fig:spin-hcb-ee-correlator}, would take exactly the same shape.
This overlap between both results indicates the correctness of our results obtained with finite bond dimension.
However, employing single\hyp site \gls{dmrg} for data production, we would consider additional checks to be necessary that will be the target for future work.
\par
The second possible mitigation to the resource problem might be introducing an additional step during the two\hyp site update, which is shown in~\cref{fig:ee-algo} for reference.
If we want to avoid the occurrence of objects with two supersite indices, each having a local dimension of $\tilde{d}$, we could compute the contraction of the left environment, the left \gls{mps} and the left local \gls{mpo} tensor separately from the analogous contractions for the right\hyp sided objects.
The updated two\hyp site tensor would then be computed from contracting the left\hyp side and the right\hyp side network.
Furthermore, left\hyp side and right\hyp side network would need to be kept separate consistently as the Krylov\hyp subspace is being build up by subsequent applications of the Hamiltonian.
This necessitates application of a  \acrlong{qr} for every two\hyp site tensor (i.e. after every application of the Hamiltonian).
By doing so, the memory cost would stay comparable to that of single\hyp site \gls{dmrg}, but at the cost of additional \acrshortpl{qr} at every update step.
Trading memory for compute time in this fashion would allow keeping the large variational Hilbert space of the two\hyp site procedure and hence avoid potential local minima.
Based on pilot results, we currently estimate a runtime that would be four to five times longer than thhat of the standard algorithm used in the present work.
\section{Energy eigenstate observable overlaps}
As discussed in~\cref{subsec:measurements}, observables computed with \gls{ee} wavefunctions are summations over the \gls{1d} matrix elements of $\qop H_1$ weighted by the \gls{rdm} of the corresponding chain supersite Hilbert spaces.
Here, we want to demonstrate the spatial behaviour of these matrix elements.
For instance, in order to compute the single\hyp particle density matrix of the system described in~\cref{subsec:hcb-rect}, we must expand
\begin{align}
	\label{appendix:fig:hcb-single-particle-correlator}
	\mel
	{
		\psi
	}
	{
		\qop b^\dagger_
		{
			x^\noprime
			c
		}
		\qop b^\nodagger_
		{
			x^\prime
			c
		}
	}
	{
		\psi
	}
	&=
	\sum_
	{
		n_y
		e^\noprime_y
		e^\prime_y
	}
	\rho^
	{
		n_y
		e^\prime_y
	}
	_
	{
		n_y
		e^\noprime_y
	}
	\mel
	{
		n e^\noprime
	}
	{
		\qop b^\dagger_{x^\noprime}
		\qop b^\nodagger_{x^\prime}
	}
	{
		n e^\prime
	}_c
	\mathcomma
\end{align}
where we introduced the single chain \gls{rdm}
\begin{align}
	\rho^
	{
		n_y
		e^\prime_y
	}
	_
	{
		n_y
		e^\noprime_y
	}
	&=
	M^
	{
		n_y
		e^\noprime_y
		\dagger
	}
	\cdot
	M^
	{
		n_y
		e^\prime_y
	}
\end{align}
and $c = W/2$ again.
We are now particularly interested in the \gls{1d} matrix elements of different eigenstates $\ket{e^\noprime}$ and $\ket{e^\prime}$ of $\qop H_1$.
In the case of $e^\noprime \neq e^\prime$ these are not observables themselves but their scaling is crucial for the \gls{ee} to be able to capture \gls{2d} physics such as off-diagonal long range order correctly.
The results for the system of size $L = 16$ at half filling $N = 8$ and chain nearest neighbour repulsion $V = 2$ are shown in~\cref{appendix:fig:1d-overlaps}.
\begin{figure}
	\centering
	\includegraphics[height = 0.95\textheight]
	{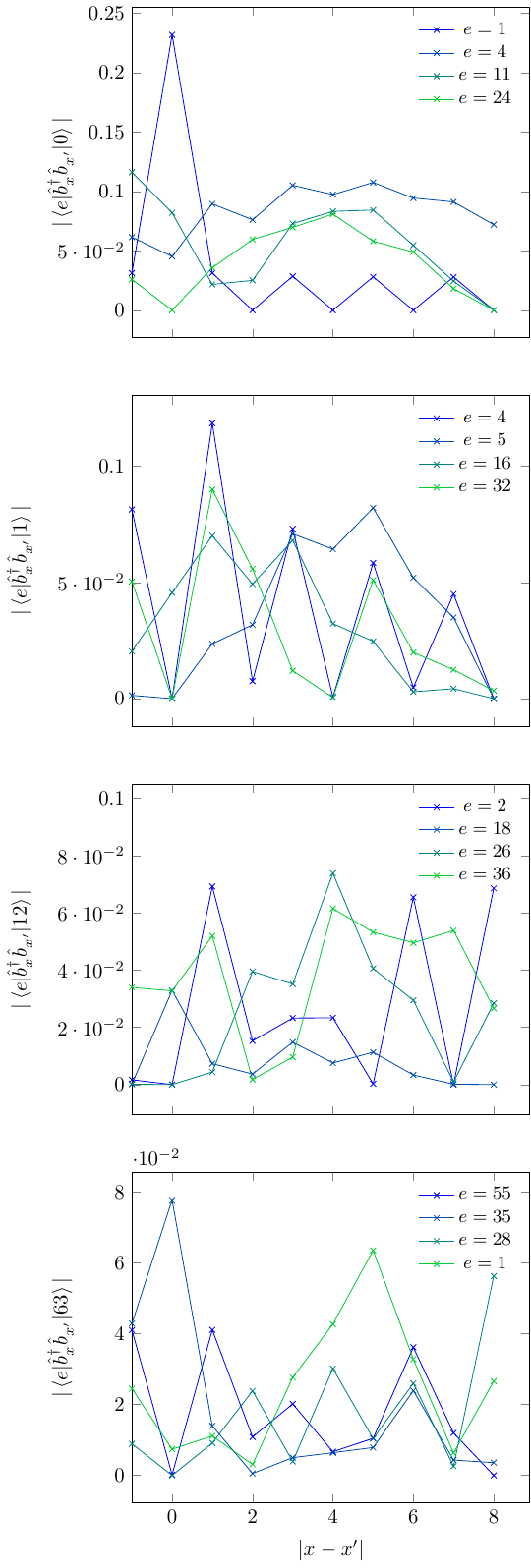}
	\caption
	{%
		\label{appendix:fig:1d-overlaps}%
		\gls{1d} single particle correlation functions for $N = 8$ \glspl{hcb} on a $L = 16$ chain with $V = 2$.
		The excited states with which the correlator was computed were manually picked to show functions which would not decay towards zero and hence could mimic the true \gls{2d} behaviour in the weighted sum~\cref{appendix:fig:hcb-single-particle-correlator}.%
	}
\end{figure}
\section{Observable scaling with supersite size}
The number of \gls{1d} states incorporated into the local supersite is crucial for the convergence of many observables and the quantum state $\ket\psi$ itself.
Here we want to demonstrate the impact that different values of the dimension $\tilde{d}$ of the retained single\hyp chain Hilbert space can have on the square\hyp lattice spin system discussed in~\cref{subsec:spins}.
As \gls{qmc} does not suffer from the sign problem for this model, we can validate convergence of the \gls{ee} framework against data that is exact for all practical purposes.
The definition of the perpendicular spin\hyp spin correlator is given in~\cref{eq:spin-hcb-perp,fig:spin-hcb-perp-corrlength}.
We emphasize that $\tilde{d}$ does not uniquely specify a basis, as one \gls{1d} state removed from one sector while another one is being added to another sector will not change the value of $\tilde{d}$.
\begin{figure}
	\centering
	\subfloat[%
		\label{appendix:fig:d-scaling-tp0.1}%
	]
	{\includegraphics[width = 0.45\textwidth]{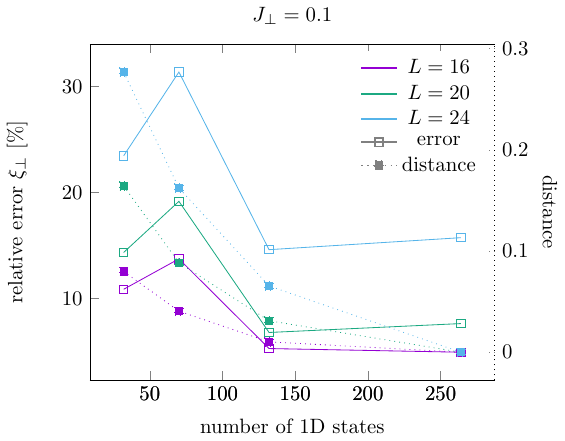}}%
	\\
	\subfloat[%
		\label{appendix:fig:d-scaling-tp0.2}%
	]
	{\includegraphics[width = 0.45\textwidth]{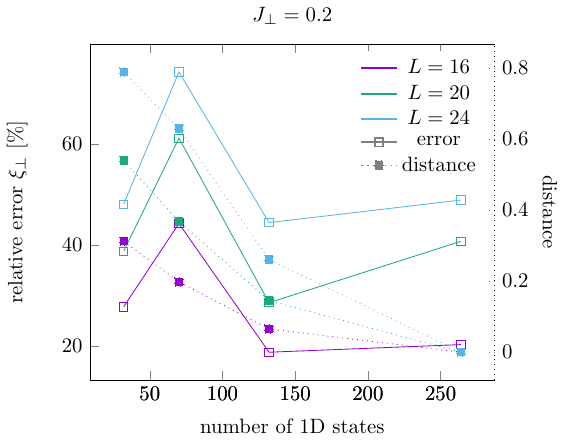}}%
	\caption{%
		\label{appendix:fig:d-scaling}%
		Relative error of perpendicular correlation length defined in~\cref{fig:spin-hcb-perp-corrlength} for the square lattice ($\square$) spin system defined in~\cref{eq:spin-hcb-sql-ee-hamiltonian} compared to exact \gls{qmc}.
		The right $y$\hyp axis shows the distance of the quantum states relative to the state with the largest local basis using dotted lines (\protect\dottedline).%
	}
\end{figure}
In addition, we introduce the distance of two states $\ket\psi$ and $|\tilde\psi\rangle$ as
\begin{align}
	f(\ket\psi, |\tilde\psi\rangle)
	&=
	\norm{
		\ket\psi
		-
		|\tilde\psi\rangle
	}^2_2
	\mathcomma
\end{align}
as a measure how much a state with large supersite Hilbert space $\ket\psi$ is varying from a state with a smaller supersite Hilbert space.
The relative error of the results compared to exact \gls{qmc} data is shown in~\cref{appendix:fig:d-scaling} for the anisotropic Heisenberg Hamiltonian on the square lattice (c.f.~\cref{subsec:spins}).
All results were obtained for a final bond dimension of $m = 256$ to maintain comparability, even though the smaller Hilbert space sizes can be computed with higher bond dimension.
We clearly recognize the distance to be a monotonically decreasing function in $\tilde{d}$ while the same does not hold for the correlation length $\xi_\perp$.
This is most likely due to higher bond dimension necessary to resolve correlations when more chain eigenstates are involved in the process.
\section{Spin to \acrlong{hcb} mapping}
\label{appendix-subsec:spin-to-hcb}
Starting from~\cref{eq:spin-hamiltonian} and identifying spin down with an empty and spin up with an occupied site we arrive at the \acrlong{hcb} Hamiltonians
\begin{align}
	\label{eq:spin-hcb-h1-hamiltonian}
	\notag
	2 H_1 =
	\sum_{xy}
	\left(
	\qop b^\dagger_{x,y} \qop b^\nodagger_{x+1,y}
	+
	\qop b^\dagger_{x+1,y} \qop b^\nodagger_{x,y}
	\right)
	\\
	+
	2/t\sum_{xy} \qop n_{x,y} \qop n_{x+1,y}
	+\const
	\mathcomma
	\\
	\label{eq:spin-hcb-sql-hperp-hamiltonian}
	\notag
	2 H_\perp =
	\sum_{xy}
	\left(
	\qop b^\dagger_{x,y} \qop b^\nodagger_{x,y+1}
	+
	\qop b^\dagger_{x,y+1} \qop b^\nodagger_{x,y}
	\right)
	\\
	+
	2/t\sum_{xy} \qop n_{x,y} \qop n_{x,y+1}
	+\const
	\mathcomma
	\\
	\label{eq:spin-hcb-triangular-hperp-hamiltonian}
	\notag
	2H^\prime_\perp =
	\sum_{xy}
	\left(
	\qop b^\dagger_{x,y} \qop b^\nodagger_{x+1,y+1}
	+
	\qop b^\dagger_{x+1,y+1} \qop b^\nodagger_{x,y}
	\right)
	\\
	+
	2/t\sum_{xy} \qop n_{x,y} \qop n_{x+1,y+1}
	+\const
	\mathcomma
\end{align}
where we introduced the ordinary creation and annihilation operators as well as the particle number operator.
This means the two models are essentially the same with the identifications
\begin{align}
	\label{eq:spin-to-hdb-identifications}
	t = J\mathcomma
	\qquad
	V = 2J\mathcomma
	\qquad
	t_\perp = J_\perp\mathcomma
	\qquad
	V_\perp = 2 J_\perp
	\mathperiod
\end{align}
It is worthwhile pointing out that the sign of the \gls{hcb} kinetic energy is positive now in all three~\cref{eq:spin-hcb-h1-hamiltonian,eq:spin-hcb-sql-hperp-hamiltonian,eq:spin-hcb-triangular-hperp-hamiltonian} which shifts the energy minimum of the single particle dispersion from zero momentum to  ${(\pi/a,\pi/a)}$.

\bibliography{literature}

\begin{thebibliography}{84}%
\makeatletter
\providecommand \@ifxundefined [1]{%
 \@ifx{#1\undefined}
}%
\providecommand \@ifnum [1]{%
 \ifnum #1\expandafter \@firstoftwo
 \else \expandafter \@secondoftwo
 \fi
}%
\providecommand \@ifx [1]{%
 \ifx #1\expandafter \@firstoftwo
 \else \expandafter \@secondoftwo
 \fi
}%
\providecommand \natexlab [1]{#1}%
\providecommand \enquote  [1]{``#1''}%
\providecommand \bibnamefont  [1]{#1}%
\providecommand \bibfnamefont [1]{#1}%
\providecommand \citenamefont [1]{#1}%
\providecommand \href@noop [0]{\@secondoftwo}%
\providecommand \href [0]{\begingroup \@sanitize@url \@href}%
\providecommand \@href[1]{\@@startlink{#1}\@@href}%
\providecommand \@@href[1]{\endgroup#1\@@endlink}%
\providecommand \@sanitize@url [0]{\catcode `\\12\catcode `\$12\catcode
  `\&12\catcode `\#12\catcode `\^12\catcode `\_12\catcode `\%12\relax}%
\providecommand \@@startlink[1]{}%
\providecommand \@@endlink[0]{}%
\providecommand \url  [0]{\begingroup\@sanitize@url \@url }%
\providecommand \@url [1]{\endgroup\@href {#1}{\urlprefix }}%
\providecommand \urlprefix  [0]{URL }%
\providecommand \Eprint [0]{\href }%
\providecommand \doibase [0]{https://doi.org/}%
\providecommand \selectlanguage [0]{\@gobble}%
\providecommand \bibinfo  [0]{\@secondoftwo}%
\providecommand \bibfield  [0]{\@secondoftwo}%
\providecommand \translation [1]{[#1]}%
\providecommand \BibitemOpen [0]{}%
\providecommand \bibitemStop [0]{}%
\providecommand \bibitemNoStop [0]{.\EOS\space}%
\providecommand \EOS [0]{\spacefactor3000\relax}%
\providecommand \BibitemShut  [1]{\csname bibitem#1\endcsname}%
\let\auto@bib@innerbib\@empty
\bibitem [{\citenamefont {Bourbonnais}\ and\ \citenamefont
  {Jérome}(2008)}]{Bourbonnais2007}%
  \BibitemOpen
  \bibfield  {author} {\bibinfo {author} {\bibfnamefont {C.}~\bibnamefont
  {Bourbonnais}}\ and\ \bibinfo {author} {\bibfnamefont {D.}~\bibnamefont
  {Jérome}},\ }\bibfield  {title} {\bibinfo {title} {Interacting {Electrons}
  in {Quasi}-{One}-{Dimensional} {Organic} {Superconductors}},\ }in\ \href@noop
  {} {\emph {\bibinfo {booktitle} {The {Physics} of {Organic} {Superconductors}
  and {Conductors}}}},\ \bibinfo {editor} {edited by\ \bibinfo {editor}
  {\bibfnamefont {A.}~\bibnamefont {Lebed}}}\ (\bibinfo  {publisher}
  {Springer},\ \bibinfo {address} {Heidelberg},\ \bibinfo {year} {2008})\ p.\
  \bibinfo {pages} {pp. 357}\BibitemShut {NoStop}%
\bibitem [{\citenamefont {Jerome}\ and\ \citenamefont
  {Bourbonnais}(2024)}]{Jerome2024}%
  \BibitemOpen
  \bibfield  {author} {\bibinfo {author} {\bibfnamefont {D.}~\bibnamefont
  {Jerome}}\ and\ \bibinfo {author} {\bibfnamefont {C.}~\bibnamefont
  {Bourbonnais}},\ }\bibfield  {title} {\bibinfo {title} {Quasi one-dimensional
  organic conductors: from {Fr\"{o}hlich} conductivity and {Peierls} insulating
  state to magnetically-mediated superconductivity, a retrospective},\ }\href
  {https://doi.org/10.5802/crphys.164} {\bibfield  {journal} {\bibinfo
  {journal} {C. R. Phys.}\ }\textbf {\bibinfo {volume} {25}},\ \bibinfo {pages}
  {17} (\bibinfo {year} {2024})}\BibitemShut {NoStop}%
\bibitem [{\citenamefont {Nagata}\ \emph {et~al.}(1998)\citenamefont {Nagata},
  \citenamefont {Uehara}, \citenamefont {Goto}, \citenamefont {Akimitsu},
  \citenamefont {Motoyama}, \citenamefont {Eisaki}, \citenamefont {Uchida},
  \citenamefont {Takahashi}, \citenamefont {Nakanishi},\ and\ \citenamefont
  {Môri}}]{Nagata1998}%
  \BibitemOpen
  \bibfield  {author} {\bibinfo {author} {\bibfnamefont {T.}~\bibnamefont
  {Nagata}}, \bibinfo {author} {\bibfnamefont {M.}~\bibnamefont {Uehara}},
  \bibinfo {author} {\bibfnamefont {J.}~\bibnamefont {Goto}}, \bibinfo {author}
  {\bibfnamefont {J.}~\bibnamefont {Akimitsu}}, \bibinfo {author}
  {\bibfnamefont {N.}~\bibnamefont {Motoyama}}, \bibinfo {author}
  {\bibfnamefont {H.}~\bibnamefont {Eisaki}}, \bibinfo {author} {\bibfnamefont
  {S.}~\bibnamefont {Uchida}}, \bibinfo {author} {\bibfnamefont
  {H.}~\bibnamefont {Takahashi}}, \bibinfo {author} {\bibfnamefont
  {T.}~\bibnamefont {Nakanishi}},\ and\ \bibinfo {author} {\bibfnamefont
  {N.}~\bibnamefont {Môri}},\ }\bibfield  {title} {\bibinfo {title}
  {Pressure-{Induced} {Dimensional} {Crossover} and {Superconductivity} in the
  {Hole}-{Doped} {Two}-{Leg} {Ladder} {Compound}
  {Sr$_{14-x}$Ca$_x$Cu$_24$O$_41$}},\ }\href
  {https://doi.org/10.1103/PhysRevLett.81.1090} {\bibfield  {journal} {\bibinfo
   {journal} {Phys. Rev. Lett.}\ }\textbf {\bibinfo {volume} {81}},\ \bibinfo
  {pages} {1090} (\bibinfo {year} {1998})}\BibitemShut {NoStop}%
\bibitem [{\citenamefont {Dagotto}(1999)}]{Dagotto1999}%
  \BibitemOpen
  \bibfield  {author} {\bibinfo {author} {\bibfnamefont {E.}~\bibnamefont
  {Dagotto}},\ }\bibfield  {title} {\bibinfo {title} {Experiments on ladders
  reveal a complex interplay between a spin-gapped normal state and
  superconductivity},\ }\href {https://doi.org/10.1088/0034-4885/62/11/202}
  {\bibfield  {journal} {\bibinfo  {journal} {Rep. Prog. Phys.}\ }\textbf
  {\bibinfo {volume} {62}},\ \bibinfo {pages} {1525} (\bibinfo {year}
  {1999})}\BibitemShut {NoStop}%
\bibitem [{\citenamefont {Bao}\ \emph {et~al.}(2015)\citenamefont {Bao},
  \citenamefont {Liu}, \citenamefont {Ma}, \citenamefont {Meng}, \citenamefont
  {Tang}, \citenamefont {Sun}, \citenamefont {Zhai}, \citenamefont {Jiang},
  \citenamefont {Bai}, \citenamefont {Feng}, \citenamefont {Xu},\ and\
  \citenamefont {Cao}}]{Bao2015}%
  \BibitemOpen
  \bibfield  {author} {\bibinfo {author} {\bibfnamefont {J.-K.}\ \bibnamefont
  {Bao}}, \bibinfo {author} {\bibfnamefont {J.-Y.}\ \bibnamefont {Liu}},
  \bibinfo {author} {\bibfnamefont {C.-W.}\ \bibnamefont {Ma}}, \bibinfo
  {author} {\bibfnamefont {Z.-H.}\ \bibnamefont {Meng}}, \bibinfo {author}
  {\bibfnamefont {Z.-T.}\ \bibnamefont {Tang}}, \bibinfo {author}
  {\bibfnamefont {Y.-L.}\ \bibnamefont {Sun}}, \bibinfo {author} {\bibfnamefont
  {H.-F.}\ \bibnamefont {Zhai}}, \bibinfo {author} {\bibfnamefont
  {H.}~\bibnamefont {Jiang}}, \bibinfo {author} {\bibfnamefont
  {H.}~\bibnamefont {Bai}}, \bibinfo {author} {\bibfnamefont {C.-M.}\
  \bibnamefont {Feng}}, \bibinfo {author} {\bibfnamefont {Z.-A.}\ \bibnamefont
  {Xu}},\ and\ \bibinfo {author} {\bibfnamefont {G.-H.}\ \bibnamefont {Cao}},\
  }\bibfield  {title} {\bibinfo {title} {Superconductivity in
  {Quasi}-{One}-{Dimensional} {K}$_2${Cr}$_3${As}$_3$ with {Significant}
  {Electron} {Correlations}},\ }\href
  {https://doi.org/10.1103/PhysRevX.5.011013} {\bibfield  {journal} {\bibinfo
  {journal} {Phys. Rev. X}\ }\textbf {\bibinfo {volume} {5}},\ \bibinfo {pages}
  {011013} (\bibinfo {year} {2015})}\BibitemShut {NoStop}%
\bibitem [{\citenamefont {Watson}\ \emph {et~al.}(2017)\citenamefont {Watson},
  \citenamefont {Feng}, \citenamefont {Nicholson}, \citenamefont {Monney},
  \citenamefont {Riley}, \citenamefont {Iwasawa}, \citenamefont {Refson},
  \citenamefont {Sacksteder}, \citenamefont {Adroja}, \citenamefont {Zhao},\
  and\ \citenamefont {Hoesch}}]{Watson2017}%
  \BibitemOpen
  \bibfield  {author} {\bibinfo {author} {\bibfnamefont {M.~D.}\ \bibnamefont
  {Watson}}, \bibinfo {author} {\bibfnamefont {Y.}~\bibnamefont {Feng}},
  \bibinfo {author} {\bibfnamefont {C.~W.}\ \bibnamefont {Nicholson}}, \bibinfo
  {author} {\bibfnamefont {C.}~\bibnamefont {Monney}}, \bibinfo {author}
  {\bibfnamefont {J.~M.}\ \bibnamefont {Riley}}, \bibinfo {author}
  {\bibfnamefont {H.}~\bibnamefont {Iwasawa}}, \bibinfo {author} {\bibfnamefont
  {K.}~\bibnamefont {Refson}}, \bibinfo {author} {\bibfnamefont
  {V.}~\bibnamefont {Sacksteder}}, \bibinfo {author} {\bibfnamefont {D.~T.}\
  \bibnamefont {Adroja}}, \bibinfo {author} {\bibfnamefont {J.}~\bibnamefont
  {Zhao}},\ and\ \bibinfo {author} {\bibfnamefont {M.}~\bibnamefont {Hoesch}},\
  }\bibfield  {title} {\bibinfo {title} {Multiband {One}-{Dimensional}
  {Electronic} {Structure} and {Spectroscopic} {Signature} of
  {Tomonaga}-{Luttinger} {Liquid} {Behavior} in {K}$_2${Cr}$_3${As}$_3$},\
  }\href {https://doi.org/10.1103/PhysRevLett.118.097002} {\bibfield  {journal}
  {\bibinfo  {journal} {Phys. Rev. Lett.}\ }\textbf {\bibinfo {volume} {118}},\
  \bibinfo {pages} {097002} (\bibinfo {year} {2017})}\BibitemShut {NoStop}%
\bibitem [{\citenamefont {Coldea}\ \emph {et~al.}(1996)\citenamefont {Coldea},
  \citenamefont {Tennant}, \citenamefont {Cowley}, \citenamefont {McMorrow},
  \citenamefont {Dorner},\ and\ \citenamefont {Tylczynski}}]{Coldea1996}%
  \BibitemOpen
  \bibfield  {author} {\bibinfo {author} {\bibfnamefont {R.}~\bibnamefont
  {Coldea}}, \bibinfo {author} {\bibfnamefont {D.~A.}\ \bibnamefont {Tennant}},
  \bibinfo {author} {\bibfnamefont {R.~A.}\ \bibnamefont {Cowley}}, \bibinfo
  {author} {\bibfnamefont {D.~F.}\ \bibnamefont {McMorrow}}, \bibinfo {author}
  {\bibfnamefont {B.}~\bibnamefont {Dorner}},\ and\ \bibinfo {author}
  {\bibfnamefont {Z.}~\bibnamefont {Tylczynski}},\ }\bibfield  {title}
  {\bibinfo {title} {Neutron scattering study of the magnetic structure of
  cs$_2$cucl$_4$},\ }\href {https://doi.org/10.1088/0953-8984/8/40/012}
  {\bibfield  {journal} {\bibinfo  {journal} {Journal of Physics: Condensed
  Matter}\ }\textbf {\bibinfo {volume} {8}},\ \bibinfo {pages} {7473} (\bibinfo
  {year} {1996})}\BibitemShut {NoStop}%
\bibitem [{\citenamefont {Coldea}\ \emph {et~al.}(2001)\citenamefont {Coldea},
  \citenamefont {Tennant}, \citenamefont {Tsvelik},\ and\ \citenamefont
  {Tylczynski}}]{Coldea2001}%
  \BibitemOpen
  \bibfield  {author} {\bibinfo {author} {\bibfnamefont {R.}~\bibnamefont
  {Coldea}}, \bibinfo {author} {\bibfnamefont {D.~A.}\ \bibnamefont {Tennant}},
  \bibinfo {author} {\bibfnamefont {A.~M.}\ \bibnamefont {Tsvelik}},\ and\
  \bibinfo {author} {\bibfnamefont {Z.}~\bibnamefont {Tylczynski}},\ }\bibfield
   {title} {\bibinfo {title} {{Experimental Realization of a 2D Fractional
  Quantum Spin Liquid}},\ }\href {https://doi.org/10.1103/PhysRevLett.86.1335}
  {\bibfield  {journal} {\bibinfo  {journal} {Phys. Rev. Lett.}\ }\textbf
  {\bibinfo {volume} {86}},\ \bibinfo {pages} {1335} (\bibinfo {year}
  {2001})}\BibitemShut {NoStop}%
\bibitem [{\citenamefont {Coldea}\ \emph {et~al.}(2002)\citenamefont {Coldea},
  \citenamefont {Tennant}, \citenamefont {Habicht}, \citenamefont {Smeibidl},
  \citenamefont {Wolters},\ and\ \citenamefont {Tylczynski}}]{Coldea2002}%
  \BibitemOpen
  \bibfield  {author} {\bibinfo {author} {\bibfnamefont {R.}~\bibnamefont
  {Coldea}}, \bibinfo {author} {\bibfnamefont {D.~A.}\ \bibnamefont {Tennant}},
  \bibinfo {author} {\bibfnamefont {K.}~\bibnamefont {Habicht}}, \bibinfo
  {author} {\bibfnamefont {P.}~\bibnamefont {Smeibidl}}, \bibinfo {author}
  {\bibfnamefont {C.}~\bibnamefont {Wolters}},\ and\ \bibinfo {author}
  {\bibfnamefont {Z.}~\bibnamefont {Tylczynski}},\ }\bibfield  {title}
  {\bibinfo {title} {Direct {Measurement} of the {Spin} {Hamiltonian} and
  {Observation} of {Condensation} of {Magnons} in the {2D} {Frustrated}
  {Quantum} {Magnet} {Cs$_2$CuCl$_4$}},\ }\href
  {https://doi.org/10.1103/PhysRevLett.88.137203} {\bibfield  {journal}
  {\bibinfo  {journal} {Phys. Rev. Lett.}\ }\textbf {\bibinfo {volume} {88}},\
  \bibinfo {pages} {137203} (\bibinfo {year} {2002})}\BibitemShut {NoStop}%
\bibitem [{\citenamefont {Tanaka}\ \emph {et~al.}(2002)\citenamefont {Tanaka},
  \citenamefont {Ono}, \citenamefont {Katori}, \citenamefont {Mitamura},
  \citenamefont {Ishikawa},\ and\ \citenamefont {Goto}}]{Tanaka2002}%
  \BibitemOpen
  \bibfield  {author} {\bibinfo {author} {\bibfnamefont {H.}~\bibnamefont
  {Tanaka}}, \bibinfo {author} {\bibfnamefont {T.}~\bibnamefont {Ono}},
  \bibinfo {author} {\bibfnamefont {H.~A.}\ \bibnamefont {Katori}}, \bibinfo
  {author} {\bibfnamefont {H.}~\bibnamefont {Mitamura}}, \bibinfo {author}
  {\bibfnamefont {F.}~\bibnamefont {Ishikawa}},\ and\ \bibinfo {author}
  {\bibfnamefont {T.}~\bibnamefont {Goto}},\ }\bibfield  {title} {\bibinfo
  {title} {{Magnetic Phase Transition and Magnetization Plateau in
  Cs$_2$CuBr$_4$}},\ }\href {https://doi.org/10.1143/PTPS.145.101} {\bibfield
  {journal} {\bibinfo  {journal} {Progress of Theoretical Physics Supplement}\
  }\textbf {\bibinfo {volume} {145}},\ \bibinfo {pages} {101} (\bibinfo {year}
  {2002})}\BibitemShut {NoStop}%
\bibitem [{\citenamefont {Ono}\ \emph {et~al.}(2003)\citenamefont {Ono},
  \citenamefont {Tanaka}, \citenamefont {Aruga~Katori}, \citenamefont
  {Ishikawa}, \citenamefont {Mitamura},\ and\ \citenamefont {Goto}}]{Ono2003}%
  \BibitemOpen
  \bibfield  {author} {\bibinfo {author} {\bibfnamefont {T.}~\bibnamefont
  {Ono}}, \bibinfo {author} {\bibfnamefont {H.}~\bibnamefont {Tanaka}},
  \bibinfo {author} {\bibfnamefont {H.}~\bibnamefont {Aruga~Katori}}, \bibinfo
  {author} {\bibfnamefont {F.}~\bibnamefont {Ishikawa}}, \bibinfo {author}
  {\bibfnamefont {H.}~\bibnamefont {Mitamura}},\ and\ \bibinfo {author}
  {\bibfnamefont {T.}~\bibnamefont {Goto}},\ }\bibfield  {title} {\bibinfo
  {title} {{Magnetization plateau in the frustrated quantum spin system
  ${\mathrm{Cs}}_{2}{\mathrm{CuBr}}_{4}$}},\ }\href
  {https://doi.org/10.1103/PhysRevB.67.104431} {\bibfield  {journal} {\bibinfo
  {journal} {Phys. Rev. B}\ }\textbf {\bibinfo {volume} {67}},\ \bibinfo
  {pages} {104431} (\bibinfo {year} {2003})}\BibitemShut {NoStop}%
\bibitem [{\citenamefont {Alicea}\ \emph {et~al.}(2006)\citenamefont {Alicea},
  \citenamefont {Motrunich},\ and\ \citenamefont {Fisher}}]{Alicea2006}%
  \BibitemOpen
  \bibfield  {author} {\bibinfo {author} {\bibfnamefont {J.}~\bibnamefont
  {Alicea}}, \bibinfo {author} {\bibfnamefont {O.~I.}\ \bibnamefont
  {Motrunich}},\ and\ \bibinfo {author} {\bibfnamefont {M.~P.~A.}\ \bibnamefont
  {Fisher}},\ }\bibfield  {title} {\bibinfo {title} {Theory of the algebraic
  vortex liquid in an anisotropic ${\text{spin-}\frac 12}$ triangular
  antiferromagnet},\ }\href {https://doi.org/10.1103/PhysRevB.73.174430}
  {\bibfield  {journal} {\bibinfo  {journal} {Phys. Rev. B}\ }\textbf {\bibinfo
  {volume} {73}},\ \bibinfo {pages} {174430} (\bibinfo {year}
  {2006})}\BibitemShut {NoStop}%
\bibitem [{\citenamefont {Weng}\ \emph {et~al.}(2006)\citenamefont {Weng},
  \citenamefont {Sheng}, \citenamefont {Weng},\ and\ \citenamefont
  {Bursill}}]{Weng2006}%
  \BibitemOpen
  \bibfield  {author} {\bibinfo {author} {\bibfnamefont {M.~Q.}\ \bibnamefont
  {Weng}}, \bibinfo {author} {\bibfnamefont {D.~N.}\ \bibnamefont {Sheng}},
  \bibinfo {author} {\bibfnamefont {Z.~Y.}\ \bibnamefont {Weng}},\ and\
  \bibinfo {author} {\bibfnamefont {R.~J.}\ \bibnamefont {Bursill}},\
  }\bibfield  {title} {\bibinfo {title} {Spin-liquid phase in an anisotropic
  triangular-lattice {Heisenberg} model: {Exact} diagonalization and
  density-matrix renormalization group calculations},\ }\href
  {https://doi.org/10.1103/PhysRevB.74.012407} {\bibfield  {journal} {\bibinfo
  {journal} {Phys. Rev. B}\ }\textbf {\bibinfo {volume} {74}},\ \bibinfo
  {pages} {012407} (\bibinfo {year} {2006})}\BibitemShut {NoStop}%
\bibitem [{\citenamefont {Yunoki}\ and\ \citenamefont
  {Sorella}(2006)}]{Yunoki2006}%
  \BibitemOpen
  \bibfield  {author} {\bibinfo {author} {\bibfnamefont {S.}~\bibnamefont
  {Yunoki}}\ and\ \bibinfo {author} {\bibfnamefont {S.}~\bibnamefont
  {Sorella}},\ }\bibfield  {title} {\bibinfo {title} {{Two spin liquid phases
  in the spatially anisotropic triangular Heisenberg model}},\ }\href
  {https://doi.org/10.1103/PhysRevB.74.014408} {\bibfield  {journal} {\bibinfo
  {journal} {Phys. Rev. B}\ }\textbf {\bibinfo {volume} {74}},\ \bibinfo
  {pages} {014408} (\bibinfo {year} {2006})}\BibitemShut {NoStop}%
\bibitem [{\citenamefont {Fjærestad}\ \emph {et~al.}(2007)\citenamefont
  {Fjærestad}, \citenamefont {Zheng}, \citenamefont {Singh}, \citenamefont
  {McKenzie},\ and\ \citenamefont {Coldea}}]{Fjaerestad2007}%
  \BibitemOpen
  \bibfield  {author} {\bibinfo {author} {\bibfnamefont {J.~O.}\ \bibnamefont
  {Fjærestad}}, \bibinfo {author} {\bibfnamefont {W.}~\bibnamefont {Zheng}},
  \bibinfo {author} {\bibfnamefont {R.~R.~P.}\ \bibnamefont {Singh}}, \bibinfo
  {author} {\bibfnamefont {R.~H.}\ \bibnamefont {McKenzie}},\ and\ \bibinfo
  {author} {\bibfnamefont {R.}~\bibnamefont {Coldea}},\ }\bibfield  {title}
  {\bibinfo {title} {Excitation spectra and ground state properties of the
  layered spin-$\frac 12$ frustrated antiferromagnets {Cs}$_2${Cu}{Cl}$_4$ and
  {Cs}$_2${Cu}{Br}$_4$},\ }\href {https://doi.org/10.1103/PhysRevB.75.174447}
  {\bibfield  {journal} {\bibinfo  {journal} {Phys. Rev. B}\ }\textbf {\bibinfo
  {volume} {75}},\ \bibinfo {pages} {174447} (\bibinfo {year}
  {2007})}\BibitemShut {NoStop}%
\bibitem [{\citenamefont {Starykh}\ and\ \citenamefont
  {Balents}(2007)}]{Starykh2007}%
  \BibitemOpen
  \bibfield  {author} {\bibinfo {author} {\bibfnamefont {O.~A.}\ \bibnamefont
  {Starykh}}\ and\ \bibinfo {author} {\bibfnamefont {L.}~\bibnamefont
  {Balents}},\ }\bibfield  {title} {\bibinfo {title} {{Ordering in spatially
  anisotropic triangular antiferromagnets}},\ }\href
  {https://doi.org/10.1103/PhysRevLett.98.077205} {\bibfield  {journal}
  {\bibinfo  {journal} {Phys. Rev. Lett.}\ }\textbf {\bibinfo {volume} {98}},\
  \bibinfo {pages} {077205} (\bibinfo {year} {2007})}\BibitemShut {NoStop}%
\bibitem [{\citenamefont {Hayashi}\ and\ \citenamefont
  {Ogata}(2007)}]{Hayashi2007}%
  \BibitemOpen
  \bibfield  {author} {\bibinfo {author} {\bibfnamefont {Y.}~\bibnamefont
  {Hayashi}}\ and\ \bibinfo {author} {\bibfnamefont {M.}~\bibnamefont
  {Ogata}},\ }\bibfield  {title} {\bibinfo {title} {Possibility of gapless spin
  liquid state by one-dimensionalization},\ }\href
  {https://doi.org/10.1143/JPSJ.76.053705} {\bibfield  {journal} {\bibinfo
  {journal} {Journal of the Physical Society of Japan}\ }\textbf {\bibinfo
  {volume} {76}},\ \bibinfo {pages} {053705} (\bibinfo {year}
  {2007})}\BibitemShut {NoStop}%
\bibitem [{\citenamefont {Pardini}\ and\ \citenamefont
  {Singh}(2008)}]{Pardini2008}%
  \BibitemOpen
  \bibfield  {author} {\bibinfo {author} {\bibfnamefont {T.}~\bibnamefont
  {Pardini}}\ and\ \bibinfo {author} {\bibfnamefont {R.~R.~P.}\ \bibnamefont
  {Singh}},\ }\bibfield  {title} {\bibinfo {title} {Magnetic order in coupled
  spin-half and spin-one {Heisenberg} chains in anisotropic triangular-lattice
  geometry},\ }\href {https://doi.org/10.1103/PhysRevB.77.214433} {\bibfield
  {journal} {\bibinfo  {journal} {Phys. Rev. B}\ }\textbf {\bibinfo {volume}
  {77}},\ \bibinfo {pages} {214433} (\bibinfo {year} {2008})}\BibitemShut
  {NoStop}%
\bibitem [{\citenamefont {Heidarian}\ \emph {et~al.}(2009)\citenamefont
  {Heidarian}, \citenamefont {Sorella},\ and\ \citenamefont
  {Becca}}]{Heidarian2009}%
  \BibitemOpen
  \bibfield  {author} {\bibinfo {author} {\bibfnamefont {D.}~\bibnamefont
  {Heidarian}}, \bibinfo {author} {\bibfnamefont {S.}~\bibnamefont {Sorella}},\
  and\ \bibinfo {author} {\bibfnamefont {F.}~\bibnamefont {Becca}},\ }\bibfield
   {title} {\bibinfo {title} {{Spin-$\frac{1}{2}$ Heisenberg model on the
  anisotropic triangular lattice: From magnetism to a one-dimensional spin
  liquid}},\ }\href {https://doi.org/10.1103/PhysRevB.80.012404} {\bibfield
  {journal} {\bibinfo  {journal} {Phys. Rev. B}\ }\textbf {\bibinfo {volume}
  {80}},\ \bibinfo {pages} {012404} (\bibinfo {year} {2009})}\BibitemShut
  {NoStop}%
\bibitem [{\citenamefont {Starykh}\ \emph {et~al.}(2010)\citenamefont
  {Starykh}, \citenamefont {Katsura},\ and\ \citenamefont
  {Balents}}]{Starykh2010}%
  \BibitemOpen
  \bibfield  {author} {\bibinfo {author} {\bibfnamefont {O.~A.}\ \bibnamefont
  {Starykh}}, \bibinfo {author} {\bibfnamefont {H.}~\bibnamefont {Katsura}},\
  and\ \bibinfo {author} {\bibfnamefont {L.}~\bibnamefont {Balents}},\
  }\bibfield  {title} {\bibinfo {title} {Extreme sensitivity of a frustrated
  quantum magnet: {Cs}$_2${CuCl}$_4$},\ }\href
  {https://doi.org/10.1103/PhysRevB.82.014421} {\bibfield  {journal} {\bibinfo
  {journal} {Phys. Rev. B}\ }\textbf {\bibinfo {volume} {82}},\ \bibinfo
  {pages} {014421} (\bibinfo {year} {2010})}\BibitemShut {NoStop}%
\bibitem [{\citenamefont {Cong}\ \emph {et~al.}(2011)\citenamefont {Cong},
  \citenamefont {Wolf}, \citenamefont {de~Souza}, \citenamefont {Kr\"uger},
  \citenamefont {Haghighirad}, \citenamefont {Gottlieb-Schoenmeyer},
  \citenamefont {Ritter}, \citenamefont {Assmus}, \citenamefont {Opahle},
  \citenamefont {Foyevtsova}, \citenamefont {Jeschke}, \citenamefont
  {Valent\'{\i}}, \citenamefont {Wiehl},\ and\ \citenamefont
  {Lang}}]{Cong2011}%
  \BibitemOpen
  \bibfield  {author} {\bibinfo {author} {\bibfnamefont {P.~T.}\ \bibnamefont
  {Cong}}, \bibinfo {author} {\bibfnamefont {B.}~\bibnamefont {Wolf}}, \bibinfo
  {author} {\bibfnamefont {M.}~\bibnamefont {de~Souza}}, \bibinfo {author}
  {\bibfnamefont {N.}~\bibnamefont {Kr\"uger}}, \bibinfo {author}
  {\bibfnamefont {A.~A.}\ \bibnamefont {Haghighirad}}, \bibinfo {author}
  {\bibfnamefont {S.}~\bibnamefont {Gottlieb-Schoenmeyer}}, \bibinfo {author}
  {\bibfnamefont {F.}~\bibnamefont {Ritter}}, \bibinfo {author} {\bibfnamefont
  {W.}~\bibnamefont {Assmus}}, \bibinfo {author} {\bibfnamefont
  {I.}~\bibnamefont {Opahle}}, \bibinfo {author} {\bibfnamefont
  {K.}~\bibnamefont {Foyevtsova}}, \bibinfo {author} {\bibfnamefont {H.~O.}\
  \bibnamefont {Jeschke}}, \bibinfo {author} {\bibfnamefont {R.}~\bibnamefont
  {Valent\'{\i}}}, \bibinfo {author} {\bibfnamefont {L.}~\bibnamefont
  {Wiehl}},\ and\ \bibinfo {author} {\bibfnamefont {M.}~\bibnamefont {Lang}},\
  }\bibfield  {title} {\bibinfo {title} {{Distinct magnetic regimes through
  site-selective atom substitution in the frustrated quantum antiferromagnet
  Cs${}_{2}$CuCl${}_{4\ensuremath{-}x}$Br${}_{x}$}},\ }\href
  {https://doi.org/10.1103/PhysRevB.83.064425} {\bibfield  {journal} {\bibinfo
  {journal} {Phys. Rev. B}\ }\textbf {\bibinfo {volume} {83}},\ \bibinfo
  {pages} {064425} (\bibinfo {year} {2011})}\BibitemShut {NoStop}%
\bibitem [{\citenamefont {Weichselbaum}\ and\ \citenamefont
  {White}(2011)}]{Weichselbaum2011}%
  \BibitemOpen
  \bibfield  {author} {\bibinfo {author} {\bibfnamefont {A.}~\bibnamefont
  {Weichselbaum}}\ and\ \bibinfo {author} {\bibfnamefont {S.~R.}\ \bibnamefont
  {White}},\ }\bibfield  {title} {\bibinfo {title} {Incommensurate correlations
  in the anisotropic triangular {Heisenberg} lattice},\ }\href
  {https://doi.org/10.1103/PhysRevB.84.245130} {\bibfield  {journal} {\bibinfo
  {journal} {Phys. Rev. B}\ }\textbf {\bibinfo {volume} {84}},\ \bibinfo
  {pages} {245130} (\bibinfo {year} {2011})}\BibitemShut {NoStop}%
\bibitem [{\citenamefont {Reuther}\ and\ \citenamefont
  {Thomale}(2011)}]{Reuther2011}%
  \BibitemOpen
  \bibfield  {author} {\bibinfo {author} {\bibfnamefont {J.}~\bibnamefont
  {Reuther}}\ and\ \bibinfo {author} {\bibfnamefont {R.}~\bibnamefont
  {Thomale}},\ }\bibfield  {title} {\bibinfo {title} {Functional
  renormalization group for the anisotropic triangular antiferromagnet},\
  }\href {https://doi.org/10.1103/PhysRevB.83.024402} {\bibfield  {journal}
  {\bibinfo  {journal} {Phys. Rev. B}\ }\textbf {\bibinfo {volume} {83}},\
  \bibinfo {pages} {024402} (\bibinfo {year} {2011})}\BibitemShut {NoStop}%
\bibitem [{\citenamefont {Thesberg}\ and\ \citenamefont
  {S\o{}rensen}(2014)}]{Thesberg2014}%
  \BibitemOpen
  \bibfield  {author} {\bibinfo {author} {\bibfnamefont {M.}~\bibnamefont
  {Thesberg}}\ and\ \bibinfo {author} {\bibfnamefont {E.~S.}\ \bibnamefont
  {S\o{}rensen}},\ }\bibfield  {title} {\bibinfo {title} {Exact diagonalization
  study of the anisotropic triangular lattice {Heisenberg} model using twisted
  boundary conditions},\ }\href {https://doi.org/10.1103/PhysRevB.90.115117}
  {\bibfield  {journal} {\bibinfo  {journal} {Phys. Rev. B}\ }\textbf {\bibinfo
  {volume} {90}},\ \bibinfo {pages} {115117} (\bibinfo {year}
  {2014})}\BibitemShut {NoStop}%
\bibitem [{\citenamefont {Ghorbani}\ \emph {et~al.}(2016)\citenamefont
  {Ghorbani}, \citenamefont {Tocchio},\ and\ \citenamefont
  {Becca}}]{PhysRevB.93.085111}%
  \BibitemOpen
  \bibfield  {author} {\bibinfo {author} {\bibfnamefont {E.}~\bibnamefont
  {Ghorbani}}, \bibinfo {author} {\bibfnamefont {L.~F.}\ \bibnamefont
  {Tocchio}},\ and\ \bibinfo {author} {\bibfnamefont {F.}~\bibnamefont
  {Becca}},\ }\bibfield  {title} {\bibinfo {title} {Variational wave functions
  for the $s=\frac{1}{2}$ {Heisenberg} model on the anisotropic triangular
  lattice: Spin liquids and spiral orders},\ }\href
  {https://doi.org/10.1103/PhysRevB.93.085111} {\bibfield  {journal} {\bibinfo
  {journal} {Phys. Rev. B}\ }\textbf {\bibinfo {volume} {93}},\ \bibinfo
  {pages} {085111} (\bibinfo {year} {2016})}\BibitemShut {NoStop}%
\bibitem [{\citenamefont {Tutsch}\ \emph {et~al.}(2019)\citenamefont {Tutsch},
  \citenamefont {Tsyplyatyev}, \citenamefont {Kuhnt}, \citenamefont {Postulka},
  \citenamefont {Wolf}, \citenamefont {Cong}, \citenamefont {Ritter},
  \citenamefont {Krellner}, \citenamefont {A\ss{}mus}, \citenamefont {Schmidt},
  \citenamefont {Thalmeier}, \citenamefont {Kopietz},\ and\ \citenamefont
  {Lang}}]{Tutsch2019}%
  \BibitemOpen
  \bibfield  {author} {\bibinfo {author} {\bibfnamefont {U.}~\bibnamefont
  {Tutsch}}, \bibinfo {author} {\bibfnamefont {O.}~\bibnamefont {Tsyplyatyev}},
  \bibinfo {author} {\bibfnamefont {M.}~\bibnamefont {Kuhnt}}, \bibinfo
  {author} {\bibfnamefont {L.}~\bibnamefont {Postulka}}, \bibinfo {author}
  {\bibfnamefont {B.}~\bibnamefont {Wolf}}, \bibinfo {author} {\bibfnamefont
  {P.~T.}\ \bibnamefont {Cong}}, \bibinfo {author} {\bibfnamefont
  {F.}~\bibnamefont {Ritter}}, \bibinfo {author} {\bibfnamefont
  {C.}~\bibnamefont {Krellner}}, \bibinfo {author} {\bibfnamefont
  {W.}~\bibnamefont {A\ss{}mus}}, \bibinfo {author} {\bibfnamefont
  {B.}~\bibnamefont {Schmidt}}, \bibinfo {author} {\bibfnamefont
  {P.}~\bibnamefont {Thalmeier}}, \bibinfo {author} {\bibfnamefont
  {P.}~\bibnamefont {Kopietz}},\ and\ \bibinfo {author} {\bibfnamefont
  {M.}~\bibnamefont {Lang}},\ }\bibfield  {title} {\bibinfo {title} {{Specific
  Heat Study of 1D and 2D Excitations in the Layered Frustrated Quantum
  Antiferromagnets
  ${\mathrm{Cs}}_{2}{\mathrm{CuCl}}_{4\ensuremath{-}x}{\mathrm{Br}}_{x}$}},\
  }\href {https://doi.org/10.1103/PhysRevLett.123.147202} {\bibfield  {journal}
  {\bibinfo  {journal} {Phys. Rev. Lett.}\ }\textbf {\bibinfo {volume} {123}},\
  \bibinfo {pages} {147202} (\bibinfo {year} {2019})}\BibitemShut {NoStop}%
\bibitem [{\citenamefont {Gonzalez}\ \emph {et~al.}(2020)\citenamefont
  {Gonzalez}, \citenamefont {Ghioldi}, \citenamefont {Gazza}, \citenamefont
  {Manuel},\ and\ \citenamefont {Trumper}}]{Gonzalez2020}%
  \BibitemOpen
  \bibfield  {author} {\bibinfo {author} {\bibfnamefont {M.~G.}\ \bibnamefont
  {Gonzalez}}, \bibinfo {author} {\bibfnamefont {E.~A.}\ \bibnamefont
  {Ghioldi}}, \bibinfo {author} {\bibfnamefont {C.~J.}\ \bibnamefont {Gazza}},
  \bibinfo {author} {\bibfnamefont {L.~O.}\ \bibnamefont {Manuel}},\ and\
  \bibinfo {author} {\bibfnamefont {A.~E.}\ \bibnamefont {Trumper}},\
  }\bibfield  {title} {\bibinfo {title} {Interplay between spatial anisotropy
  and next-nearest-neighbor exchange interactions in the triangular
  {Heisenberg} model},\ }\href {https://doi.org/10.1103/PhysRevB.102.224410}
  {\bibfield  {journal} {\bibinfo  {journal} {Phys. Rev. B}\ }\textbf {\bibinfo
  {volume} {102}},\ \bibinfo {pages} {224410} (\bibinfo {year}
  {2020})}\BibitemShut {NoStop}%
\bibitem [{\citenamefont {Szasz}\ and\ \citenamefont
  {Motruk}(2021)}]{Szasz2021}%
  \BibitemOpen
  \bibfield  {author} {\bibinfo {author} {\bibfnamefont {A.}~\bibnamefont
  {Szasz}}\ and\ \bibinfo {author} {\bibfnamefont {J.}~\bibnamefont {Motruk}},\
  }\bibfield  {title} {\bibinfo {title} {Phase diagram of the anisotropic
  triangular lattice {Hubbard} model},\ }\href
  {https://doi.org/10.1103/PhysRevB.103.235132} {\bibfield  {journal} {\bibinfo
   {journal} {Phys. Rev. B}\ }\textbf {\bibinfo {volume} {103}},\ \bibinfo
  {pages} {235132} (\bibinfo {year} {2021})}\BibitemShut {NoStop}%
\bibitem [{\citenamefont {Gonzalez}\ \emph {et~al.}(2022)\citenamefont
  {Gonzalez}, \citenamefont {Bernu}, \citenamefont {Pierre},\ and\
  \citenamefont {Messio}}]{Gonzalez2022}%
  \BibitemOpen
  \bibfield  {author} {\bibinfo {author} {\bibfnamefont {M.}~\bibnamefont
  {Gonzalez}}, \bibinfo {author} {\bibfnamefont {B.}~\bibnamefont {Bernu}},
  \bibinfo {author} {\bibfnamefont {L.}~\bibnamefont {Pierre}},\ and\ \bibinfo
  {author} {\bibfnamefont {L.}~\bibnamefont {Messio}},\ }\bibfield  {title}
  {\bibinfo {title} {Ground-state and thermodynamic properties of the
  ${\text{spin-}\frac{1}{2}}$ {Heisenberg} model on the anisotropic triangular
  lattice},\ }\href {https://doi.org/10.21468/SciPostPhys.12.3.112} {\bibfield
  {journal} {\bibinfo  {journal} {SciPost Physics}\ }\textbf {\bibinfo {volume}
  {12}},\ \bibinfo {pages} {112} (\bibinfo {year} {2022})}\BibitemShut
  {NoStop}%
\bibitem [{\citenamefont {Yu}\ \emph {et~al.}(2023)\citenamefont {Yu},
  \citenamefont {Li}, \citenamefont {Iskakov},\ and\ \citenamefont
  {Gull}}]{Yu2023}%
  \BibitemOpen
  \bibfield  {author} {\bibinfo {author} {\bibfnamefont {Y.}~\bibnamefont
  {Yu}}, \bibinfo {author} {\bibfnamefont {S.}~\bibnamefont {Li}}, \bibinfo
  {author} {\bibfnamefont {S.}~\bibnamefont {Iskakov}},\ and\ \bibinfo {author}
  {\bibfnamefont {E.}~\bibnamefont {Gull}},\ }\bibfield  {title} {\bibinfo
  {title} {Magnetic phases of the anisotropic triangular lattice {Hubbard}
  model},\ }\href {https://doi.org/10.1103/PhysRevB.107.075106} {\bibfield
  {journal} {\bibinfo  {journal} {Phys. Rev. B}\ }\textbf {\bibinfo {volume}
  {107}},\ \bibinfo {pages} {075106} (\bibinfo {year} {2023})}\BibitemShut
  {NoStop}%
\bibitem [{\citenamefont {Anderson}(2002)}]{Anderson2002}%
  \BibitemOpen
  \bibfield  {author} {\bibinfo {author} {\bibfnamefont {P.~W.}\ \bibnamefont
  {Anderson}},\ }\bibfield  {title} {\bibinfo {title} {Physics of the pseudogap
  phase of high {$T_c$} cuprates, or, {RVB} meets umklapp},\ }\href
  {https://doi.org/10.1016/S0022-3697(02)00247-0} {\bibfield  {journal}
  {\bibinfo  {journal} {J. Phys. Chem. Solids}\ }\textbf {\bibinfo {volume}
  {63}},\ \bibinfo {pages} {2145} (\bibinfo {year} {2002})}\BibitemShut
  {NoStop}%
\bibitem [{\citenamefont {Scalapino}(2012)}]{Scalapino2012a}%
  \BibitemOpen
  \bibfield  {author} {\bibinfo {author} {\bibfnamefont {D.~J.}\ \bibnamefont
  {Scalapino}},\ }\bibfield  {title} {\bibinfo {title} {A common thread: {The}
  pairing interaction for unconventional superconductors},\ }\href
  {https://doi.org/10.1103/RevModPhys.84.1383} {\bibfield  {journal} {\bibinfo
  {journal} {Rev. Mod. Phys.}\ }\textbf {\bibinfo {volume} {84}},\ \bibinfo
  {pages} {1383} (\bibinfo {year} {2012})}\BibitemShut {NoStop}%
\bibitem [{\citenamefont {Stewart}(2017)}]{Stewart2017}%
  \BibitemOpen
  \bibfield  {author} {\bibinfo {author} {\bibfnamefont {G.~R.}\ \bibnamefont
  {Stewart}},\ }\bibfield  {title} {\bibinfo {title} {Unconventional
  superconductivity},\ }\href {https://doi.org/10.1080/00018732.2017.1331615}
  {\bibfield  {journal} {\bibinfo  {journal} {Adv. Phys.}\ }\textbf {\bibinfo
  {volume} {66}},\ \bibinfo {pages} {75} (\bibinfo {year} {2017})}\BibitemShut
  {NoStop}%
\bibitem [{\citenamefont {Shimizu}\ \emph {et~al.}(2003)\citenamefont
  {Shimizu}, \citenamefont {Miyagawa}, \citenamefont {Kanoda}, \citenamefont
  {Maesato},\ and\ \citenamefont {Saito}}]{Shimizu2003}%
  \BibitemOpen
  \bibfield  {author} {\bibinfo {author} {\bibfnamefont {Y.}~\bibnamefont
  {Shimizu}}, \bibinfo {author} {\bibfnamefont {K.}~\bibnamefont {Miyagawa}},
  \bibinfo {author} {\bibfnamefont {K.}~\bibnamefont {Kanoda}}, \bibinfo
  {author} {\bibfnamefont {M.}~\bibnamefont {Maesato}},\ and\ \bibinfo {author}
  {\bibfnamefont {G.}~\bibnamefont {Saito}},\ }\bibfield  {title} {\bibinfo
  {title} {Spin {Liquid} {State} in an {Organic} {Mott} {Insulator} with
  {Triangular} {Lattice}},\ }\href
  {https://doi.org/10.1103/PhysRevLett.91.107001} {\bibfield  {journal}
  {\bibinfo  {journal} {Phys. Rev. Lett.}\ }\textbf {\bibinfo {volume} {91}},\
  \bibinfo {pages} {107001} (\bibinfo {year} {2003})}\BibitemShut {NoStop}%
\bibitem [{\citenamefont {Itou}\ \emph {et~al.}(2008)\citenamefont {Itou},
  \citenamefont {Oyamada}, \citenamefont {Maegawa}, \citenamefont {Tamura},\
  and\ \citenamefont {Kato}}]{Itou2008}%
  \BibitemOpen
  \bibfield  {author} {\bibinfo {author} {\bibfnamefont {T.}~\bibnamefont
  {Itou}}, \bibinfo {author} {\bibfnamefont {A.}~\bibnamefont {Oyamada}},
  \bibinfo {author} {\bibfnamefont {S.}~\bibnamefont {Maegawa}}, \bibinfo
  {author} {\bibfnamefont {M.}~\bibnamefont {Tamura}},\ and\ \bibinfo {author}
  {\bibfnamefont {R.}~\bibnamefont {Kato}},\ }\bibfield  {title} {\bibinfo
  {title} {{Quantum spin liquid in the spin-$\frac{1}{2}$ triangular
  antiferromagnet
  $\mathrm{Et}{\mathrm{Me}}_{3}\mathrm{Sb}{[\mathrm{Pd}{(\text{dmit})}_{2}]}_{2}$}},\
  }\href {https://doi.org/10.1103/PhysRevB.77.104413} {\bibfield  {journal}
  {\bibinfo  {journal} {Phys. Rev. B}\ }\textbf {\bibinfo {volume} {77}},\
  \bibinfo {pages} {104413} (\bibinfo {year} {2008})}\BibitemShut {NoStop}%
\bibitem [{\citenamefont {Itou}\ \emph {et~al.}(2009)\citenamefont {Itou},
  \citenamefont {Oyamada}, \citenamefont {Maegawa}, \citenamefont {Tamura},\
  and\ \citenamefont {Kato}}]{Itou2009}%
  \BibitemOpen
  \bibfield  {author} {\bibinfo {author} {\bibfnamefont {T.}~\bibnamefont
  {Itou}}, \bibinfo {author} {\bibfnamefont {A.}~\bibnamefont {Oyamada}},
  \bibinfo {author} {\bibfnamefont {S.}~\bibnamefont {Maegawa}}, \bibinfo
  {author} {\bibfnamefont {M.}~\bibnamefont {Tamura}},\ and\ \bibinfo {author}
  {\bibfnamefont {R.}~\bibnamefont {Kato}},\ }\bibfield  {title} {\bibinfo
  {title} {{$^13$C NMR study of the spin-liquid state in the triangular quantum
  antiferromagnet EtMe$_3$Sb[Pd(dmit)$_2$]$_2$}},\ }\href
  {https://doi.org/10.1088/1742-6596/145/1/012039} {\bibfield  {journal}
  {\bibinfo  {journal} {Journal of Physics: Conference Series}\ }\textbf
  {\bibinfo {volume} {145}},\ \bibinfo {pages} {012039} (\bibinfo {year}
  {2009})}\BibitemShut {NoStop}%
\bibitem [{\citenamefont {Giamarchi}(2003)}]{BookGiamarchi2003}%
  \BibitemOpen
  \bibfield  {author} {\bibinfo {author} {\bibfnamefont {T.}~\bibnamefont
  {Giamarchi}},\ }\href@noop {} {\emph {\bibinfo {title} {Quantum {Physics} in
  {One} {Dimension}}}}\ (\bibinfo  {publisher} {Oxford University Press},\
  \bibinfo {year} {2003})\BibitemShut {NoStop}%
\bibitem [{\citenamefont {Schollwöck}(2011)}]{SCHOLLWOCK201196}%
  \BibitemOpen
  \bibfield  {author} {\bibinfo {author} {\bibfnamefont {U.}~\bibnamefont
  {Schollwöck}},\ }\bibfield  {title} {\bibinfo {title} {The density-matrix
  renormalization group in the age of matrix product states},\ }\href
  {https://doi.org/https://doi.org/10.1016/j.aop.2010.09.012} {\bibfield
  {journal} {\bibinfo  {journal} {Annals of Physics}\ }\textbf {\bibinfo
  {volume} {326}},\ \bibinfo {pages} {96} (\bibinfo {year} {2011})}\BibitemShut
  {NoStop}%
\bibitem [{\citenamefont {Paeckel}\ \emph {et~al.}(2019)\citenamefont
  {Paeckel}, \citenamefont {Köhler}, \citenamefont {Swoboda}, \citenamefont
  {Manmana}, \citenamefont {Schollwöck},\ and\ \citenamefont
  {Hubig}}]{Paeckel2019a}%
  \BibitemOpen
  \bibfield  {author} {\bibinfo {author} {\bibfnamefont {S.}~\bibnamefont
  {Paeckel}}, \bibinfo {author} {\bibfnamefont {T.}~\bibnamefont {Köhler}},
  \bibinfo {author} {\bibfnamefont {A.}~\bibnamefont {Swoboda}}, \bibinfo
  {author} {\bibfnamefont {S.~R.}\ \bibnamefont {Manmana}}, \bibinfo {author}
  {\bibfnamefont {U.}~\bibnamefont {Schollwöck}},\ and\ \bibinfo {author}
  {\bibfnamefont {C.}~\bibnamefont {Hubig}},\ }\bibfield  {title} {\bibinfo
  {title} {Time-evolution methods for matrix-product states},\ }\href
  {https://doi.org/10.1016/j.aop.2019.167998} {\bibfield  {journal} {\bibinfo
  {journal} {Ann. Phys.}\ }\textbf {\bibinfo {volume} {411}},\ \bibinfo {pages}
  {167998} (\bibinfo {year} {2019})}\BibitemShut {NoStop}%
\bibitem [{\citenamefont {Schulz}(1996)}]{Schulz1996}%
  \BibitemOpen
  \bibfield  {author} {\bibinfo {author} {\bibfnamefont {H.~J.}\ \bibnamefont
  {Schulz}},\ }\bibfield  {title} {\bibinfo {title} {{Dynamics of Coupled
  Quantum Spin Chains}},\ }\href {https://doi.org/10.1103/PhysRevLett.77.2790}
  {\bibfield  {journal} {\bibinfo  {journal} {Phys. Rev. Lett.}\ }\textbf
  {\bibinfo {volume} {77}},\ \bibinfo {pages} {2790} (\bibinfo {year}
  {1996})}\BibitemShut {NoStop}%
\bibitem [{\citenamefont {Giamarchi}\ and\ \citenamefont
  {Tsvelik}(1999)}]{Giamarchi1999}%
  \BibitemOpen
  \bibfield  {author} {\bibinfo {author} {\bibfnamefont {T.}~\bibnamefont
  {Giamarchi}}\ and\ \bibinfo {author} {\bibfnamefont {A.~M.}\ \bibnamefont
  {Tsvelik}},\ }\bibfield  {title} {\bibinfo {title} {Coupled ladders in a
  magnetic field},\ }\href {https://doi.org/10.1103/PhysRevB.59.11398}
  {\bibfield  {journal} {\bibinfo  {journal} {Phys. Rev. B}\ }\textbf {\bibinfo
  {volume} {59}},\ \bibinfo {pages} {11398} (\bibinfo {year}
  {1999})}\BibitemShut {NoStop}%
\bibitem [{\citenamefont {Sandvik}(1999)}]{Sandvik1999}%
  \BibitemOpen
  \bibfield  {author} {\bibinfo {author} {\bibfnamefont {A.~W.}\ \bibnamefont
  {Sandvik}},\ }\bibfield  {title} {\bibinfo {title} {{Multichain Mean-Field
  Theory of Quasi-One-Dimensional Quantum Spin Systems}},\ }\href
  {https://doi.org/10.1103/PhysRevLett.83.3069} {\bibfield  {journal} {\bibinfo
   {journal} {Phys. Rev. Lett.}\ }\textbf {\bibinfo {volume} {83}},\ \bibinfo
  {pages} {3069} (\bibinfo {year} {1999})}\BibitemShut {NoStop}%
\bibitem [{\citenamefont {Dupont}\ \emph {et~al.}(2018)\citenamefont {Dupont},
  \citenamefont {Capponi}, \citenamefont {Laflorencie},\ and\ \citenamefont
  {Orignac}}]{Dupont2018}%
  \BibitemOpen
  \bibfield  {author} {\bibinfo {author} {\bibfnamefont {M.}~\bibnamefont
  {Dupont}}, \bibinfo {author} {\bibfnamefont {S.}~\bibnamefont {Capponi}},
  \bibinfo {author} {\bibfnamefont {N.}~\bibnamefont {Laflorencie}},\ and\
  \bibinfo {author} {\bibfnamefont {E.}~\bibnamefont {Orignac}},\ }\bibfield
  {title} {\bibinfo {title} {Dynamical response and dimensional crossover for
  spatially anisotropic antiferromagnets},\ }\href
  {https://doi.org/10.1103/PhysRevB.98.094403} {\bibfield  {journal} {\bibinfo
  {journal} {Phys. Rev. B}\ }\textbf {\bibinfo {volume} {98}},\ \bibinfo
  {pages} {094403} (\bibinfo {year} {2018})}\BibitemShut {NoStop}%
\bibitem [{\citenamefont {Karakonstantakis}\ \emph {et~al.}(2011)\citenamefont
  {Karakonstantakis}, \citenamefont {Berg}, \citenamefont {White},\ and\
  \citenamefont {Kivelson}}]{Karakonstantakis2011}%
  \BibitemOpen
  \bibfield  {author} {\bibinfo {author} {\bibfnamefont {G.}~\bibnamefont
  {Karakonstantakis}}, \bibinfo {author} {\bibfnamefont {E.}~\bibnamefont
  {Berg}}, \bibinfo {author} {\bibfnamefont {S.~R.}\ \bibnamefont {White}},\
  and\ \bibinfo {author} {\bibfnamefont {S.~A.}\ \bibnamefont {Kivelson}},\
  }\bibfield  {title} {\bibinfo {title} {Enhanced pairing in the checkerboard
  {Hubbard} ladder},\ }\href {https://doi.org/10.1103/PhysRevB.83.054508}
  {\bibfield  {journal} {\bibinfo  {journal} {Phys. Rev. B}\ }\textbf {\bibinfo
  {volume} {83}},\ \bibinfo {pages} {054508} (\bibinfo {year}
  {2011})}\BibitemShut {NoStop}%
\bibitem [{\citenamefont {Bollmark}\ \emph {et~al.}(2023)\citenamefont
  {Bollmark}, \citenamefont {Köhler}, \citenamefont {Pizzino}, \citenamefont
  {Yang}, \citenamefont {Hofmann}, \citenamefont {Shi}, \citenamefont {Zhang},
  \citenamefont {Giamarchi},\ and\ \citenamefont {Kantian}}]{Bollmark2023}%
  \BibitemOpen
  \bibfield  {author} {\bibinfo {author} {\bibfnamefont {G.}~\bibnamefont
  {Bollmark}}, \bibinfo {author} {\bibfnamefont {T.}~\bibnamefont {Köhler}},
  \bibinfo {author} {\bibfnamefont {L.}~\bibnamefont {Pizzino}}, \bibinfo
  {author} {\bibfnamefont {Y.}~\bibnamefont {Yang}}, \bibinfo {author}
  {\bibfnamefont {J.~S.}\ \bibnamefont {Hofmann}}, \bibinfo {author}
  {\bibfnamefont {H.}~\bibnamefont {Shi}}, \bibinfo {author} {\bibfnamefont
  {S.}~\bibnamefont {Zhang}}, \bibinfo {author} {\bibfnamefont
  {T.}~\bibnamefont {Giamarchi}},\ and\ \bibinfo {author} {\bibfnamefont
  {A.}~\bibnamefont {Kantian}},\ }\bibfield  {title} {\bibinfo {title} {Solving
  {2D} and {3D} {Lattice} {Models} of {Correlated} {Fermions}—{Combining}
  {Matrix} {Product} {States} with {Mean}-{Field} {Theory}},\ }\href
  {https://doi.org/https://doi.org/10.1103/PhysRevX.13.011039} {\bibfield
  {journal} {\bibinfo  {journal} {Phys. Rev. X}\ }\textbf {\bibinfo {volume}
  {13}},\ \bibinfo {pages} {011039} (\bibinfo {year} {2023})}\BibitemShut
  {NoStop}%
\bibitem [{\citenamefont {Bollmark}\ \emph {et~al.}(2025)\citenamefont
  {Bollmark}, \citenamefont {Köhler},\ and\ \citenamefont
  {Kantian}}]{Bollmark2025}%
  \BibitemOpen
  \bibfield  {author} {\bibinfo {author} {\bibfnamefont {G.}~\bibnamefont
  {Bollmark}}, \bibinfo {author} {\bibfnamefont {T.}~\bibnamefont {Köhler}},\
  and\ \bibinfo {author} {\bibfnamefont {A.}~\bibnamefont {Kantian}},\
  }\bibfield  {title} {\bibinfo {title} {Resolving competition of charge
  density wave and superconducting phases using the matrix product state plus
  mean field algorithm},\ }\href {https://doi.org/10.1103/PhysRevB.111.125141}
  {\bibfield  {journal} {\bibinfo  {journal} {Physical Review B}\ }\textbf
  {\bibinfo {volume} {111}},\ \bibinfo {pages} {125141} (\bibinfo {year}
  {2025})}\BibitemShut {NoStop}%
\bibitem [{\citenamefont {R\"uegg}\ \emph {et~al.}(2008)\citenamefont
  {R\"uegg}, \citenamefont {Kiefer}, \citenamefont {Thielemann}, \citenamefont
  {McMorrow}, \citenamefont {Zapf}, \citenamefont {Normand}, \citenamefont
  {Zvonarev}, \citenamefont {Bouillot}, \citenamefont {Kollath}, \citenamefont
  {Giamarchi}, \citenamefont {Capponi}, \citenamefont {Poilblanc},
  \citenamefont {Biner},\ and\ \citenamefont {Kr\"amer}}]{ruegg2008}%
  \BibitemOpen
  \bibfield  {author} {\bibinfo {author} {\bibfnamefont {C.}~\bibnamefont
  {R\"uegg}}, \bibinfo {author} {\bibfnamefont {K.}~\bibnamefont {Kiefer}},
  \bibinfo {author} {\bibfnamefont {B.}~\bibnamefont {Thielemann}}, \bibinfo
  {author} {\bibfnamefont {D.~F.}\ \bibnamefont {McMorrow}}, \bibinfo {author}
  {\bibfnamefont {V.}~\bibnamefont {Zapf}}, \bibinfo {author} {\bibfnamefont
  {B.}~\bibnamefont {Normand}}, \bibinfo {author} {\bibfnamefont {M.~B.}\
  \bibnamefont {Zvonarev}}, \bibinfo {author} {\bibfnamefont {P.}~\bibnamefont
  {Bouillot}}, \bibinfo {author} {\bibfnamefont {C.}~\bibnamefont {Kollath}},
  \bibinfo {author} {\bibfnamefont {T.}~\bibnamefont {Giamarchi}}, \bibinfo
  {author} {\bibfnamefont {S.}~\bibnamefont {Capponi}}, \bibinfo {author}
  {\bibfnamefont {D.}~\bibnamefont {Poilblanc}}, \bibinfo {author}
  {\bibfnamefont {D.}~\bibnamefont {Biner}},\ and\ \bibinfo {author}
  {\bibfnamefont {K.~W.}\ \bibnamefont {Kr\"amer}},\ }\bibfield  {title}
  {\bibinfo {title} {{Thermodynamics of the Spin Luttinger Liquid in a Model
  Ladder Material}},\ }\href
  {https://link.aps.org/doi/10.1103/PhysRevLett.101.247202} {\bibfield
  {journal} {\bibinfo  {journal} {Phys. Rev. Lett.}\ }\textbf {\bibinfo
  {volume} {101}},\ \bibinfo {pages} {247202} (\bibinfo {year}
  {2008})}\BibitemShut {NoStop}%
\bibitem [{\citenamefont {Klanjšek}\ \emph {et~al.}(2008)\citenamefont
  {Klanjšek}, \citenamefont {Mayaffre}, \citenamefont {Berthier},
  \citenamefont {Horvatić}, \citenamefont {Chiari}, \citenamefont {Piovesana},
  \citenamefont {Bouillot}, \citenamefont {Kollath}, \citenamefont {Orignac},
  \citenamefont {Citro},\ and\ \citenamefont {Giamarchi}}]{Klanjsek2008}%
  \BibitemOpen
  \bibfield  {author} {\bibinfo {author} {\bibfnamefont {M.}~\bibnamefont
  {Klanjšek}}, \bibinfo {author} {\bibfnamefont {H.}~\bibnamefont {Mayaffre}},
  \bibinfo {author} {\bibfnamefont {C.}~\bibnamefont {Berthier}}, \bibinfo
  {author} {\bibfnamefont {M.}~\bibnamefont {Horvatić}}, \bibinfo {author}
  {\bibfnamefont {B.}~\bibnamefont {Chiari}}, \bibinfo {author} {\bibfnamefont
  {O.}~\bibnamefont {Piovesana}}, \bibinfo {author} {\bibfnamefont
  {P.}~\bibnamefont {Bouillot}}, \bibinfo {author} {\bibfnamefont
  {C.}~\bibnamefont {Kollath}}, \bibinfo {author} {\bibfnamefont
  {E.}~\bibnamefont {Orignac}}, \bibinfo {author} {\bibfnamefont
  {R.}~\bibnamefont {Citro}},\ and\ \bibinfo {author} {\bibfnamefont
  {T.}~\bibnamefont {Giamarchi}},\ }\bibfield  {title} {\bibinfo {title}
  {Controlling {Luttinger} {Liquid} {Physics} in {Spin} {Ladders} under a
  {Magnetic} {Field}},\ }\href {https://doi.org/10.1103/PhysRevLett.101.137207}
  {\bibfield  {journal} {\bibinfo  {journal} {Phys. Rev. Lett.}\ }\textbf
  {\bibinfo {volume} {101}},\ \bibinfo {pages} {137207} (\bibinfo {year}
  {2008})}\BibitemShut {NoStop}%
\bibitem [{\citenamefont {Thielemann}\ \emph {et~al.}(2009)\citenamefont
  {Thielemann}, \citenamefont {R\"uegg}, \citenamefont {Kiefer}, \citenamefont
  {R\o{}nnow}, \citenamefont {Normand}, \citenamefont {Bouillot}, \citenamefont
  {Kollath}, \citenamefont {Orignac}, \citenamefont {Citro}, \citenamefont
  {Giamarchi}, \citenamefont {L\"auchli}, \citenamefont {Biner}, \citenamefont
  {Kr\"amer}, \citenamefont {Wolff-Fabris}, \citenamefont {Zapf}, \citenamefont
  {Jaime}, \citenamefont {Stahn}, \citenamefont {Christensen}, \citenamefont
  {Grenier}, \citenamefont {McMorrow},\ and\ \citenamefont
  {Mesot}}]{thielemann2009}%
  \BibitemOpen
  \bibfield  {author} {\bibinfo {author} {\bibfnamefont {B.}~\bibnamefont
  {Thielemann}}, \bibinfo {author} {\bibfnamefont {C.}~\bibnamefont {R\"uegg}},
  \bibinfo {author} {\bibfnamefont {K.}~\bibnamefont {Kiefer}}, \bibinfo
  {author} {\bibfnamefont {H.~M.}\ \bibnamefont {R\o{}nnow}}, \bibinfo {author}
  {\bibfnamefont {B.}~\bibnamefont {Normand}}, \bibinfo {author} {\bibfnamefont
  {P.}~\bibnamefont {Bouillot}}, \bibinfo {author} {\bibfnamefont
  {C.}~\bibnamefont {Kollath}}, \bibinfo {author} {\bibfnamefont
  {E.}~\bibnamefont {Orignac}}, \bibinfo {author} {\bibfnamefont
  {R.}~\bibnamefont {Citro}}, \bibinfo {author} {\bibfnamefont
  {T.}~\bibnamefont {Giamarchi}}, \bibinfo {author} {\bibfnamefont {A.~M.}\
  \bibnamefont {L\"auchli}}, \bibinfo {author} {\bibfnamefont {D.}~\bibnamefont
  {Biner}}, \bibinfo {author} {\bibfnamefont {K.~W.}\ \bibnamefont {Kr\"amer}},
  \bibinfo {author} {\bibfnamefont {F.}~\bibnamefont {Wolff-Fabris}}, \bibinfo
  {author} {\bibfnamefont {V.~S.}\ \bibnamefont {Zapf}}, \bibinfo {author}
  {\bibfnamefont {M.}~\bibnamefont {Jaime}}, \bibinfo {author} {\bibfnamefont
  {J.}~\bibnamefont {Stahn}}, \bibinfo {author} {\bibfnamefont {N.~B.}\
  \bibnamefont {Christensen}}, \bibinfo {author} {\bibfnamefont
  {B.}~\bibnamefont {Grenier}}, \bibinfo {author} {\bibfnamefont {D.~F.}\
  \bibnamefont {McMorrow}},\ and\ \bibinfo {author} {\bibfnamefont
  {J.}~\bibnamefont {Mesot}},\ }\bibfield  {title} {\bibinfo {title}
  {Field-controlled magnetic order in the quantum spin-ladder system
  ${(\mathrm{Hpip})}_{2}${CuBr}$_{4}$},\ }\href
  {https://link.aps.org/doi/10.1103/PhysRevB.79.020408} {\bibfield  {journal}
  {\bibinfo  {journal} {Phys. Rev. B}\ }\textbf {\bibinfo {volume} {79}},\
  \bibinfo {pages} {020408} (\bibinfo {year} {2009})}\BibitemShut {NoStop}%
\bibitem [{\citenamefont {Bouillot}\ \emph {et~al.}(2011)\citenamefont
  {Bouillot}, \citenamefont {Kollath}, \citenamefont {Läuchli}, \citenamefont
  {Zvonarev}, \citenamefont {Thielemann}, \citenamefont {Rüegg}, \citenamefont
  {Orignac}, \citenamefont {Citro}, \citenamefont {Klanjšek}, \citenamefont
  {Berthier}, \citenamefont {Horvatić},\ and\ \citenamefont
  {Giamarchi}}]{Bouillot2011}%
  \BibitemOpen
  \bibfield  {author} {\bibinfo {author} {\bibfnamefont {P.}~\bibnamefont
  {Bouillot}}, \bibinfo {author} {\bibfnamefont {C.}~\bibnamefont {Kollath}},
  \bibinfo {author} {\bibfnamefont {A.~M.}\ \bibnamefont {Läuchli}}, \bibinfo
  {author} {\bibfnamefont {M.}~\bibnamefont {Zvonarev}}, \bibinfo {author}
  {\bibfnamefont {B.}~\bibnamefont {Thielemann}}, \bibinfo {author}
  {\bibfnamefont {C.}~\bibnamefont {Rüegg}}, \bibinfo {author} {\bibfnamefont
  {E.}~\bibnamefont {Orignac}}, \bibinfo {author} {\bibfnamefont
  {R.}~\bibnamefont {Citro}}, \bibinfo {author} {\bibfnamefont
  {M.}~\bibnamefont {Klanjšek}}, \bibinfo {author} {\bibfnamefont
  {C.}~\bibnamefont {Berthier}}, \bibinfo {author} {\bibfnamefont
  {M.}~\bibnamefont {Horvatić}},\ and\ \bibinfo {author} {\bibfnamefont
  {T.}~\bibnamefont {Giamarchi}},\ }\bibfield  {title} {\bibinfo {title}
  {Statics and dynamics of weakly coupled antiferromagnetic spin-$\frac{1}{2}$
  ladders in a magnetic field},\ }\href
  {https://doi.org/10.1103/PhysRevB.83.054407} {\bibfield  {journal} {\bibinfo
  {journal} {Phys. Rev. B}\ }\textbf {\bibinfo {volume} {83}},\ \bibinfo
  {pages} {054407} (\bibinfo {year} {2011})}\BibitemShut {NoStop}%
\bibitem [{\citenamefont {Bollmark}\ \emph {et~al.}(2020)\citenamefont
  {Bollmark}, \citenamefont {Laflorencie},\ and\ \citenamefont
  {Kantian}}]{Bollmark2020a}%
  \BibitemOpen
  \bibfield  {author} {\bibinfo {author} {\bibfnamefont {G.}~\bibnamefont
  {Bollmark}}, \bibinfo {author} {\bibfnamefont {N.}~\bibnamefont
  {Laflorencie}},\ and\ \bibinfo {author} {\bibfnamefont {A.}~\bibnamefont
  {Kantian}},\ }\bibfield  {title} {\bibinfo {title} {Dimensional crossover and
  phase transitions in coupled chains : {Density} matrix renormalization group
  results},\ }\href {https://doi.org/10.1103/PhysRevB.102.195145} {\bibfield
  {journal} {\bibinfo  {journal} {Phys. Rev. B}\ }\textbf {\bibinfo {volume}
  {102}},\ \bibinfo {pages} {1} (\bibinfo {year} {2020})}\BibitemShut {NoStop}%
\bibitem [{\citenamefont {Jeong}\ \emph {et~al.}(2017)\citenamefont {Jeong},
  \citenamefont {Mayaffre}, \citenamefont {Berthier}, \citenamefont
  {Schmidiger}, \citenamefont {Zheludev},\ and\ \citenamefont
  {Horvati\ifmmode~\acute{c}\else \'{c}\fi{}}}]{jeong2017}%
  \BibitemOpen
  \bibfield  {author} {\bibinfo {author} {\bibfnamefont {M.}~\bibnamefont
  {Jeong}}, \bibinfo {author} {\bibfnamefont {H.}~\bibnamefont {Mayaffre}},
  \bibinfo {author} {\bibfnamefont {C.}~\bibnamefont {Berthier}}, \bibinfo
  {author} {\bibfnamefont {D.}~\bibnamefont {Schmidiger}}, \bibinfo {author}
  {\bibfnamefont {A.}~\bibnamefont {Zheludev}},\ and\ \bibinfo {author}
  {\bibfnamefont {M.}~\bibnamefont {Horvati\ifmmode~\acute{c}\else
  \'{c}\fi{}}},\ }\bibfield  {title} {\bibinfo {title} {{Magnetic-Order
  Crossover in Coupled Spin Ladders}},\ }\href
  {https://doi.org/10.1103/PhysRevLett.118.167206} {\bibfield  {journal}
  {\bibinfo  {journal} {Phys. Rev. Lett.}\ }\textbf {\bibinfo {volume} {118}},\
  \bibinfo {pages} {167206} (\bibinfo {year} {2017})}\BibitemShut {NoStop}%
\bibitem [{\citenamefont {Schmidiger}\ \emph {et~al.}(2012)\citenamefont
  {Schmidiger}, \citenamefont {Bouillot}, \citenamefont {M\"uhlbauer},
  \citenamefont {Gvasaliya}, \citenamefont {Kollath}, \citenamefont
  {Giamarchi},\ and\ \citenamefont {Zheludev}}]{schmidiger2012}%
  \BibitemOpen
  \bibfield  {author} {\bibinfo {author} {\bibfnamefont {D.}~\bibnamefont
  {Schmidiger}}, \bibinfo {author} {\bibfnamefont {P.}~\bibnamefont
  {Bouillot}}, \bibinfo {author} {\bibfnamefont {S.}~\bibnamefont
  {M\"uhlbauer}}, \bibinfo {author} {\bibfnamefont {S.}~\bibnamefont
  {Gvasaliya}}, \bibinfo {author} {\bibfnamefont {C.}~\bibnamefont {Kollath}},
  \bibinfo {author} {\bibfnamefont {T.}~\bibnamefont {Giamarchi}},\ and\
  \bibinfo {author} {\bibfnamefont {A.}~\bibnamefont {Zheludev}},\ }\bibfield
  {title} {\bibinfo {title} {{Spectral and Thermodynamic Properties of a
  Strong-Leg Quantum Spin Ladder}},\ }\href
  {https://doi.org/10.1103/PhysRevLett.108.167201} {\bibfield  {journal}
  {\bibinfo  {journal} {Phys. Rev. Lett.}\ }\textbf {\bibinfo {volume} {108}},\
  \bibinfo {pages} {167201} (\bibinfo {year} {2012})}\BibitemShut {NoStop}%
\bibitem [{\citenamefont {Jeong}\ \emph {et~al.}(2013)\citenamefont {Jeong},
  \citenamefont {Mayaffre}, \citenamefont {Berthier}, \citenamefont
  {Schmidiger}, \citenamefont {Zheludev},\ and\ \citenamefont
  {Horvati\ifmmode~\acute{c}\else \'{c}\fi{}}}]{jeong2013}%
  \BibitemOpen
  \bibfield  {author} {\bibinfo {author} {\bibfnamefont {M.}~\bibnamefont
  {Jeong}}, \bibinfo {author} {\bibfnamefont {H.}~\bibnamefont {Mayaffre}},
  \bibinfo {author} {\bibfnamefont {C.}~\bibnamefont {Berthier}}, \bibinfo
  {author} {\bibfnamefont {D.}~\bibnamefont {Schmidiger}}, \bibinfo {author}
  {\bibfnamefont {A.}~\bibnamefont {Zheludev}},\ and\ \bibinfo {author}
  {\bibfnamefont {M.}~\bibnamefont {Horvati\ifmmode~\acute{c}\else
  \'{c}\fi{}}},\ }\bibfield  {title} {\bibinfo {title} {{Attractive
  Tomonaga-Luttinger Liquid in a Quantum Spin Ladder}},\ }\href
  {https://doi.org/10.1103/PhysRevLett.111.106404} {\bibfield  {journal}
  {\bibinfo  {journal} {Phys. Rev. Lett.}\ }\textbf {\bibinfo {volume} {111}},\
  \bibinfo {pages} {106404} (\bibinfo {year} {2013})}\BibitemShut {NoStop}%
\bibitem [{\citenamefont {Furuya}\ \emph {et~al.}(2016)\citenamefont {Furuya},
  \citenamefont {Dupont}, \citenamefont {Capponi}, \citenamefont
  {Laflorencie},\ and\ \citenamefont {Giamarchi}}]{Furuya2016}%
  \BibitemOpen
  \bibfield  {author} {\bibinfo {author} {\bibfnamefont {S.~C.}\ \bibnamefont
  {Furuya}}, \bibinfo {author} {\bibfnamefont {M.}~\bibnamefont {Dupont}},
  \bibinfo {author} {\bibfnamefont {S.}~\bibnamefont {Capponi}}, \bibinfo
  {author} {\bibfnamefont {N.}~\bibnamefont {Laflorencie}},\ and\ \bibinfo
  {author} {\bibfnamefont {T.}~\bibnamefont {Giamarchi}},\ }\bibfield  {title}
  {\bibinfo {title} {Dimensional modulation of spontaneous magnetic order in
  quasi-two-dimensional quantum antiferromagnets},\ }\href
  {https://doi.org/10.1103/PhysRevB.94.144403} {\bibfield  {journal} {\bibinfo
  {journal} {Phys. Rev. B}\ }\textbf {\bibinfo {volume} {94}},\ \bibinfo
  {pages} {144403} (\bibinfo {year} {2016})}\BibitemShut {NoStop}%
\bibitem [{\citenamefont {Mukhopadhyay}\ \emph {et~al.}(2012)\citenamefont
  {Mukhopadhyay}, \citenamefont {Klanj\ifmmode~\check{s}\else \v{s}\fi{}ek},
  \citenamefont {Grbi\ifmmode~\acute{c}\else \'{c}\fi{}}, \citenamefont
  {Blinder}, \citenamefont {Mayaffre}, \citenamefont {Berthier}, \citenamefont
  {Horvati\ifmmode~\acute{c}\else \'{c}\fi{}}, \citenamefont {Continentino},
  \citenamefont {Paduan-Filho}, \citenamefont {Chiari},\ and\ \citenamefont
  {Piovesana}}]{mukhopadhyay2012}%
  \BibitemOpen
  \bibfield  {author} {\bibinfo {author} {\bibfnamefont {S.}~\bibnamefont
  {Mukhopadhyay}}, \bibinfo {author} {\bibfnamefont {M.}~\bibnamefont
  {Klanj\ifmmode~\check{s}\else \v{s}\fi{}ek}}, \bibinfo {author}
  {\bibfnamefont {M.~S.}\ \bibnamefont {Grbi\ifmmode~\acute{c}\else
  \'{c}\fi{}}}, \bibinfo {author} {\bibfnamefont {R.}~\bibnamefont {Blinder}},
  \bibinfo {author} {\bibfnamefont {H.}~\bibnamefont {Mayaffre}}, \bibinfo
  {author} {\bibfnamefont {C.}~\bibnamefont {Berthier}}, \bibinfo {author}
  {\bibfnamefont {M.}~\bibnamefont {Horvati\ifmmode~\acute{c}\else
  \'{c}\fi{}}}, \bibinfo {author} {\bibfnamefont {M.~A.}\ \bibnamefont
  {Continentino}}, \bibinfo {author} {\bibfnamefont {A.}~\bibnamefont
  {Paduan-Filho}}, \bibinfo {author} {\bibfnamefont {B.}~\bibnamefont
  {Chiari}},\ and\ \bibinfo {author} {\bibfnamefont {O.}~\bibnamefont
  {Piovesana}},\ }\bibfield  {title} {\bibinfo {title} {{Quantum-Critical Spin
  Dynamics in Quasi-One-Dimensional Antiferromagnets}},\ }\href
  {https://doi.org/10.1103/PhysRevLett.109.177206} {\bibfield  {journal}
  {\bibinfo  {journal} {Phys. Rev. Lett.}\ }\textbf {\bibinfo {volume} {109}},\
  \bibinfo {pages} {177206} (\bibinfo {year} {2012})}\BibitemShut {NoStop}%
\bibitem [{\citenamefont {Blinder}\ \emph {et~al.}(2017)\citenamefont
  {Blinder}, \citenamefont {Dupont}, \citenamefont {Mukhopadhyay},
  \citenamefont {Grbi\ifmmode~\acute{c}\else \'{c}\fi{}}, \citenamefont
  {Laflorencie}, \citenamefont {Capponi}, \citenamefont {Mayaffre},
  \citenamefont {Berthier}, \citenamefont {Paduan-Filho},\ and\ \citenamefont
  {Horvati\ifmmode~\acute{c}\else \'{c}\fi{}}}]{blinder2017}%
  \BibitemOpen
  \bibfield  {author} {\bibinfo {author} {\bibfnamefont {R.}~\bibnamefont
  {Blinder}}, \bibinfo {author} {\bibfnamefont {M.}~\bibnamefont {Dupont}},
  \bibinfo {author} {\bibfnamefont {S.}~\bibnamefont {Mukhopadhyay}}, \bibinfo
  {author} {\bibfnamefont {M.~S.}\ \bibnamefont {Grbi\ifmmode~\acute{c}\else
  \'{c}\fi{}}}, \bibinfo {author} {\bibfnamefont {N.}~\bibnamefont
  {Laflorencie}}, \bibinfo {author} {\bibfnamefont {S.}~\bibnamefont
  {Capponi}}, \bibinfo {author} {\bibfnamefont {H.}~\bibnamefont {Mayaffre}},
  \bibinfo {author} {\bibfnamefont {C.}~\bibnamefont {Berthier}}, \bibinfo
  {author} {\bibfnamefont {A.}~\bibnamefont {Paduan-Filho}},\ and\ \bibinfo
  {author} {\bibfnamefont {M.}~\bibnamefont {Horvati\ifmmode~\acute{c}\else
  \'{c}\fi{}}},\ }\bibfield  {title} {\bibinfo {title} {Nuclear magnetic
  resonance study of the magnetic-field-induced ordered phase in the
  {NiCl}$_{2}${-}4{SC}({NH}$_{2}$)$_{2}$ compound},\ }\href
  {https://doi.org/10.1103/PhysRevB.95.020404} {\bibfield  {journal} {\bibinfo
  {journal} {Phys. Rev. B}\ }\textbf {\bibinfo {volume} {95}},\ \bibinfo
  {pages} {020404(R)} (\bibinfo {year} {2017})}\BibitemShut {NoStop}%
\bibitem [{\citenamefont {Biermann}\ \emph {et~al.}(2001)\citenamefont
  {Biermann}, \citenamefont {Georges}, \citenamefont {Lichtenstein},\ and\
  \citenamefont {Giamarchi}}]{Biermann2001}%
  \BibitemOpen
  \bibfield  {author} {\bibinfo {author} {\bibfnamefont {S.}~\bibnamefont
  {Biermann}}, \bibinfo {author} {\bibfnamefont {A.}~\bibnamefont {Georges}},
  \bibinfo {author} {\bibfnamefont {A.}~\bibnamefont {Lichtenstein}},\ and\
  \bibinfo {author} {\bibfnamefont {T.}~\bibnamefont {Giamarchi}},\ }\bibfield
  {title} {\bibinfo {title} {Deconfinement {Transition} and {Luttinger} to
  {Fermi} {Liquid} {Crossover} in {Quasi}-{One}-{Dimensional} {Systems}},\
  }\href {https://doi.org/10.1103/PhysRevLett.87.276405} {\bibfield  {journal}
  {\bibinfo  {journal} {Phys. Rev. Lett.}\ }\textbf {\bibinfo {volume} {87}},\
  \bibinfo {pages} {276405} (\bibinfo {year} {2001})}\BibitemShut {NoStop}%
\bibitem [{\citenamefont {Berthod}\ \emph {et~al.}(2006)\citenamefont
  {Berthod}, \citenamefont {Giamarchi}, \citenamefont {Biermann},\ and\
  \citenamefont {Georges}}]{Berthod2006}%
  \BibitemOpen
  \bibfield  {author} {\bibinfo {author} {\bibfnamefont {C.}~\bibnamefont
  {Berthod}}, \bibinfo {author} {\bibfnamefont {T.}~\bibnamefont {Giamarchi}},
  \bibinfo {author} {\bibfnamefont {S.}~\bibnamefont {Biermann}},\ and\
  \bibinfo {author} {\bibfnamefont {A.}~\bibnamefont {Georges}},\ }\bibfield
  {title} {\bibinfo {title} {Breakup of the {Fermi} {Surface} {Near} the {Mott}
  {Transition} in {Low}-{Dimensional} {Systems}},\ }\href
  {https://doi.org/10.1103/PhysRevLett.97.136401} {\bibfield  {journal}
  {\bibinfo  {journal} {Phys. Rev. Lett.}\ }\textbf {\bibinfo {volume} {97}},\
  \bibinfo {pages} {136401} (\bibinfo {year} {2006})}\BibitemShut {NoStop}%
\bibitem [{\citenamefont {Eisert}\ \emph {et~al.}(2010)\citenamefont {Eisert},
  \citenamefont {Cramer},\ and\ \citenamefont {Plenio}}]{Eisert2010}%
  \BibitemOpen
  \bibfield  {author} {\bibinfo {author} {\bibfnamefont {J.}~\bibnamefont
  {Eisert}}, \bibinfo {author} {\bibfnamefont {M.}~\bibnamefont {Cramer}},\
  and\ \bibinfo {author} {\bibfnamefont {M.~B.}\ \bibnamefont {Plenio}},\
  }\bibfield  {title} {\bibinfo {title} {Colloquium: Area laws for the
  entanglement entropy},\ }\href {https://doi.org/10.1103/RevModPhys.82.277}
  {\bibfield  {journal} {\bibinfo  {journal} {Rev. Mod. Phys.}\ }\textbf
  {\bibinfo {volume} {82}},\ \bibinfo {pages} {277} (\bibinfo {year}
  {2010})}\BibitemShut {NoStop}%
\bibitem [{\citenamefont {Stoudenmire}\ and\ \citenamefont
  {White}(2012)}]{Stoudenmire2012}%
  \BibitemOpen
  \bibfield  {author} {\bibinfo {author} {\bibfnamefont {E.}~\bibnamefont
  {Stoudenmire}}\ and\ \bibinfo {author} {\bibfnamefont {S.~R.}\ \bibnamefont
  {White}},\ }\bibfield  {title} {\bibinfo {title} {Studying
  {Two}-{Dimensional} {Systems} with the {Density} {Matrix} {Renormalization}
  {Group}},\ }\href {https://doi.org/10.1146/annurev-conmatphys-020911-125018}
  {\bibfield  {journal} {\bibinfo  {journal} {Annu. Rev. Condens. Matter
  Phys.}\ }\textbf {\bibinfo {volume} {3}},\ \bibinfo {pages} {111} (\bibinfo
  {year} {2012})}\BibitemShut {NoStop}%
\bibitem [{\citenamefont {Kantian}\ \emph {et~al.}(2019)\citenamefont
  {Kantian}, \citenamefont {Dolfi}, \citenamefont {Troyer},\ and\ \citenamefont
  {Giamarchi}}]{PhysRevB.100.075138}%
  \BibitemOpen
  \bibfield  {author} {\bibinfo {author} {\bibfnamefont {A.}~\bibnamefont
  {Kantian}}, \bibinfo {author} {\bibfnamefont {M.}~\bibnamefont {Dolfi}},
  \bibinfo {author} {\bibfnamefont {M.}~\bibnamefont {Troyer}},\ and\ \bibinfo
  {author} {\bibfnamefont {T.}~\bibnamefont {Giamarchi}},\ }\bibfield  {title}
  {\bibinfo {title} {Understanding repulsively mediated superconductivity of
  correlated electrons via massively parallel density matrix renormalization
  group},\ }\href {https://doi.org/10.1103/PhysRevB.100.075138} {\bibfield
  {journal} {\bibinfo  {journal} {Phys. Rev. B}\ }\textbf {\bibinfo {volume}
  {100}},\ \bibinfo {pages} {075138} (\bibinfo {year} {2019})}\BibitemShut
  {NoStop}%
\bibitem [{\citenamefont {Sandvik}(2010)}]{Sandvik2010}%
  \BibitemOpen
  \bibfield  {author} {\bibinfo {author} {\bibfnamefont {A.~W.}\ \bibnamefont
  {Sandvik}},\ }\bibfield  {title} {\bibinfo {title} {Computational studies of
  quantum spin systems},\ }\href {https://doi.org/10.1063/1.3518900} {\bibfield
   {journal} {\bibinfo  {journal} {AIP Conference Proceedings}\ }\textbf
  {\bibinfo {volume} {1297}},\ \bibinfo {pages} {135} (\bibinfo {year}
  {2010})}\BibitemShut {NoStop}%
\bibitem [{\citenamefont {L{\"a}uchli}(2010)}]{lauchli2010numerical}%
  \BibitemOpen
  \bibfield  {author} {\bibinfo {author} {\bibfnamefont {A.~M.}\ \bibnamefont
  {L{\"a}uchli}},\ }\bibfield  {title} {\bibinfo {title} {Numerical simulations
  of frustrated systems},\ }in\ \href@noop {} {\emph {\bibinfo {booktitle}
  {Introduction to Frustrated Magnetism: Materials, Experiments, Theory}}}\
  (\bibinfo  {publisher} {Springer},\ \bibinfo {year} {2010})\ pp.\ \bibinfo
  {pages} {481--511}\BibitemShut {NoStop}%
\bibitem [{\citenamefont {Bauer}\ \emph {et~al.}(2011)\citenamefont {Bauer},
  \citenamefont {Carr}, \citenamefont {Evertz}, \citenamefont {Feiguin},
  \citenamefont {Freire}, \citenamefont {Fuchs}, \citenamefont {Gamper},
  \citenamefont {Gukelberger}, \citenamefont {Gull}, \citenamefont {Guertler},
  \citenamefont {Hehn}, \citenamefont {Igarashi}, \citenamefont {Isakov},
  \citenamefont {Koop}, \citenamefont {Ma}, \citenamefont {Mates},
  \citenamefont {Matsuo}, \citenamefont {Parcollet}, \citenamefont {Pawlowski},
  \citenamefont {Picon}, \citenamefont {Pollet}, \citenamefont {Santos},
  \citenamefont {Scarola}, \citenamefont {Schollw\"{o}ck}, \citenamefont
  {Silva}, \citenamefont {Surer}, \citenamefont {Todo}, \citenamefont {Trebst},
  \citenamefont {Troyer}, \citenamefont {Wall}, \citenamefont {Werner},\ and\
  \citenamefont {Wessel}}]{Bauer_2011}%
  \BibitemOpen
  \bibfield  {author} {\bibinfo {author} {\bibfnamefont {B.}~\bibnamefont
  {Bauer}}, \bibinfo {author} {\bibfnamefont {L.~D.}\ \bibnamefont {Carr}},
  \bibinfo {author} {\bibfnamefont {H.~G.}\ \bibnamefont {Evertz}}, \bibinfo
  {author} {\bibfnamefont {A.}~\bibnamefont {Feiguin}}, \bibinfo {author}
  {\bibfnamefont {J.}~\bibnamefont {Freire}}, \bibinfo {author} {\bibfnamefont
  {S.}~\bibnamefont {Fuchs}}, \bibinfo {author} {\bibfnamefont
  {L.}~\bibnamefont {Gamper}}, \bibinfo {author} {\bibfnamefont
  {J.}~\bibnamefont {Gukelberger}}, \bibinfo {author} {\bibfnamefont
  {E.}~\bibnamefont {Gull}}, \bibinfo {author} {\bibfnamefont {S.}~\bibnamefont
  {Guertler}}, \bibinfo {author} {\bibfnamefont {A.}~\bibnamefont {Hehn}},
  \bibinfo {author} {\bibfnamefont {R.}~\bibnamefont {Igarashi}}, \bibinfo
  {author} {\bibfnamefont {S.~V.}\ \bibnamefont {Isakov}}, \bibinfo {author}
  {\bibfnamefont {D.}~\bibnamefont {Koop}}, \bibinfo {author} {\bibfnamefont
  {P.~N.}\ \bibnamefont {Ma}}, \bibinfo {author} {\bibfnamefont
  {P.}~\bibnamefont {Mates}}, \bibinfo {author} {\bibfnamefont
  {H.}~\bibnamefont {Matsuo}}, \bibinfo {author} {\bibfnamefont
  {O.}~\bibnamefont {Parcollet}}, \bibinfo {author} {\bibfnamefont
  {G.}~\bibnamefont {Pawlowski}}, \bibinfo {author} {\bibfnamefont {J.~D.}\
  \bibnamefont {Picon}}, \bibinfo {author} {\bibfnamefont {L.}~\bibnamefont
  {Pollet}}, \bibinfo {author} {\bibfnamefont {E.}~\bibnamefont {Santos}},
  \bibinfo {author} {\bibfnamefont {V.~W.}\ \bibnamefont {Scarola}}, \bibinfo
  {author} {\bibfnamefont {U.}~\bibnamefont {Schollw\"{o}ck}}, \bibinfo
  {author} {\bibfnamefont {C.}~\bibnamefont {Silva}}, \bibinfo {author}
  {\bibfnamefont {B.}~\bibnamefont {Surer}}, \bibinfo {author} {\bibfnamefont
  {S.}~\bibnamefont {Todo}}, \bibinfo {author} {\bibfnamefont {S.}~\bibnamefont
  {Trebst}}, \bibinfo {author} {\bibfnamefont {M.}~\bibnamefont {Troyer}},
  \bibinfo {author} {\bibfnamefont {M.~L.}\ \bibnamefont {Wall}}, \bibinfo
  {author} {\bibfnamefont {P.}~\bibnamefont {Werner}},\ and\ \bibinfo {author}
  {\bibfnamefont {S.}~\bibnamefont {Wessel}},\ }\bibfield  {title} {\bibinfo
  {title} {{The ALPS project release 2.0: open source software for strongly
  correlated systems}},\ }\href
  {https://doi.org/10.1088/1742-5468/2011/05/P05001} {\bibfield  {journal}
  {\bibinfo  {journal} {Journal of Statistical Mechanics: Theory and
  Experiment}\ }\textbf {\bibinfo {volume} {2011}},\ \bibinfo {pages} {P05001}
  (\bibinfo {year} {2011})}\BibitemShut {NoStop}%
\bibitem [{\citenamefont {Gaenko}\ \emph {et~al.}(2017)\citenamefont {Gaenko},
  \citenamefont {Antipov}, \citenamefont {Carcassi}, \citenamefont {Chen},
  \citenamefont {Chen}, \citenamefont {Dong}, \citenamefont {Gamper},
  \citenamefont {Gukelberger}, \citenamefont {Igarashi}, \citenamefont
  {Iskakov}, \citenamefont {K\"{o}nz}, \citenamefont {LeBlanc}, \citenamefont
  {Levy}, \citenamefont {Ma}, \citenamefont {Paki}, \citenamefont {Shinaoka},
  \citenamefont {Todo}, \citenamefont {Troyer},\ and\ \citenamefont
  {Gull}}]{GAENKO2017235}%
  \BibitemOpen
  \bibfield  {author} {\bibinfo {author} {\bibfnamefont {A.}~\bibnamefont
  {Gaenko}}, \bibinfo {author} {\bibfnamefont {A.}~\bibnamefont {Antipov}},
  \bibinfo {author} {\bibfnamefont {G.}~\bibnamefont {Carcassi}}, \bibinfo
  {author} {\bibfnamefont {T.}~\bibnamefont {Chen}}, \bibinfo {author}
  {\bibfnamefont {X.}~\bibnamefont {Chen}}, \bibinfo {author} {\bibfnamefont
  {Q.}~\bibnamefont {Dong}}, \bibinfo {author} {\bibfnamefont {L.}~\bibnamefont
  {Gamper}}, \bibinfo {author} {\bibfnamefont {J.}~\bibnamefont {Gukelberger}},
  \bibinfo {author} {\bibfnamefont {R.}~\bibnamefont {Igarashi}}, \bibinfo
  {author} {\bibfnamefont {S.}~\bibnamefont {Iskakov}}, \bibinfo {author}
  {\bibfnamefont {M.}~\bibnamefont {K\"{o}nz}}, \bibinfo {author}
  {\bibfnamefont {J.}~\bibnamefont {LeBlanc}}, \bibinfo {author} {\bibfnamefont
  {R.}~\bibnamefont {Levy}}, \bibinfo {author} {\bibfnamefont {P.}~\bibnamefont
  {Ma}}, \bibinfo {author} {\bibfnamefont {J.}~\bibnamefont {Paki}}, \bibinfo
  {author} {\bibfnamefont {H.}~\bibnamefont {Shinaoka}}, \bibinfo {author}
  {\bibfnamefont {S.}~\bibnamefont {Todo}}, \bibinfo {author} {\bibfnamefont
  {M.}~\bibnamefont {Troyer}},\ and\ \bibinfo {author} {\bibfnamefont
  {E.}~\bibnamefont {Gull}},\ }\bibfield  {title} {\bibinfo {title} {{Updated
  core libraries of the ALPS project}},\ }\href
  {https://doi.org/https://doi.org/10.1016/j.cpc.2016.12.009} {\bibfield
  {journal} {\bibinfo  {journal} {Computer Physics Communications}\ }\textbf
  {\bibinfo {volume} {213}},\ \bibinfo {pages} {235} (\bibinfo {year}
  {2017})}\BibitemShut {NoStop}%
\bibitem [{\citenamefont {Becca}\ and\ \citenamefont
  {Sorella}(2017)}]{becca2017quantum}%
  \BibitemOpen
  \bibfield  {author} {\bibinfo {author} {\bibfnamefont {F.}~\bibnamefont
  {Becca}}\ and\ \bibinfo {author} {\bibfnamefont {S.}~\bibnamefont
  {Sorella}},\ }\href@noop {} {\emph {\bibinfo {title} {Quantum Monte Carlo
  approaches for correlated systems}}}\ (\bibinfo  {publisher} {Cambridge
  University Press},\ \bibinfo {year} {2017})\BibitemShut {NoStop}%
\bibitem [{\citenamefont {Cirac}\ \emph {et~al.}(2021)\citenamefont {Cirac},
  \citenamefont {P\'erez-Garc\'{\i}a}, \citenamefont {Schuch},\ and\
  \citenamefont {Verstraete}}]{Cirac2021}%
  \BibitemOpen
  \bibfield  {author} {\bibinfo {author} {\bibfnamefont {J.~I.}\ \bibnamefont
  {Cirac}}, \bibinfo {author} {\bibfnamefont {D.}~\bibnamefont
  {P\'erez-Garc\'{\i}a}}, \bibinfo {author} {\bibfnamefont {N.}~\bibnamefont
  {Schuch}},\ and\ \bibinfo {author} {\bibfnamefont {F.}~\bibnamefont
  {Verstraete}},\ }\bibfield  {title} {\bibinfo {title} {Matrix product states
  and projected entangled pair states: Concepts, symmetries, theorems},\ }\href
  {https://doi.org/10.1103/RevModPhys.93.045003} {\bibfield  {journal}
  {\bibinfo  {journal} {Rev. Mod. Phys.}\ }\textbf {\bibinfo {volume} {93}},\
  \bibinfo {pages} {045003} (\bibinfo {year} {2021})}\BibitemShut {NoStop}%
\bibitem [{\citenamefont {He}\ \emph {et~al.}(2019)\citenamefont {He},
  \citenamefont {Qin}, \citenamefont {Shi}, \citenamefont {Lu},\ and\
  \citenamefont {Zhang}}]{He2019}%
  \BibitemOpen
  \bibfield  {author} {\bibinfo {author} {\bibfnamefont {Y.~Y.}\ \bibnamefont
  {He}}, \bibinfo {author} {\bibfnamefont {M.}~\bibnamefont {Qin}}, \bibinfo
  {author} {\bibfnamefont {H.}~\bibnamefont {Shi}}, \bibinfo {author}
  {\bibfnamefont {Z.~Y.}\ \bibnamefont {Lu}},\ and\ \bibinfo {author}
  {\bibfnamefont {S.}~\bibnamefont {Zhang}},\ }\bibfield  {title} {\bibinfo
  {title} {Finite-temperature auxiliary-field quantum {Monte} {Carlo}:
  {Self}-consistent constraint and systematic approach to low temperatures},\
  }\bibfield  {journal} {\bibinfo  {journal} {Phys. Rev. B}\ }\textbf {\bibinfo
  {volume} {99}},\ \href {https://doi.org/10.1103/PhysRevB.99.045108}
  {10.1103/PhysRevB.99.045108} (\bibinfo {year} {2019})\BibitemShut {NoStop}%
\bibitem [{\citenamefont {James}\ and\ \citenamefont
  {Konik}(2013)}]{James2013}%
  \BibitemOpen
  \bibfield  {author} {\bibinfo {author} {\bibfnamefont {A.~J.~A.}\
  \bibnamefont {James}}\ and\ \bibinfo {author} {\bibfnamefont {R.~M.}\
  \bibnamefont {Konik}},\ }\bibfield  {title} {\bibinfo {title} {Understanding
  the entanglement entropy and spectra of {2D} quantum systems through arrays
  of coupled {1D} chains},\ }\href {https://doi.org/10.1103/PhysRevB.87.241103}
  {\bibfield  {journal} {\bibinfo  {journal} {Phys. Rev. B}\ }\textbf {\bibinfo
  {volume} {87}},\ \bibinfo {pages} {241103} (\bibinfo {year}
  {2013})}\BibitemShut {NoStop}%
\bibitem [{\citenamefont {James}\ \emph {et~al.}(2018)\citenamefont {James},
  \citenamefont {Konik}, \citenamefont {Lecheminant}, \citenamefont
  {Robinson},\ and\ \citenamefont {Tsvelik}}]{James2018}%
  \BibitemOpen
  \bibfield  {author} {\bibinfo {author} {\bibfnamefont {A.~J.~A.}\
  \bibnamefont {James}}, \bibinfo {author} {\bibfnamefont {R.~M.}\ \bibnamefont
  {Konik}}, \bibinfo {author} {\bibfnamefont {P.}~\bibnamefont {Lecheminant}},
  \bibinfo {author} {\bibfnamefont {N.~J.}\ \bibnamefont {Robinson}},\ and\
  \bibinfo {author} {\bibfnamefont {A.~M.}\ \bibnamefont {Tsvelik}},\
  }\bibfield  {title} {\bibinfo {title} {Non-perturbative methodologies for
  low-dimensional strongly-correlated systems: {From} non-{Abelian}
  bosonization to truncated spectrum methods},\ }\href
  {https://doi.org/10.1088/1361-6633/aa91ea} {\bibfield  {journal} {\bibinfo
  {journal} {Rep. Prog. Phys.}\ }\textbf {\bibinfo {volume} {81}},\ \bibinfo
  {pages} {046002} (\bibinfo {year} {2018})}\BibitemShut {NoStop}%
\bibitem [{\citenamefont {Vidal}\ \emph {et~al.}(2003)\citenamefont {Vidal},
  \citenamefont {Latorre}, \citenamefont {Rico},\ and\ \citenamefont
  {Kitaev}}]{Vidal2003}%
  \BibitemOpen
  \bibfield  {author} {\bibinfo {author} {\bibfnamefont {G.}~\bibnamefont
  {Vidal}}, \bibinfo {author} {\bibfnamefont {J.~I.}\ \bibnamefont {Latorre}},
  \bibinfo {author} {\bibfnamefont {E.}~\bibnamefont {Rico}},\ and\ \bibinfo
  {author} {\bibfnamefont {A.}~\bibnamefont {Kitaev}},\ }\bibfield  {title}
  {\bibinfo {title} {{Entanglement in Quantum Critical Phenomena}},\ }\href
  {https://doi.org/10.1103/PhysRevLett.90.227902} {\bibfield  {journal}
  {\bibinfo  {journal} {Phys. Rev. Lett.}\ }\textbf {\bibinfo {volume} {90}},\
  \bibinfo {pages} {227902} (\bibinfo {year} {2003})}\BibitemShut {NoStop}%
\bibitem [{\citenamefont {Calabrese}\ and\ \citenamefont
  {Cardy}(2004)}]{Calabrese2004}%
  \BibitemOpen
  \bibfield  {author} {\bibinfo {author} {\bibfnamefont {P.}~\bibnamefont
  {Calabrese}}\ and\ \bibinfo {author} {\bibfnamefont {J.}~\bibnamefont
  {Cardy}},\ }\bibfield  {title} {\bibinfo {title} {Entanglement entropy and
  quantum field theory},\ }\href
  {https://doi.org/10.1088/1742-5468/2004/06/P06002} {\bibfield  {journal}
  {\bibinfo  {journal} {Journal of Statistical Mechanics: Theory and
  Experiment}\ }\textbf {\bibinfo {volume} {2004}},\ \bibinfo {pages} {P06002}
  (\bibinfo {year} {2004})}\BibitemShut {NoStop}%
\bibitem [{\citenamefont {Hubig}(2017)}]{ediss21348}%
  \BibitemOpen
  \bibfield  {author} {\bibinfo {author} {\bibfnamefont {C.}~\bibnamefont
  {Hubig}},\ }\href {http://nbn-resolving.de/urn:nbn:de:bvb:19-213484}
  {\bibinfo {title} {Symmetry-protected tensor networks}} (\bibinfo {year}
  {2017})\BibitemShut {NoStop}%
\bibitem [{\citenamefont {Singh}\ \emph {et~al.}(2010)\citenamefont {Singh},
  \citenamefont {Pfeifer},\ and\ \citenamefont {Vidal}}]{PhysRevA.82.050301}%
  \BibitemOpen
  \bibfield  {author} {\bibinfo {author} {\bibfnamefont {S.}~\bibnamefont
  {Singh}}, \bibinfo {author} {\bibfnamefont {R.~N.~C.}\ \bibnamefont
  {Pfeifer}},\ and\ \bibinfo {author} {\bibfnamefont {G.}~\bibnamefont
  {Vidal}},\ }\bibfield  {title} {\bibinfo {title} {Tensor network
  decompositions in the presence of a global symmetry},\ }\href
  {https://doi.org/10.1103/PhysRevA.82.050301} {\bibfield  {journal} {\bibinfo
  {journal} {Phys. Rev. A}\ }\textbf {\bibinfo {volume} {82}},\ \bibinfo
  {pages} {050301} (\bibinfo {year} {2010})}\BibitemShut {NoStop}%
\bibitem [{\citenamefont {Singh}\ \emph {et~al.}(2011)\citenamefont {Singh},
  \citenamefont {Pfeifer},\ and\ \citenamefont {Vidal}}]{PhysRevB.83.115125}%
  \BibitemOpen
  \bibfield  {author} {\bibinfo {author} {\bibfnamefont {S.}~\bibnamefont
  {Singh}}, \bibinfo {author} {\bibfnamefont {R.~N.~C.}\ \bibnamefont
  {Pfeifer}},\ and\ \bibinfo {author} {\bibfnamefont {G.}~\bibnamefont
  {Vidal}},\ }\bibfield  {title} {\bibinfo {title} {Tensor network states and
  algorithms in the presence of a global {U(1)} symmetry},\ }\href
  {https://doi.org/10.1103/PhysRevB.83.115125} {\bibfield  {journal} {\bibinfo
  {journal} {Phys. Rev. B}\ }\textbf {\bibinfo {volume} {83}},\ \bibinfo
  {pages} {115125} (\bibinfo {year} {2011})}\BibitemShut {NoStop}%
\bibitem [{\citenamefont {Schollw\"ock}(2005)}]{RevModPhys.77.259}%
  \BibitemOpen
  \bibfield  {author} {\bibinfo {author} {\bibfnamefont {U.}~\bibnamefont
  {Schollw\"ock}},\ }\bibfield  {title} {\bibinfo {title} {The density-matrix
  renormalization group},\ }\href {https://doi.org/10.1103/RevModPhys.77.259}
  {\bibfield  {journal} {\bibinfo  {journal} {Rev. Mod. Phys.}\ }\textbf
  {\bibinfo {volume} {77}},\ \bibinfo {pages} {259} (\bibinfo {year}
  {2005})}\BibitemShut {NoStop}%
\bibitem [{\citenamefont {Sylju\aa{}sen}\ and\ \citenamefont
  {Sandvik}(2002)}]{Syljuasen2002}%
  \BibitemOpen
  \bibfield  {author} {\bibinfo {author} {\bibfnamefont {O.~F.}\ \bibnamefont
  {Sylju\aa{}sen}}\ and\ \bibinfo {author} {\bibfnamefont {A.~W.}\ \bibnamefont
  {Sandvik}},\ }\bibfield  {title} {\bibinfo {title} {{Quantum Monte Carlo with
  directed loops}},\ }\href {https://doi.org/10.1103/PhysRevE.66.046701}
  {\bibfield  {journal} {\bibinfo  {journal} {Phys. Rev. E}\ }\textbf {\bibinfo
  {volume} {66}},\ \bibinfo {pages} {046701} (\bibinfo {year}
  {2002})}\BibitemShut {NoStop}%
\bibitem [{\citenamefont {Weihong}\ \emph {et~al.}(1999)\citenamefont
  {Weihong}, \citenamefont {McKenzie},\ and\ \citenamefont
  {Singh}}]{Weihong1999}%
  \BibitemOpen
  \bibfield  {author} {\bibinfo {author} {\bibfnamefont {Z.}~\bibnamefont
  {Weihong}}, \bibinfo {author} {\bibfnamefont {R.~H.}\ \bibnamefont
  {McKenzie}},\ and\ \bibinfo {author} {\bibfnamefont {R.~R.~P.}\ \bibnamefont
  {Singh}},\ }\bibfield  {title} {\bibinfo {title} {Phase diagram for a class
  of spin-half {Heisenberg} models interpolating between the square-lattice,
  the triangular-lattice and the linear chain limits},\ }\href
  {https://doi.org/10.1103/PhysRevB.59.14367} {\bibfield  {journal} {\bibinfo
  {journal} {Phys. Rev. B}\ }\textbf {\bibinfo {volume} {59}},\ \bibinfo
  {pages} {14367} (\bibinfo {year} {1999})}\BibitemShut {NoStop}%
\bibitem [{\citenamefont {Bourbonnais}\ and\ \citenamefont
  {Sedeki}(2011)}]{Bourbonnais2011}%
  \BibitemOpen
  \bibfield  {author} {\bibinfo {author} {\bibfnamefont {C.}~\bibnamefont
  {Bourbonnais}}\ and\ \bibinfo {author} {\bibfnamefont {A.}~\bibnamefont
  {Sedeki}},\ }\bibfield  {title} {\bibinfo {title} {Superconductivity and
  antiferromagnetism as interfering orders in organic conductors},\ }\href
  {https://doi.org/10.1016/j.crhy.2011.04.005} {\bibfield  {journal} {\bibinfo
  {journal} {C. R. Phys.}\ }\textbf {\bibinfo {volume} {12}},\ \bibinfo {pages}
  {532} (\bibinfo {year} {2011})}\BibitemShut {NoStop}%
\bibitem [{\citenamefont {Qin}\ \emph {et~al.}(2020)\citenamefont {Qin},
  \citenamefont {Chung}, \citenamefont {Shi}, \citenamefont {Vitali},
  \citenamefont {Hubig}, \citenamefont {Schollw\"ock}, \citenamefont {White},\
  and\ \citenamefont {Zhang}}]{PhysRevX.10.031016}%
  \BibitemOpen
  \bibfield  {author} {\bibinfo {author} {\bibfnamefont {M.}~\bibnamefont
  {Qin}}, \bibinfo {author} {\bibfnamefont {C.-M.}\ \bibnamefont {Chung}},
  \bibinfo {author} {\bibfnamefont {H.}~\bibnamefont {Shi}}, \bibinfo {author}
  {\bibfnamefont {E.}~\bibnamefont {Vitali}}, \bibinfo {author} {\bibfnamefont
  {C.}~\bibnamefont {Hubig}}, \bibinfo {author} {\bibfnamefont
  {U.}~\bibnamefont {Schollw\"ock}}, \bibinfo {author} {\bibfnamefont {S.~R.}\
  \bibnamefont {White}},\ and\ \bibinfo {author} {\bibfnamefont
  {S.}~\bibnamefont {Zhang}} (\bibinfo {collaboration} {Simons Collaboration on
  the Many-Electron Problem}),\ }\bibfield  {title} {\bibinfo {title} {{Absence
  of Superconductivity in the Pure Two-Dimensional Hubbard Model}},\ }\href
  {https://doi.org/10.1103/PhysRevX.10.031016} {\bibfield  {journal} {\bibinfo
  {journal} {Phys. Rev. X}\ }\textbf {\bibinfo {volume} {10}},\ \bibinfo
  {pages} {031016} (\bibinfo {year} {2020})}\BibitemShut {NoStop}%
\bibitem [{\citenamefont {Holzner}\ \emph {et~al.}(2011)\citenamefont
  {Holzner}, \citenamefont {Weichselbaum}, \citenamefont {McCulloch},
  \citenamefont {Schollwöck},\ and\ \citenamefont {von Delft}}]{Holzner2011}%
  \BibitemOpen
  \bibfield  {author} {\bibinfo {author} {\bibfnamefont {A.}~\bibnamefont
  {Holzner}}, \bibinfo {author} {\bibfnamefont {A.}~\bibnamefont
  {Weichselbaum}}, \bibinfo {author} {\bibfnamefont {I.~P.}\ \bibnamefont
  {McCulloch}}, \bibinfo {author} {\bibfnamefont {U.}~\bibnamefont
  {Schollwöck}},\ and\ \bibinfo {author} {\bibfnamefont {J.}~\bibnamefont {von
  Delft}},\ }\bibfield  {title} {\bibinfo {title} {Chebyshev matrix product
  state approach for spectral functions},\ }\href
  {https://doi.org/10.1103/PhysRevB.83.195115} {\bibfield  {journal} {\bibinfo
  {journal} {Phys. Rev. B}\ }\textbf {\bibinfo {volume} {83}},\ \bibinfo
  {pages} {195115} (\bibinfo {year} {2011})}\BibitemShut {NoStop}%
\bibitem [{\citenamefont {Hubig}\ \emph {et~al.}()\citenamefont {Hubig},
  \citenamefont {Lachenmaier}, \citenamefont {Linden}, \citenamefont
  {Reinhard}, \citenamefont {Stenzel}, \citenamefont {Swoboda}, \citenamefont
  {Grundner},\ and\ \citenamefont {Mardazad}}]{hubig:_syten_toolk}%
  \BibitemOpen
  \bibfield  {author} {\bibinfo {author} {\bibfnamefont {C.}~\bibnamefont
  {Hubig}}, \bibinfo {author} {\bibfnamefont {F.}~\bibnamefont {Lachenmaier}},
  \bibinfo {author} {\bibfnamefont {N.-O.}\ \bibnamefont {Linden}}, \bibinfo
  {author} {\bibfnamefont {T.}~\bibnamefont {Reinhard}}, \bibinfo {author}
  {\bibfnamefont {L.}~\bibnamefont {Stenzel}}, \bibinfo {author} {\bibfnamefont
  {A.}~\bibnamefont {Swoboda}}, \bibinfo {author} {\bibfnamefont
  {M.}~\bibnamefont {Grundner}},\ and\ \bibinfo {author} {\bibfnamefont
  {S.}~\bibnamefont {Mardazad}},\ }\href {https://syten.eu} {\bibinfo {title}
  {The \textsc{SyTen} toolkit}}\BibitemShut {NoStop}%
\bibitem [{\citenamefont {Hubig}\ \emph {et~al.}(2015)\citenamefont {Hubig},
  \citenamefont {McCulloch}, \citenamefont {Schollw\"ock},\ and\ \citenamefont
  {Wolf}}]{PhysRevB.91.155115}%
  \BibitemOpen
  \bibfield  {author} {\bibinfo {author} {\bibfnamefont {C.}~\bibnamefont
  {Hubig}}, \bibinfo {author} {\bibfnamefont {I.~P.}\ \bibnamefont
  {McCulloch}}, \bibinfo {author} {\bibfnamefont {U.}~\bibnamefont
  {Schollw\"ock}},\ and\ \bibinfo {author} {\bibfnamefont {F.~A.}\ \bibnamefont
  {Wolf}},\ }\bibfield  {title} {\bibinfo {title} {Strictly single-site {DMRG}
  algorithm with subspace expansion},\ }\href
  {https://doi.org/10.1103/PhysRevB.91.155115} {\bibfield  {journal} {\bibinfo
  {journal} {Phys. Rev. B}\ }\textbf {\bibinfo {volume} {91}},\ \bibinfo
  {pages} {155115} (\bibinfo {year} {2015})}\BibitemShut {NoStop}%
\end{thebibliography}%
\end{document}